\documentclass[published]{aastex631}

\newcommand{\rev}[1]{\textcolor{black}{#1}}

\usepackage{wrapfig}
\usepackage{subfigure}

\begin{document}

\title{A Catalogue of Galactic Supernova Remnants and Supernova Remnant Candidates from the EMU/POSSUM Radio Sky Surveys. I.}

\author[0009-0003-2088-9433]{B. D. Ball}
\affiliation{Department of Physics, University of Alberta, Edmonton, Alberta, T6G 2E1, Canada}
\correspondingauthor{Brianna D. Ball}
\email{bdball@ualberta.ca}

\author[0000-0001-5953-0100]{R. Kothes}
\affiliation{Dominion Radio Astrophysical Observatory, Herzberg Astronomy \& Astrophysics, National Research Council \\Canada, P.O. Box 248, Penticton, BC V2A 6J9, Canada}
\affiliation{Department of Physics, University of Alberta, Edmonton, Alberta, T6G 2E1, Canada}

\author[0000-0002-5204-2259]{E. Rosolowsky}
\affiliation{Department of Physics, University of Alberta, Edmonton, Alberta, T6G 2E1, Canada}

\author[0000-0002-7239-2248]{C. Burger-Scheidlin}
\affiliation{Astronomy \& Astrophysics Section, School of Cosmic Physics, Dublin Institute for Advanced Studies, DIAS Dunsink Observatory, Dublin D15 XR2R, Ireland}

\author[0000-0002-4990-9288]{M. D. Filipovi\'c}
\affiliation{Western Sydney University, Locked Bag 1797, Penrith South DC, NSW 2751, Australia}

\author[0000-0001-6109-8548]{S. Lazarevi\'c}
\affiliation{Western Sydney University, Locked Bag 1797, Penrith South DC, NSW 2751, Australia}
\affiliation{CSIRO Space and Astronomy, Australia Telescope National Facility, PO Box 76, Epping, NSW 1710, Australia}
\affiliation{Astronomical Observatory, Volgina 7, 11060 Belgrade, Serbia}

\author[0009-0009-7061-0553]{Z. J. Smeaton}
\affiliation{Western Sydney University, Locked Bag 1797, Penrith South DC, NSW 2751, Australia}

\author[0000-0003-1173-6964]{W. Becker}
\affiliation{Max-Planck-Institut f\"ur extraterrestrische Physik, Giessenbachstra{\ss}e, 85741 Garching, Germany\\}
\affiliation{Max-Planck-Institut f\"ur Radioastronomie, Auf dem H\"ugel 69, 53121 Bonn, Germany}

\author[0000-0002-3973-8403]{E. Carretti}
\affiliation{INAF -- Istituto di Radioastronomia, Via Gobetti 101, 40129, Bologna, Italy}

\author[0000-0002-3382-9558]{B. M. Gaensler}
\affiliation{Department of Astronomy and Astrophysics, University of California Santa Cruz, 1156 High Street, Santa Cruz, CA 95064, USA}
\affiliation{Dunlap Institute for Astronomy and Astrophysics, University of Toronto, 50 St George Street, Toronto, ON M5S 3H4, Canada}
\affiliation{David A.\ Dunlap Department of Astronomy and Astrophysics, University of Toronto, 50 St George Street, Toronto, ON M5S 3H4, Canada}

\author[0000-0002-6097-2747]{A. M. Hopkins}
\affiliation{School of Mathematical and Physical Sciences, 12 Wally’s Walk, Macquarie University, NSW 2109, Australia}

\author[0000-0002-4814-958X]{D. Leahy}
\affiliation{Department of Physics and Astronomy, University of Calgary, Calgary, Alberta, T2N 1N4, Canada}

\author[0000-0001-8749-1436]{M. Tahani}
\affiliation{Banting and KIPAC Fellowships: Kavli Institute for Particle Astrophysics \& Cosmology (KIPAC), Stanford University, Stanford, CA 94305, USA}

\author[0000-0001-7722-8458]{J. L. West}
\affiliation{Dominion Radio Astrophysical Observatory, Herzberg Astronomy \& Astrophysics, National Research Council \\Canada, P.O. Box 248, Penticton, BC V2A 6J9, Canada}

\author[0000-0002-6243-7879]{C. S. Anderson}
\affiliation{Research School of Astronomy \& Astrophysics, The Australian National University, Canberra ACT 2611, Australia}

\author[0000-0001-5126-1719]{S. Loru}
\affiliation{INAF -- Osservatorio Astrofisico di Catania, Via Santa Sofia 78, 95123 Catania, Italy}

\author[0000-0003-0742-2006]{Y. K. Ma}
\affiliation{Research School of Astronomy \& Astrophysics, The Australian National University, Canberra ACT 2611, Australia}

\author[0000-0003-2730-957X]{N. M. McClure-Griffiths}
\affiliation{Research School of Astronomy \& Astrophysics, The Australian National University, Canberra ACT 2611, Australia}

\author[0000-0001-9033-4140]{M. J. Michałowski}
\affiliation{Astronomical Observatory Institute, Faculty of Physics and Astronomy, Adam Mickiewicz University, ul. S loneczna 36, 60-286 Pozna\'n, Poland}

\begin{abstract}
We use data from the EMU \rev{(Evolutionary Map of the Universe)} and POSSUM \rev{(Polarization Sky Survey of the Universe's Magnetism)} radio southern sky surveys, conducted with the Australian Square Kilometre Array Pathfinder (ASKAP), to compile a catalogue of Galactic supernova remnants (SNRs) and candidate SNRs within the region of \(277.5^\circ \leq \ell \leq 311.7^\circ\) Galactic longitude, \(|b| \leq 5.4^\circ\) Galactic latitude, as well as an additional field along the Galactic plane, approximately \(315.5^\circ \leq \ell \leq 323.0^\circ\) Galactic longitude, \rev{\(-4.5^\circ \leq b \leq 1.5^\circ\) Galactic latitude.} In the areas studied, there are 44 known SNRs and 46 SNR candidates that have been previously identified in the radio. We confirm eight of these candidates as SNRs based on evidence of linear polarization or through the calculation of nonthermal spectral indices. Additionally, we identify possible radio counterparts for seven SNR candidates that were previously only identified in X-rays (four) or optical (three). We also present \rev{six new SNRs and 37 new SNR candidates.} The results of this study demonstrate the utility of ASKAP for discovering new and potential SNRs and refining the classification of previously identified candidates. In particular, we find that the EMU and POSSUM surveys are particularly well suited for observing high-latitude SNRs and confirming SNR candidates with polarization. The region studied in this work represents approximately one-quarter of the Galactic plane, by longitude, that will eventually be surveyed by EMU/POSSUM and we expect that the ongoing surveys will continue to uncover new SNRs and SNR candidates. 
\end{abstract}


\section{Introduction} \label{sec:intro}

Supernovae and their remnants are the mechanism by which a star's energy and the heavy elements produced in its core are dispersed back into the interstellar medium (ISM). They play an important role in the stellar feedback cycle, as the heavy elements are recycled to form the next generation of stars. These large releases of matter and energy also impact large-scale galactic dynamics and evolution \citep{Fierlinger2016}. 

Within our own Galaxy, it is widely documented that there is a significant discrepancy between the number of supernova remnants (SNRs) that have been observed at radio frequencies and the number that models predict we should be able to detect \citep{Brogan2006,Helfand2006,Green2014,Green2015,Ranasinghe2022,Ball2023}, though there is debate over the size of this discrepancy. 
Interestingly, there is no discrepancy for X-ray-detected SNRs \citep{Leahy2020}, which sample the younger part of the SNR population.
The Galactic supernova rate is well established, with a supernova expected to occur approximately once every 30--50 years \citep{Tammann1994, Li2011}. However, the exact size of the radio-observable Galactic SNR population is difficult to determine because of the effect of environmental variations on radio-observable lifetime and differences in observational capabilities across the Galactic plane. Most estimates agree that there should be at least 1000 Galactic SNRs visible in the radio at any given time, and potentially many more, with estimates ranging up to a few thousand \citep{Ranasinghe2022}. Currently, only 300--400 sources have been confidently classified as Galactic SNRs \citep{Green2024,Ferrand2012}, though many more candidate SNRs have been proposed but require further observations to be confirmed \citep{Whiteoak1996,Duncan1997,Brogan2006,Kothes2006,Green2014,HurleyWalker2019b,Dokara2021,Ball2023,Anderson2024}. Based on \cite{Green2024_cat}, there are at least 364 radio SNR candidates and 56 SNR candidates at other wavelengths that have been proposed but not yet confirmed. Compiling a more complete catalogue of Galactic SNRs is required to better understand this population of objects and the significant influence they may have on the evolution of the Galaxy.

While SNRs can produce emission across the entire electromagnetic spectrum, they are most commonly discovered in the radio, particularly within our Galaxy where line-of-sight extinction effects can be significant at shorter wavelengths. Radio observations allow us to probe deeper into the inner Galaxy and thus, the development of more sophisticated radio telescopes plays an important role in discovering new SNRs. In particular, improvements in the resolution and sensitivity of radio telescopes have continued to result in detections of new sources. We expect the missing Galactic SNR population to be mostly comprised of medium to low surface brightness sources. These are particularly difficult to detect when they are located near other sources of radio emission, typically \ion{H}{2} regions or brighter SNRs. We also suspect there are SNRs at high latitudes that have not yet been detected, since these latitudes are not covered in surveys that focus only on the Galactic plane. Improvements in resolution may also allow for the detection of SNRs of small angular size, though we do not expect to find these in large numbers as the young Galactic SNR population is believed to be mostly complete \citep{Leahy2020,Ranasinghe2021}. Confirmation of previously identified candidates also represents a significant challenge to resolving the Galactic SNR population discrepancy. Doing so generally relies on either the detection of linear polarization or making observations across a range of frequencies so that a spectral index can be reliably determined. The EMU (Evolutionary Map of the Universe) \citep{Norris2011, Norris2021,Hopkins_inprep} and POSSUM (Polarization Sky Survey of the Universe's Magnetism) \citep{Gaensler2010,Gaensler_inprep} surveys have already proved to be useful tools for confirming sources as SNRs, particularly at high latitudes with remnants like G288.8$-$6.3 \citep{Filipovic2023} and \rev{G278.9+1.3} \citep{Filipovic2024}. When possible, we look for evidence of polarization or calculate spectral indices with the goal of confirming candidates as SNRs. 

Here, we present a catalogue of known Galactic SNRs, known SNR candidates, and new SNR candidates observed in the first part of the EMU and POSSUM radio sky surveys, which are being conducted with the Australian Square Kilometre Array Pathfinder (ASKAP) \citep{Hotan2021}. When complete, these surveys will cover almost the entire southern sky, encompassing the Galactic plane spanning approximately \(220^\circ \leq \ell \leq 18^\circ\) Galactic longitude. This work is an extension of \cite{Ball2023}, a catalogue of the known SNRs and SNR candidates in the EMU/POSSUM Galactic pilot field, to the full EMU/POSSUM surveys. These surveys are currently ongoing, and observations of the Galactic plane are not yet complete. This work covers the region of \(277.5^\circ \leq \ell \leq 311.7^\circ\) Galactic longitude, \(|b| \leq 5.4^\circ\) Galactic latitude, as well as an additional field of the Galactic plane of \(315.5^\circ \leq \ell \leq 323.0^\circ\) Galactic longitude, \(-4.5^\circ \leq b \leq 1.5^\circ\) Galactic latitude. This represents approximately one-quarter of the portion of the Galactic plane, by degrees longitude, that will eventually be covered by the EMU and POSSUM surveys.

In Section \ref{sec:obs} we describe the data used in this paper. Section \ref{sec:methods} outlines the techniques we use to identify, and in some cases confirm, SNR candidates. Section \ref{sec:results} covers the results, including lists of known SNRs, known SNR candidates, and new SNRs/SNR candidates within the surveyed region. In Section \ref{sec:disc} we discuss the implications of these results and present updated statistics. Section \ref{sec:conc} summarizes our conclusions.

\section{Observations} \label{sec:obs}

\begin{figure}
    \centering
    \subfigure{\includegraphics[width=\textwidth]{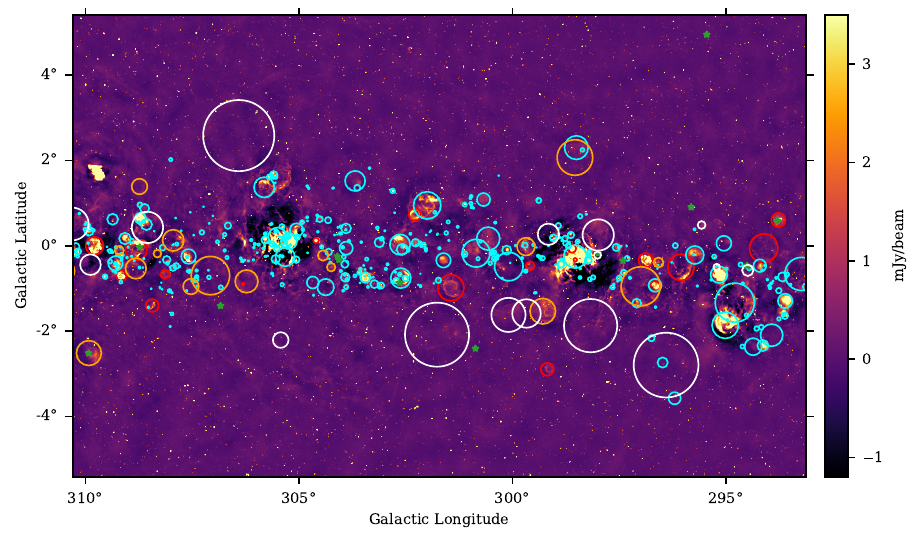}}
    \subfigure{\includegraphics[width=\textwidth]{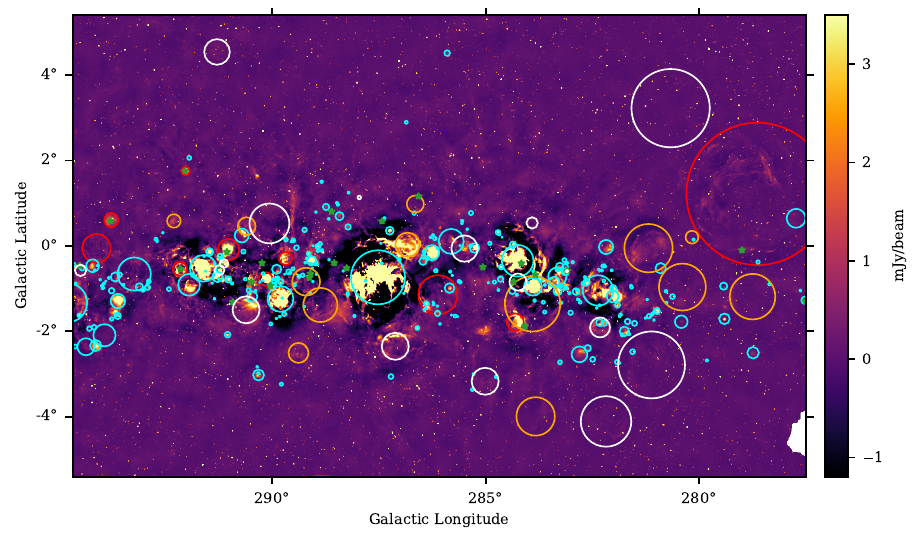}}
    \caption{ASKAP 943 MHz radio continuum maps of the \rev{contiguous} surveyed region of the Galactic plane. Annotations indicate the locations of known SNRs (red), known SNR candidates (orange), new SNR candidates (white), known \ion{H}{2} regions (cyan), and young (characteristic age $\leq$ 500 kyrs) pulsars (green stars).}
    \label{fig:field}
\end{figure}

\subsection{EMU/POSSUM Observations with ASKAP}

\begin{wraptable}{l}{8cm}
    \centering
    \begin{tabular}{c|c|c|c}
        \hline
        Field & SB & Longitude & Latitude \\
        \hline
        EMU0936-60 & 53568 & 280.91 & $-$6.11 \\
        EMU0954-55 & 51428 & 279.74	& $-$1.06 \\
        EMU1003-64 & 46974 & 286.34	& $-$7.59 \\
        EMU1005-51 & 54774 & 278.30	& 3.64 \\
        EMU1017-60 & 46953 & 284.93	& $-$2.97 \\
        EMU1029-55 & 46915 & 283.90	& 1.83 \\
        EMU1050-64 & 54771 & 290.54	& $-$5.00 \\
        EMU1058-60 & 60336 & 289.36	& $-$0.47 \\
        EMU1106-55 & 64412 & 288.41	& 4.14 \\
        EMU1136-64 & 46948 & 295.12	& $-$3.18 \\
        EMU1139-60 & 60586 & 294.13	& 1.31 \\
        EMU1141-55 & 47138 & 293.21	& 5.82 \\
        EMU1220-60 & 61083 & 299.12	& 2.31 \\
        EMU1223-64 & 55363 & 299.94	& $-$2.21 \\
        EMU1301-60 & 45830 & 304.20	& 2.52 \\
        EMU1309-64 & 62225 & 304.86	& $-$2.11 \\
        EMU1342-60 & 54095 & 309.26	& 1.91 \\
        EMU1356-64 & 53310 & 309.72	& $-$2.90 \\
        EMU1442-64 & 45821 & 314.36	& $-$4.54 \\
        EMU1505-60 & 46964 & 318.76	& $-$1.63 \\
        \hline
    \end{tabular}
    \caption{EMU/POSSUM fields used in this work. The fields are named based on the right ascension and declination of the field centre. We also list scheduling block (SB) numbers and the field centre in Galactic coordinates. }
    \label{tab:fields}
\end{wraptable}

We use data from ASKAP observations \citep{Hotan2021}, specifically the EMU \citep{Norris2011, Norris2021,Hopkins_inprep} and POSSUM \citep{Gaensler2010,Gaensler_inprep} radio sky surveys. EMU and POSSUM are being observed commensally and will eventually cover the entire southern sky. ASKAP is an interferometer comprised of 36 12-meter dishes with baselines of up to 6 km. Each ASKAP antenna is equipped with a phased array feed, allowing it to simultaneously form 36 beams. This gives ASKAP a wide field of view, approximately 30 deg$^2$, making it a particularly effective tool for conducting large-scale surveys. 

The Stokes I images are from the EMU survey, with a bandwidth of 288~MHz that is centred at 943~MHz. Each field is formed from a mosaic of 36 beams and convolved to a common angular resolution of 15$\arcsec$ with a median rms noise of 30 $\mu$Jy beam$^{-1}$ \citep{Hopkins_inprep}. Imaging is performed using the standard ASKAPsoft pipeline \citep{Guzman2019} as described by \cite{Norris2021}. In total, we use data from 20 EMU fields. We use the Stokes Q and U frequency cubes with 1~MHz channels from POSSUM to produce polarized intensity (PI) maps, convolved to $18\arcsec$ resolution. The PI maps presented here were made by de-rotating the Q and U data in each frequency channel for each rotation measure (RM) and taking the peak of the Faraday depth function, as described by \cite{Ball2023}.

This work uses data from 20 EMU/POSSUM fields, listed in Table~\ref{tab:fields}. This represents approximately one-quarter of the Galactic plane, by degrees longitude, that will eventually be covered by these surveys. These observations took place between November 24, 2022 and August 7, 2024. All of these fields, except EMU1505-60, are adjacent to one another and form a contiguous region. The image shown in Figure~\ref{fig:field} is a mosaic of the connected fields. Figure~\ref{fig:field_pi} shows the same region in PI. All data are available on the CSIRO ASKAP Data Science Archive (CASDA)\footnote{\url{https://research.csiro.au/casda/}}.

\subsection{Mid-Infrared from WISE}
To identify SNRs and SNR candidates, we rely on the comparison of radio and mid-infrared (MIR) fluxes. This technique has been used in many other surveys to identify SNR candidates \citep{Whiteoak1996,Brogan2006,Green2014,Dokara2021,Ball2023,Anderson2024}. Comparing radio and MIR fluxes helps to distinguish SNRs from thermal sources of radio emission, most of which are \ion{H}{2} regions. \ion{H}{2} regions produce strong MIR emission through warm dust and polycyclic aromatic hydrocarbons (PAHs). SNRs can also produce MIR emission through similar processes but this emission is relatively weak so they are generally not detected at infrared frequencies. We compare the 943 MHz radio observations from ASKAP to 12 and 22 $\mu$m MIR data from WISE (Wide-field Infrared Survey Explorer) \citep{Wright2010}, with $6.5\arcsec$ and $12\arcsec$ resolution, respectively. These wavelengths are good tracers of hot dust and PAHs, which are found abundantly in \ion{H}{2} regions \citep{Anderson2014} but are largely absent in SNRs. 

\subsection{Ancillary Radio Data}

Spectral indices can help to confirm candidates as SNRs by distinguishing them from thermal sources of radio emission. Calculating spectral indices requires determining reliable flux densities at two or more radio frequencies. This can be done by dividing the 288 MHz wide band we observe with ASKAP into smaller bands or by using data from other radio surveys. In this paper, we calculate spectral indices for several candidates utilizing data from the following surveys. 

The SARAO MeerKAT Galactic Plane Survey (SMGPS) is a radio continuum survey of the Galactic plane centred at 1360~MHz. It covers Galactic longitudes of \(251^\circ \leq \ell \leq 358^\circ\) and \(2^\circ \leq \ell \leq 61^\circ\) at latitudes of \(|b|\leq 1.5^\circ\) \citep{Goedhart2024}. When possible, we use flux densities calculated with SMGPS data to determine spectral indices for SNR candidates. However, there are challenges with relying only on fluxes determined from ASKAP and MeerKAT, later described in Section~\ref{sec:known_indices}, so this is only possible for a few candidates.

When possible, we also use data from the Sydney University Molonglo Sky Survey (SUMSS), a radio survey conducted with The Molonglo Observatory Synthesis Telescope (MOST) centred at 843~MHz \citep{Bock1999}. However, most of our candidates are not visible in SUMSS due to limitations in resolution and sensitivity so this is only possible for a few cases. For one candidate, we calculate a 198~MHz flux using data from the GLEAM (GaLactic and Extra-galactic All-sky MWA) survey \citep{HurleyWalker2019c}.

\section{Methods} \label{sec:methods}

\begin{figure}
    \centering
    \subfigure{\includegraphics[width=\textwidth]{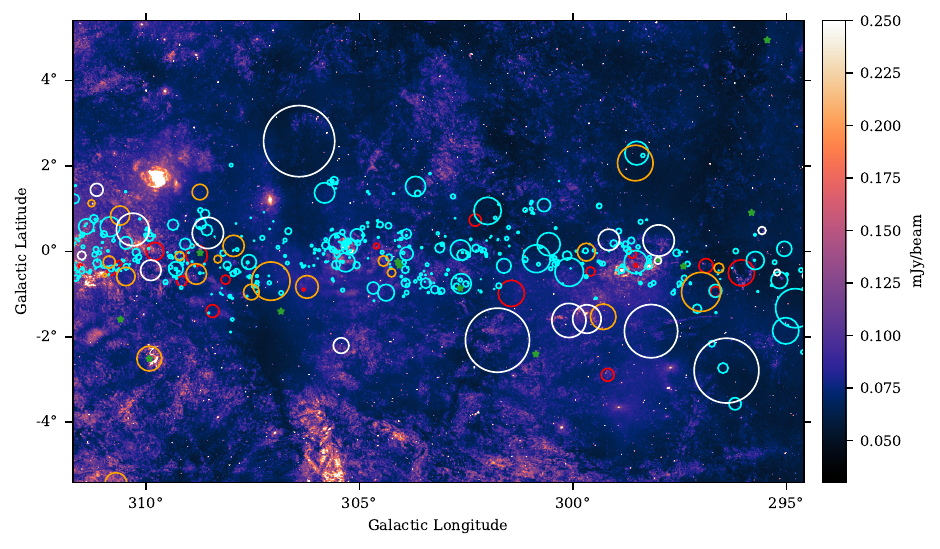}}
    \subfigure{\includegraphics[width=\textwidth]{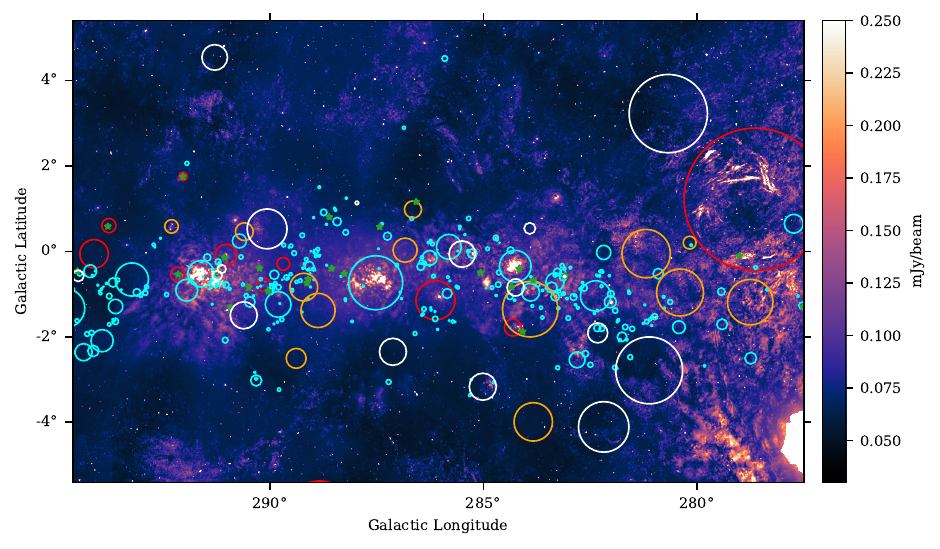}}
    \caption{ASKAP 943 MHz PI maps of the same region shown in Figure~\ref{fig:field}. Annotations indicate the locations of known SNRs (red), known SNR candidates (orange), new SNR candidates (white), known \ion{H}{2} regions (cyan), and young (characteristic age $\leq$ 500 kyrs) pulsars (green stars).}
    \label{fig:field_pi}
\end{figure}

\subsection{Identifying SNRs and SNR Candidates}

SNRs produce most of their radio emission through nonthermal synchrotron radiation. Particles are accelerated to relativistic speeds at the shock front through diffusive shock acceleration (DSA) and interact with the ambient magnetic field, which has been compressed and amplified by the shock \citep{vanderlaan1962,Filipovic1}. Since most of the radio emission is produced in a thin shell at the shock front, we typically expect SNRs to appear as well-defined limb-brightened structures in the radio. Some SNRs exhibit a centre-filled morphology, where the majority of the emission is produced by a pulsar wind nebula (PWN), but this is less common. These sources can be difficult to identify by morphology alone as they are more visually similar to thermal sources of radio emission. In other cases, we observe a composite morphology where both the shell and the PWN are visible. \rev{SNRs that do not exhibit shell-like features are more likely to be missed with this methodology, unless they show polarization or are associated with a young pulsar.}

\begin{figure}
    \centering
    \includegraphics[width=\textwidth]{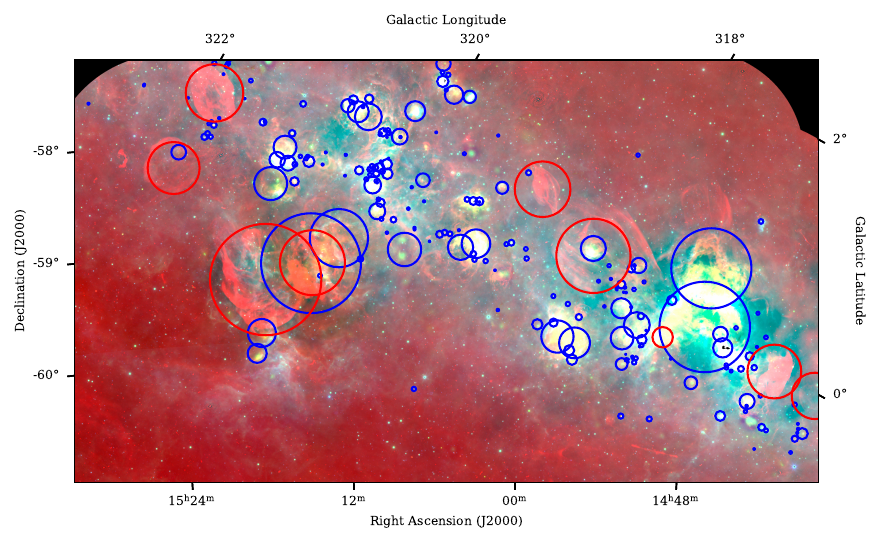}
    \caption{Part of field EMU1505$-$60 with 943~MHz radio from ASKAP in red, 12~$\mu$m MIR from WISE in blue, and 22~$\mu$m MIR from WISE in green. Annotations indicate the locations of known SNRs (red) and known \ion{H}{2} regions (blue). To identify SNR candidates, we look for shell-like structures in the radio that lack MIR counterparts.}
    \label{fig:emu1505-60}
\end{figure}

The primary challenge with identifying new SNRs and SNR candidates is confusion with other radio-emitting nebulae, predominantly \ion{H}{2} regions. We use the WISE \ion{H}{2} region catalogue \citep{Anderson2014} to rule out all known \ion{H}{2} regions, which make up the majority of extended radio sources in the Galactic plane. We compile a list of SNR candidates by scanning the unidentified sources of radio emission for characteristic SNR morphology, i.e. shell-like structures, that lack MIR counterparts. As shown in Figure~\ref{fig:emu1505-60}, this technique effectively distinguishes SNRs from \ion{H}{2} regions. While \ion{H}{2} regions are expected to be very bright in the MIR, SNRs are relatively faint and often completely undetectable \citep{Whiteoak1996,Pinheiro2011}. This is because \ion{H}{2} regions are abundant in hot dust and gas, as well as large MIR-emitting molecules like PAHs, while most of the dust found in SNRs is relatively cold or destroyed by the passage of the reverse shock \citep{Priestly2022}. Some SNRs do have MIR counterparts, with detections being more common in younger SNRs where the dust has not yet been destroyed by the passage of the reverse shock \citep{Saken1992,Reach2006,Chawner2020}. Thus, while the absence of an MIR counterpart indicates that a source is not an \ion{H}{2} region, the presence of an MIR counterpart is inconclusive. \rev{Additionally, if an SNR is spatially coincident with an HII region, the absence of an MIR counterpart may be difficult to establish, making this technique less effective for identifying SNRs in the denser regions of the Galactic plane.}

After compiling a list of SNR candidates, we then look for further evidence of nonthermal synchrotron emission, such as linear polarization or a steep, negative spectral index \citep{Filipovic2}.

\subsubsection{Young Pulsars}
A young pulsar is evidence that a supernova has recently occurred; therefore, if a candidate SNR is spatially coincident with a young pulsar, this increases our confidence that the source may be an SNR. This is particularly useful in looking for PWNe, which can be more difficult to visually distinguish from other radio sources than shell-type SNRs. We use the ATNF pulsar catalogue (Version 2.3.0\footnote{\url{https://www.atnf.csiro.au/research/pulsar/psrcat/index.php?version=2.3.0}}) \citep{Manchester2005} to identify the locations of young pulsars and look for radio features that may be associated. The radio-observable lifetimes of SNRs span a wide range and depend on properties of the environment, such as local ISM density, as well as the frequency of observations. Generally, the average visible lifetime is of the order of tens of thousands of years \citep{Frail1994,Sarbadhicary2017}, though it can be longer in certain conditions, for example, if the expansion is occurring in a low-density environment. Thus, we restrict our search to pulsars younger than 500 kiloyears (kyrs). Above this age, it is highly unlikely that we would be able to detect the host SNR. We inspect the regions surrounding these young pulsars for SNR-like morphologies. 

\subsubsection{Spectral Indices}
The spectral index describes how flux density changes with frequency and can be used to differentiate sources of thermal and nonthermal emission. We define the spectral index $\alpha$ as \(S(\nu) \propto \nu^\alpha\) where \(S(\nu)\) is the flux density at frequency $\nu$. The radio spectral index of synchrotron emission reflects the energy spectrum of the relativistic particles. At the SNR shock front, the particles are accelerated through DSA, which typically results in a radio spectral index of \(\alpha = -0.5\). Indeed, most shell-type SNRs have indices within the range \(\alpha = -0.5 \pm 0.2\) \citep{Dubner2017,Bozzetto2023,Cotton2024}. PWNe often have flatter indices, typically in the range \(-0.3 \leq \alpha \leq 0.0\), though they can be as steep as \(-0.7\) \citep{Kothes2017}. Thermally emitting \ion{H}{2} regions typically have flat spectral indices of \(\alpha = -0.1\), characteristic of optically thin free-free emission. They can also have steep positive spectral indices of up to \(\alpha = +2\), moving toward the optically thick regime at low frequencies.

Determining a spectral index requires reliable flux densities for at least two different frequencies, ideally more. One possible approach is to break the 288~MHz wide band we observe with ASKAP into smaller bands, and determine flux densities for each of the channels. However, a small frequency range means that small uncertainties in the flux densities can result in significant uncertainties in the index. Thus, this method is only reliable when the source is bright relative to its background. This is a significant obstacle for this study since we expect new SNRs and SNR candidates to be quite faint. Alternatively, we may rely on data from other surveys to determine an index. The recent SMGPS \citep{Goedhart2024} covers this region of the Galactic plane and may be used to determine indices for some sources. However, because the survey is focused on the Galactic plane, there is limited latitude coverage, and many of our candidates are not covered. Additionally, the SMGPS is centred at 1360~MHz so the frequency range is still quite small. We attempt to determine spectral indices for known SNRs and candidate SNRs that are covered by both surveys. We explore the feasibility of this approach and discuss the challenges we encounter in Section~\ref{sec:known_indices}. We also use data from SUMSS \citep{Bock1999} and GLEAM \citep{HurleyWalker2019c} when possible, but we are limited by the resolution and sensitivity of these surveys. 

\subsubsection{Polarization}

In the radio, SNRs primarily produce nonthermal synchrotron emission, which is expected to be linearly polarized \citep{Rybicki1986}. Thermally emitting sources, like \ion{H}{2} regions, should not show evidence of polarization, though we do expect to see instrumental effects for sources that are very bright. The maximum theoretical degree of linear polarization we can observe from an SNR is dependent on its spectral index. For a typical SNR with an index of \(\alpha = -0.5\), we expect the intrinsic degree of polarization to be around 70\% \citep{Longair2011}. However, observed polarization fractions are generally lower than this value. Turbulence in the magnetic field of the emitting region can result in wavelength-independent depolarization through destructive interference \citep{Burn1966}. A similar effect can occur within the beam of the telescope, but we do not expect this to be a significant source of depolarization here as the sources we study are well resolved. Since we are observing at a low frequency, we expect the most significant source of depolarization to be differential Faraday rotation \citep{Longair2011}. This wavelength-dependent effect can be produced by foreground magnetic fields or by the SNR itself due to the intermixing of synchrotron-emitting and Faraday-rotating plasmas \citep{Burn1966}. Because of these depolarizing processes, failure to detect polarization should not be considered evidence that a source is not an SNR, particularly for sources within the Galactic plane. In this study, we note that we most often see evidence of linear polarization in high-latitude SNRs. This indicates that the Galactic plane likely has a significant depolarizing effect. Within the densest parts of the Galactic plane, we can typically only detect polarization from the brightest SNRs. 

It is also important to distinguish between real and instrumental polarization. Bright \ion{H}{2} regions can appear to be polarized due to the presence of instrumental effects, as illustrated in Figure~\ref{fig:field_pi}. Instrumental polarization is produced due to leakage from Stokes I to Stokes Q and U. The typical instrumental polarization from ASKAP fields is typically around 0.2\% \citep{Gaensler_inprep}. Polarization that appears smooth and that exactly matches the total power structure is more likely to be instrumental. For evidence of real polarization, we look for structures with a speckled appearance, which indicates a change in the RM on small scales. This effect cannot be produced instrumentally. This also helps to distinguish polarized emission from an extended source from the large-scale diffuse Galactic features, which can also be seen in Figure~\ref{fig:field_pi}. 

Because of all of these factors, we do not expect to detect polarization from most of our SNR candidates, particularly those that are very faint or located within the Galactic plane. These expectations are consistent with the results of previous work \citep{Dokara2018,Ball2023}. However, when we do detect real polarization from an extended radio source, this can be taken as strong evidence that the source is an SNR. For sources that we find to be polarized, we calculate the peak percentage polarization by dividing the PI flux by the Stokes I flux for each pixel in the polarized region, taking the pixel with the largest value. If the peak percentage polarization is $>$1\%, we can be confident that the polarization is not instrumental. Because of issues with missing flux in the Stokes I data, we sometimes find peak percentage polarization greater than the maximum theoretical value of $\sim$70\% and sometimes even greater than 100\%. In these cases, we list the peak percentage polarization as $>$70\% as we only intend to use these values to prove that the polarization is not instrumental. These values can be found listed in Tables~\ref{tab:knownSNRs}, \ref{tab:known_cands}, \ref{tab:xray_opt_cands}, and \ref{tab:newcands} for sources that we observe to be polarized. 

\subsection{Flux Integration} \label{sec:flux}

Flux density calculations are performed using the same method as described by \cite{Ball2023}, utilizing the \textsc{Polygon\_Flux} software developed by \cite{HurleyWalker2019} to manually define the source boundary, perform background subtraction, and remove bright point sources when necessary. Calculations are performed at least six times per source, with slightly different definitions of the source perimeter and sampled background region. The values provided are the median flux densities from these calculations, and the uncertainties are determined by the range between the extrema. Thus, these errors reflect the uncertainties in the definitions of the source boundary and the sample of the background fluctuations. The flux densities calculated for known SNRs at 943~MHz are provided in Table~\ref{tab:knownSNRs}, where they may be compared to the 1~GHz value listed in the \cite{Green2024_cat} catalogue. Comparing our calculated flux densities to the catalogued values allows us to assess their reliability. 

For many of the known and new SNR candidates, flux densities cannot be reliably determined. This may be because only a partial shell is visible, the source is too faint, or because it is located in a complex region with multiple overlapping bright sources. In many cases, we must also contend with the effect of missing short spacings, resulting in missing flux from the extended diffuse emission. For the frequency range observed with EMU, the sensitivity of ASKAP is limited to spatial scales of around 43$\arcmin$ to 60$\arcmin$ \citep{Hopkins_inprep}. For sources that are around this size or larger, we expect to find missing flux. When possible, we provide flux densities (or limits) for known and new SNR candidates in Tables~\ref{tab:known_cands} and \ref{tab:newcands}. 

\subsection{SNR Candidate Classification}
 
 We include in our catalogue all sources that we identified as possible SNR candidates. We categorize each candidate as a new SNR, strong candidate, or weak candidate based on the strength of the evidence that it is an SNR. We classify our candidates using the following system:
\begin{itemize}
    \item New SNRs: These are sources that have a characteristic SNR radio morphology, no MIR counterpart, and that show evidence of linear polarization or a steep negative spectral index.
    \item Probable SNRs or strong candidates: These sources exhibit a characteristic SNR radio morphology with no MIR counterpart. In some cases, they may have a spatial coincidence with a young pulsar. In a few cases, we were able to calculate a nonthermal spectral index but with only two data points, so the reliability may be questionable. These sources will need further data to confidently classify them as SNRs.
    \item Possible SNRs or weak candidates: Sources that exhibit a radio morphology that is ``SNR-like" with no MIR counterpart. They may be located in complex regions with many overlapping sources, so it is difficult to determine the boundaries of the SNR candidate. In many cases, we find only a faint partial shell. Further data will be needed to determine whether these sources are SNRs.
\end{itemize}

\section{Results} \label{sec:results}

Figure \ref{fig:field} shows the \rev{contiguous} surveyed region of the Galactic plane discussed in this paper, with annotations indicating the locations of known SNRs, known SNR candidates, our new SNR candidates, known \ion{H}{2} regions, and young pulsars. Around the dense clusters of bright \ion{H}{2} regions, we see prominent negative bowls, evidence of missing faint diffuse structures. These image artifacts present a significant challenge for detecting SNR candidates in these densely occupied regions, especially as we are primarily expecting to find low surface brightness sources.

Here, we present the known SNRs, known SNR candidates, and new SNRs/SNR candidates detected with the EMU and POSSUM surveys in the studied region. We find 44 known radio SNRs \citep{Green2024} and 46 previously identified radio SNR candidates. We also find possible radio counterparts for four known X-ray SNRs/SNR candidates \citep{Ferrand2012} and three known optical SNR candidates. Of the known radio SNR candidates, we conclude that at least eight should now be confirmed as SNRs based on evidence of linear polarization and/or nonthermal spectral indices. We also present six new SNRs and 37 new SNR candidates, though we consider 31 of the candidates to be ``weak candidates" as they are very faint or have ambiguous morphologies. For completeness, we include all detected radio shell-like structures that lack MIR counterparts. Many of these candidates will require further study to determine if they are SNRs.

\subsection{Known Supernova Remnants} \label{sec:known_snrs}

\begin{table}[]
    \centering
    \begin{tabular}{cccccccc}
    \hline
    Name & RA & Dec & Size & Flux Density & Flux Density & Pol & Peak \%\\
    & (J2000) & (J2000) & [$'$] & [Jy at 1 GHz]& [Jy at 943 MHz] & & Pol\\
    \hline
    \rev{\textbf{G278.9+1.3}} & \rev{\textbf{09 59 51}} & \rev{\textbf{$-$53 20}} & \textbf{200$\times$194} & 30? & N/A & Y & $>$70\%\\
    G284.3$-$1.8 & 10 18 15 & $-$59 00 & 24? & 11? & 14 $\pm$ 2 & Y & 7\%\\
    \textbf{G286.1$-$1.1} & \textbf{10 32 54} & \textbf{$-$59 25} & \textbf{55} & 1.4? & N/A & N & \\
    G288.8$-$6.3 & \rev{10 30 20} & \rev{$-$65 15} & \rev{108$\times$96} & 11 & 7 $\pm$ 2* & Y & $>$70\%\\
    G289.7$-$0.3 & 11 01 15 & $-$60 18 & 18$\times$14 & 6.2 & 6.1 $\pm$ 0.3 & N & \\
    G290.1$-$0.8 & 11 03 05 & $-$60 56 & 19$\times$14 & 42 & 42 $\pm$ 1 & N & \\
    \textbf{G291.0+0.1} & \textbf{11 12 17} & \textbf{$-$60 29} & \textbf{30$\times$24} & 16 & 22 $\pm$ 2 & N\\
    G292.0+1.8 & 11 24 36 & $-$59 16 & 12$\times$8 & 15 & 14.5 $\pm$ 0.2 & Y & 7\%\\
    G292.2$-$0.5 & 11 19 20 & $-$61 28 & 20$\times$15 & 7 & 5.2 $\pm$ 0.3* & N \\
    G293.8+0.6 & 11 35 00 & $-$60 54 & 20 & 5? & 4.2 $\pm$ 0.1 & Y & 7\%\\
    G294.1$-$0.0 & 11 36 10 & $-$61 38 & 40	& $>$2? & 2.9 $\pm$ 0.5* & N\\
    G296.1$-$0.5 & 11 51 10 & $-$62 34 & 37$\times$25 & 8? & N/A & Y & 32\%\\
    G296.7$-$0.9 & 11 55 30 & $-$63 08 & 15$\times$8 & 3 & 3.1 $\pm$ 0.1 & Y & 8\%\\
    G296.8$-$0.3 & 11 58 30 & $-$62 35 & 20$\times$14 & 9 & 9.5 $\pm$ 0.2 & Y & 4\%\\
    G298.5$-$0.3 & 12 12 40  & $-$62 52 & 5? & 5? & N/A & N\\
    G298.6$-$0.0 & 12 13 41 & $-$62 37 & 12$\times$9 & 5? & 7.8 $\pm$ 0.6 & N\\
    G299.2$-$2.9 & 12 15 13 & $-$65 30 & 18$\times$11 & 0.5? & 0.66 $\pm$ 0.09 & Y & 41\%\\
    G299.6$-$0.5 & 12 21 45 & $-$63 09 & 13 & 1.0? & 1.3 $\pm$ 0.1 & N\\
    G301.4$-$1.0 & 12 37 55 & $-$63 49 & 37$\times$23 & 2.1? & 3.5 $\pm$ 0.3 & N\\
    G302.3+0.7 & 12 45 55 & $-$62 08 & 17 & 5? & 4.3 $\pm$ 0.6 & N\\
    G304.6+0.1 & 13 05 59 & $-$62 42 & 8 & 14 & 15.2 $\pm$ 0.3 & Y & 2\%\\
    G306.3$-$0.9 & 13 21 50 & $-$63 34 & 4 & 0.16? & 0.164 $\pm$ 0.003 & N\\
    G308.1$-$0.7 & 13 37 37 & $-$63 04 & 13 & 1.2? & 1.6 $\pm$ 0.1 & N\\
    \textbf{G308.5$-$1.4} & \textbf{13 41 37} & \textbf{$-$63 41} & \textbf{18$\times$14} & 0.4? & 1.1 $\pm$ 0.1 & N\\
    G308.8$-$0.1 & 13 42 30 & $-$62 23 & 30$\times$20? & 15? & 16 $\pm$ 2* & Y & 3\%\\
    G309.2$-$0.6 & 13 46 31 & $-$62 54 & 15$\times$12 & 7? & 6.2 $\pm$ 0.1 & Y & 5\%\\
    G309.8+0.0 & 13 50 30 & $-$62 05 & 25$\times$19 & 17 & 14 $\pm$ 1 & N\\
    G310.6$-$0.3 & 13 58 00 & $-$62 09 & 8 & 5? & 5.0 $\pm$ 0.1 & N\\
    G310.6$-$1.6 & 14 00 45 & $-$63 26 & 2.5 & ? & 0.281 $\pm$ 0.003 & Y & 4\%\\
    G310.8$-$0.4 & 14 00 00 & $-$62 17 & 12 & 6? & $<$9 & Y & 2\%\\
    G311.5$-$0.3 & 14 05 38 & $-$61 58 & 5 & 3? & 3.3 $\pm$ 0.2 & Y & 2\%\\
    G312.4$-$0.4 & 14 13 00 & $-$61 44 & 38 & 45 & N/A & N\\
    G312.5$-$3.0 & 14 21 00 & $-$64 12 & 20×18 & 3.5? & 3.5 $\pm$ 0.2 & Y & 30\%\\
    G315.4$-$0.3 & 14 35 55 & $-$60 36 & 24×13 & 8 & N/A & N\\
    G315.4$-$2.3 & 14 43 00 & $-$62 30 & 42 & 49 & 32 $\pm$ 2* & Y& 26\%\\
    G315.9$-$0.0 & 14 38 25 & $-$60 11 & 25×14 & 0.8? & 0.75 $\pm$ 0.04* & N\\
    G316.3$-$0.0 & 14 41 30 & $-$60 00 & 29×14 & 20? & 23.1 $\pm$ 0.4 & Y & 5\%\\
    G317.3$-$0.2 & 14 49 40 & $-$59 46 & 11 & 4.7? & 5.1 $\pm$ 0.3 & Y & 1\%\\
    G318.2+0.1 & 14 54 50 & $-$59 04 & 40×35 & $>$3.9? & N/A & Y & 23\%\\
    G318.9+0.4 & 14 58 30 & $-$58 29 & 30×14 & 4? & 6.2 $\pm$ 0.3 & N\\
    G320.4$-$1.2 & 15 14 30 & $-$59 08 & 35 & 60? & 59 $\pm$ 1 & Y & 23\%\\
    G320.6$-$1.6 & 15 17 50 & $-$59 16 & 60×30 & ? & 14 $\pm$ 3* & Y & 18\%\\
    G321.9$-$0.3 & 15 20 40 & $-$57 34 & 31×23 & 13 & 12.7 $\pm$ 0.5 & N\\
    G321.9$-$1.1 & 15 23 45 & $-$58 13 & 28 & $>$3.4? & 3.9 $\pm$ 0.3 & N\\
    \hline
    \end{tabular}
    \caption{Known SNRs in the surveyed region. RA, Dec, size, and 1~GHz flux density values are taken from \citet{Green2024_cat}. Bolded values have been updated from those provided in the catalogue. Catalogue values with a high degree of uncertainty are indicated with a ``?". *~indicates there is likely missing flux. N/A indicates that a flux density could not be determined. The ``Pol" column indicates if we detect linear polarization that can clearly be associated with the SNR (Y) or not (N). The peak \% pol indicates the maximum polarization percentage observed.}
    \label{tab:knownSNRs}
\end{table}

The list of known SNRs in Table~\ref{tab:knownSNRs} is taken from \cite{Green2024_cat}. When possible, we calculate a flux density at 943~MHz. These can be compared to the given catalogue values at 1~GHz, which are not directly measured but instead derived from the observed spectrum based on published flux values. For the 943~MHz values, ``*"s indicate that the flux density value provided is likely too low, generally as a consequence of missing short spacings resulting in missing flux. N/A indicates that a flux density could not be determined. This is because the source was too faint, it was not fully covered in the EMU survey, or it was located in a complex region with overlapping sources. When possible, we provide upper limits. For four of the entries, we provide updates to the catalogue values for right ascension (RA), declination (Dec), and size. This is because the ASKAP observations have revealed additional shell structures that were not previously detected. The names of these sources are also updated to reflect the new values. The ``Pol" column indicates whether or not we detect linear polarization that can clearly be associated with the SNR. We note that we typically detect polarization when the source is either very bright or at a high latitude above or below the Galactic plane. 

We provide images and comments on specific sources only for the 4 SNRs that required updates to the RA, Dec, and size values. We also provide an image and description for G288.8$-$6.3, an SNR newly confirmed using the EMU survey \citep{Filipovic2023}, and, for the first time, present the source in polarization using data obtained from POSSUM. 

\begin{figure}
    \centering
    \subfigure[\rev{G278.9+1.3} (Previously G279.0+1.1)]{\includegraphics[width=0.42\textwidth]{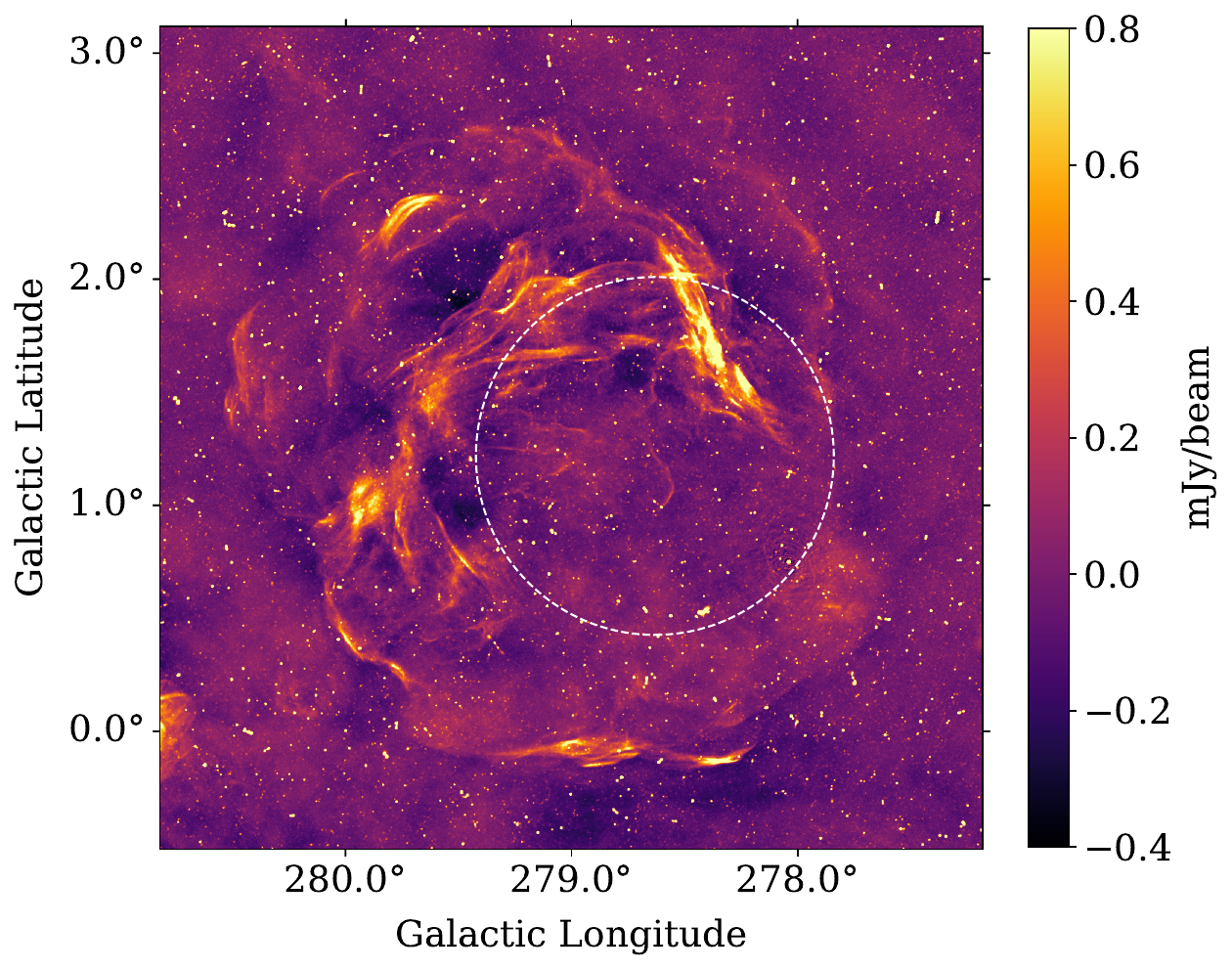}\label{fig:g278.94+1.35}}
    \subfigure[G286.1$-$1.1 (Previously G286.5$-$1.2)]{\includegraphics[width=0.45\textwidth]{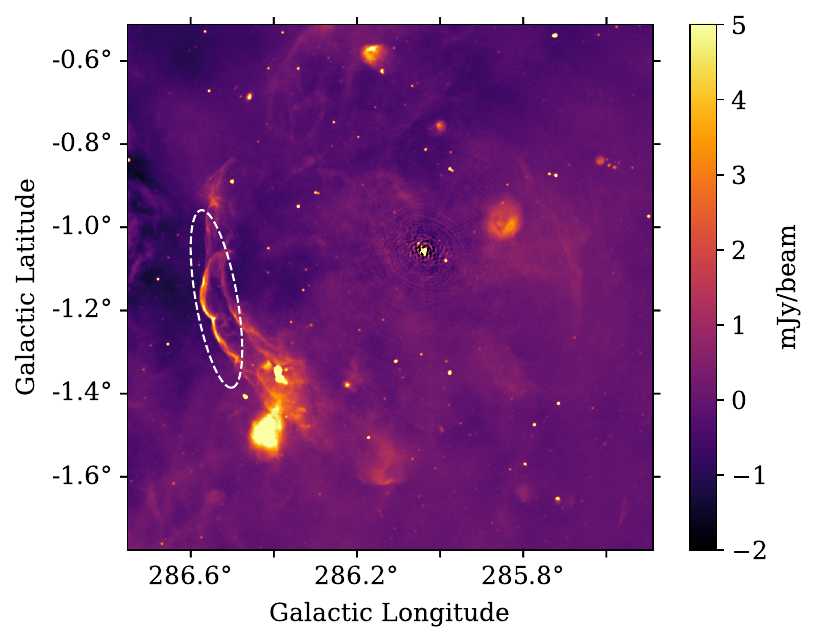}\label{fig:g286.1-1.1}}
    \subfigure[G291.0+0.1 (Previously G291.0$-$0.1)]{\includegraphics[width=0.45\textwidth]{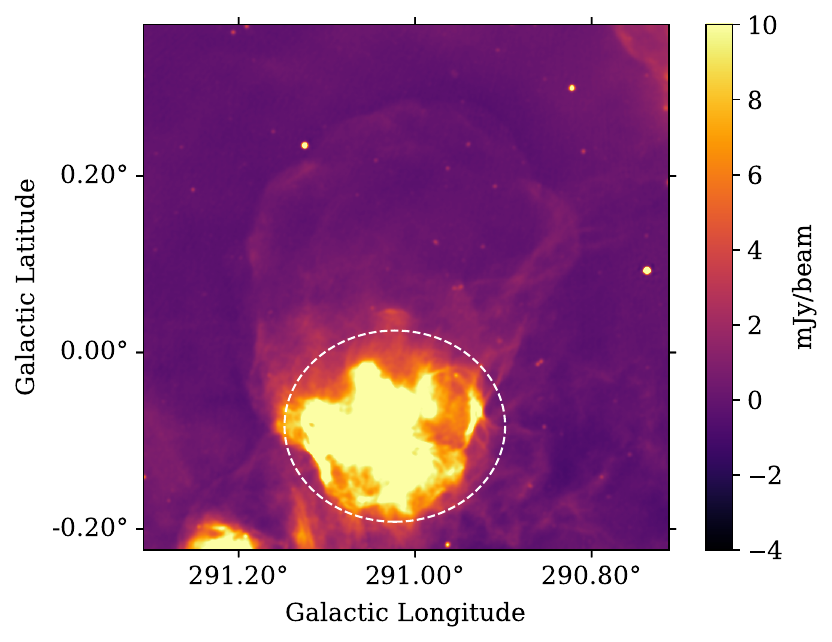}\label{fig:g291.0+0.1}}
    \subfigure[G308.5$-$1.4 (Previously G308.4$-$1.4)]{\includegraphics[width=0.45\textwidth]{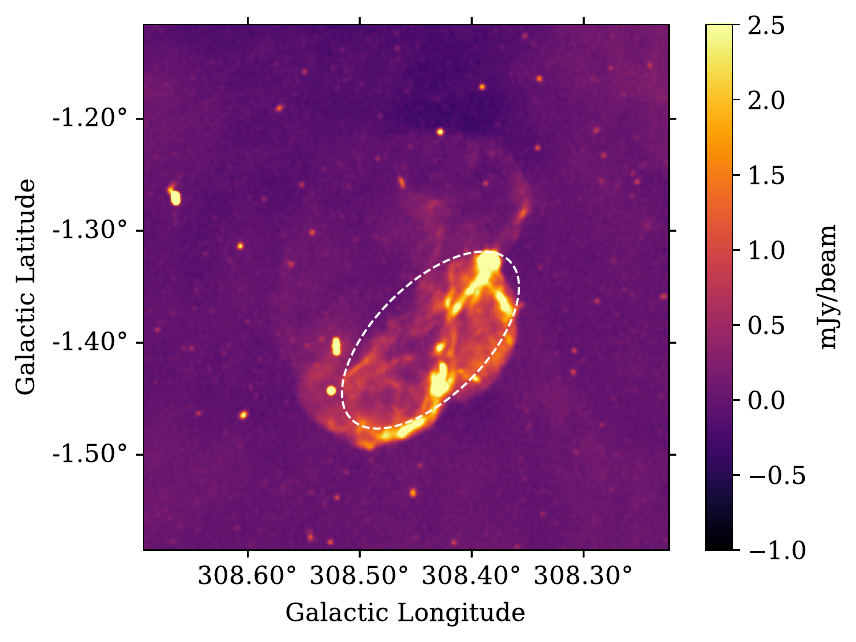}\label{fig:g308.5-1.4}}
    \caption{ASKAP 943 MHz radio continuum images of known SNRs with updated values. The white dashed lines indicate the previous extent of the SNRs, as described by \cite{Green2024_cat}.}
    \label{fig:known_snrs}
\end{figure}

\paragraph{\rev{G278.9+1.3} (Previously G279.0+1.1, Figure~\ref{fig:g278.94+1.35})} This source was first identified as an SNR by \cite{Woermann1988}. ASKAP observations have revealed faint filaments, indicating that this SNR is much larger than previously believed, with an angular size of over 3$^\circ$. RA, Dec, and size values in Table~\ref{tab:knownSNRs} have been updated to reflect the full extent of this SNR. The values provided are taken from \cite{Filipovic2024}, who studied this source in greater depth using EMU observations.

\paragraph{G286.1$-$1.1 (Previously G286.5$-$1.2, Figure~\ref{fig:g286.1-1.1})} Previously, only the brightest part of the two eastern filaments was identified as an SNR \citep{Whiteoak1996}. EMU observations have revealed the faint western shell, indicating the true size of the SNR to be much larger than previously thought. RA, Dec, and size values in Table~\ref{tab:knownSNRs} have been updated to reflect this. There is another filament extending up from the northern end of the eastern filament, but it is unclear if this is also related.

\paragraph{G291.0+0.1 (Previously G291.0$-$0.1, Figure~\ref{fig:g291.0+0.1})} Previously listed as two distinct sources, an SNR and an SNR candidate. The faint northern shell was proposed as SNR candidate G291.0+0.1 by \cite{Green2014}. The brighter southern emission has been identified as an SNR (G291.0$-$0.1) \citep{Roger1986}. Based on our observations, it appears that the two sources are related and comprise a single SNR. This SNR has recently been studied with the SMGPS in \cite{Loru2024}, who similarly find that the two components are likely related. The values provided in Table~\ref{tab:knownSNRs} are for the full source. 

\paragraph{G308.5$-$1.4 (Previously G308.4$-$1.4, Figure~\ref{fig:g308.5-1.4})} There has been some debate over the extent and morphology of this source. This is because the bright emission has an unusual overlapping arc morphology. \cite{DeHorta2013} argue that only the bright western ellipse comprises the SNR (which they call G308.3$-$1.4) with the other emission coming from background or foreground sources. This argument is based on the lack of an X-ray counterpart for the bright southeastern emission \citep{Prinz2012}. However, ASKAP observations have revealed a faint filament extending to the northeast that appears to be connected to the brighter filament. With the addition of this filament, the structure appears more consistent with a typical SNR morphology. The values provided in Table~\ref{tab:knownSNRs} represent the full source with this faint filament included.

\begin{figure}
    \centering
    \subfigure{\includegraphics[width=0.45\textwidth]{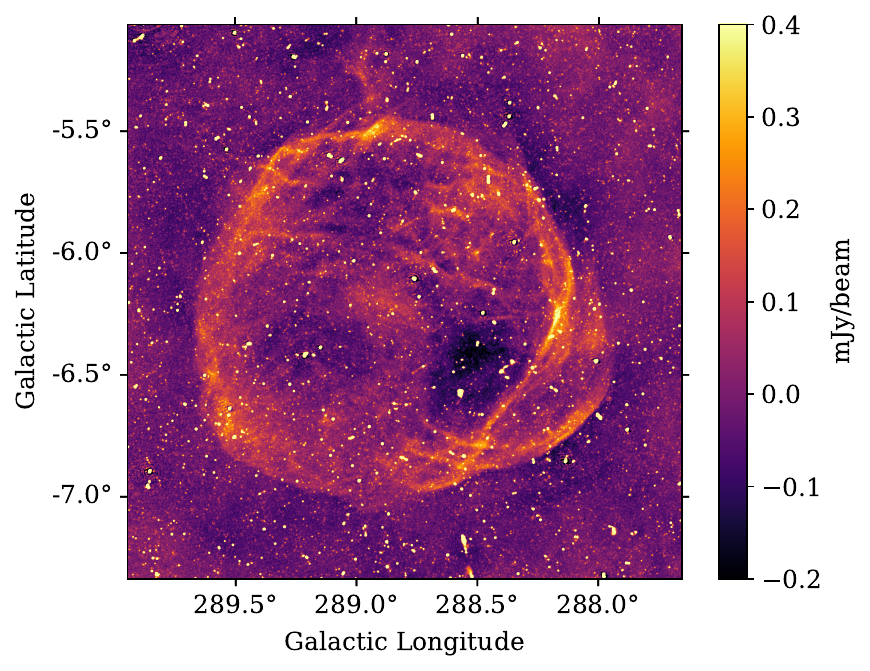}}
    \subfigure{\includegraphics[width=0.45\textwidth]{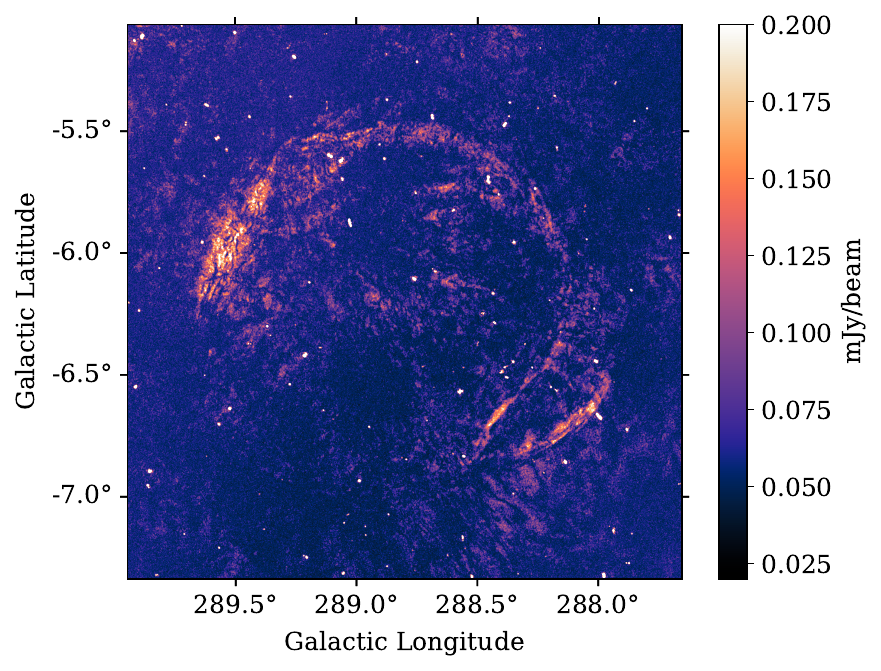}}
    \caption{ASKAP 943 MHz images of G288.8$-$6.3 in total power (left) and PI (right).}
    \label{fig:g288.8-6.3}
\end{figure}

\paragraph{G288.8$-$6.3 (Figure~\ref{fig:g288.8-6.3})} A large, high-latitude SNR. The outer shell is well defined and there are several internal filaments visible as well. This source has been previously described and identified as an SNR (using the EMU survey) by \cite{Filipovic2023} and was added to the October 2024 version of Green's SNR catalogue. Here, we demonstrate that the source is highly polarized using data from POSSUM. The outer shell is clearly polarized, particularly along the northern half, with the strongest polarization coming from the northeastern and southwestern filaments. 

\subsubsection{Determining Spectral Indices with ASKAP and MeerKAT} \label{sec:known_indices}

\begin{table}[]
    \centering
    \begin{tabular}{cccccc}
        \hline
         Name & Size & 943~MHz Flux & 1360~MHz Flux & Spectral Index  & Spectral Index \\
          & [$'$] & [Jy] & [Jy] & (Calculated) & (Catalogue)\\
        \hline
        SNRs \\
        \hline
        G290.1$-$0.8 & 19$\times$14 & 42 $\pm$ 2 & 26 $\pm$ 1 & $-$1.26 $\pm$ 0.07 & $-$0.4\\
        G304.6+0.1 & 8 & 15.2 $\pm$ 0.8 & 10.7 $\pm$ 0.5 & $-$0.93 $\pm$ 0.07 & $-$0.5\\
        G306.3$-$0.9 & 4 & 0.164 $\pm$ 0.008 & 0.118 $\pm$ 0.006 & $-$0.87 $\pm$ 0.07 & $-$0.5?\\
        G309.2$-$0.6 & 15$\times$12 & 6.2 $\pm$ 0.3 & 4.6 $\pm$ 0.2 & $-$0.82 $\pm$ 0.07 & $-$0.5?\\
        G310.6$-$0.3 & 8 & 5.0 $\pm$ 0.3 & 3.4 $\pm$ 0.2 & $-$1.03 $\pm$ 0.07 & ?\\
        G311.5$-$0.3 & 5 & 3.3 $\pm$ 0.2 & 2.6 $\pm$ 0.1 & $-$0.63 $\pm$ 0.07 & $-$0.5\\
        \hline
        \ion{H}{2} Regions & & & & & Expected Index\\
        \hline
        G279.4$-$1.7 & 8 & 0.54 $\pm$ 0.03 & 0.51 $\pm$ 0.03 & $-$0.15 $\pm$ 0.08 & $-$0.1\\
        G280.9$-$0.5 & 8 & 0.24 $\pm$ 0.01 & 0.21 $\pm$ 0.01 & $-$0.35 $\pm$ 0.06 & $-$0.1 \\
        G286.3$-$0.2 & 20 & 31 $\pm$ 2 & 22 $\pm$ 1 & $-$0.91 $\pm$ 0.08 & $-$0.1\\
        G301.6$-$0.3 & 14 & 2.1 $\pm$ 0.1 & 1.65 $\pm$ 0.08 & $-$0.64 $\pm$ 0.07 & $-$0.1\\ 
        G319.2+0.2 & 5 & 0.43 $\pm$ 0.02 & 0.37 $\pm$ 0.02 & $-$0.40 $\pm$ 0.07 & $-$0.1 \\
        \hline
    \end{tabular}
    \caption{Spectral indices of a selection of known SNRs and \ion{H}{2} regions. Catalogue values for SNRs are taken from \cite{Green2024_cat}. We assume a minimum of 5$\%$ uncertainties for flux values.}
    \label{tab:knownSNR_indices}
\end{table}

We calculate spectral indices for some of the known SNRs using fluxes from ASKAP and MeerKAT, and compare these to catalogued values to test the reliability of this approach. We also calculate indices for a few \ion{H}{2} regions to check if they are reliable and meaningfully distinct from the SNR values. The results of these calculations can be found in Table~\ref{tab:knownSNR_indices}. The flux values are determined using the method described in Section~\ref{sec:flux}. For each source, we use the same source boundary and background region definitions for the flux calculations with both data sets. The errors in the spectral indices are based only on the uncertainties in the fluxes and likely underestimate the true uncertainty. The catalogue values for the SNR spectral indices are taken from \cite{Green2024_cat}.

In general, we find significant differences between the calculated and expected results. These differences tend to be more significant for larger sources. We believe the 943~MHz fluxes to be reasonably accurate as they are generally consistent with the 1~GHz values from the \cite{Green2024_cat} catalogue, as shown in Table~\ref{tab:knownSNRs}. We also note that the indices we calculate are universally steeper than the expected values. We believe this is probably because many of the 1360~MHz flux values we determine are too low, as a consequence of missing short spacings. We also note that due to the relatively small frequency range, any uncertainties in the flux values can have a significant impact on the calculated index. Because the indices we calculated this way are universally too steep, it may be more appropriate to take them as lower limits.

For smaller sources ($<$10'), we do find that the indices determined for \ion{H}{2} regions are consistently flatter than those determined for the known SNRs, though they are all still steeper than the expected index values. For larger sources, which are likely more affected by missing flux, we can no longer reliably distinguish between the two types of sources. Thus, while the indices themselves are likely unreliable, they do seem to discriminate between thermal and nonthermal emission for sources smaller than $\sim$10'. 

Based on these results, indices calculated with only these two flux values should be considered with a degree of skepticism. Thus, in the current analysis, we only argue for the confirmation of a source as an SNR based on a steep negative spectral index if we can calculate flux densities for at least three distinct frequencies.

\subsection{Previously Identified SNR Candidates}

In Table~\ref{tab:known_cands}, we list the 46 SNR candidates that have been previously identified as such at radio frequencies. Of these, we believe we can confirm eight sources as SNRs based on polarization and/or spectral indices. \rev{For two of the strong candidates, G299.7$-$0.0 and G311.0$-$0.6, we provide spectral indices that were determined using flux values from only ASKAP and MeerKAT data. We do not confirm them as SNRs due to the issues outlined in Section~\ref{sec:known_indices}.} Descriptions of all of the sources can be found in Section~\ref{sec:new_snrs_cands}. 

\begin{table}[]
    \centering
    \begin{tabular}{ccccccccc}
    \hline
       Name & RA & Dec & Size & Flux Density & Spectral & Pol. & Peak \% & Ref.\\
        & (J2000) & (J2000) & [$'$] & [Jy at 943 MHz] & Index & & Pol & \\
    \hline
        \multicolumn{8}{l}{Newly Confirmed SNRs} \\
    \hline
        G292.3+0.6 & 11 23 34 & $-$60 27 38 & 19×12 & 0.8 $\pm$ 0.1 & $-$0.54 $\pm$ 0.13 & Y & 65\% & 1\\
        G299.3$-$1.5 & 12 17 55 & $-$64 09 53 & 36$\times$29 & 6.4 $\pm$ 0.4 & & Y & 53\% & $^a$2,3\\
        G309.8$-$2.6 & 13 57 00 & $-$64 29 20 & 35$\times$15 & 2.8 $\pm$ 0.4 & & Y & $>$70\% & $^a$2,4\\
        G310.7$-$5.4 & 14 12 18 & $-$67 05 03 & 31$\times$29 & 1.3 $\pm$ 0.2 & & Y & $>$70\% & $^a$3,5\\
        G318.9$-$0.5 & 15 01 21 & $-$59 15 29 & 17×12 & 1.3 $\pm$ 0.1* & \(-0.61 \pm 0.16\) & N & & 1\\
        G319.9$-$0.7 & 15 09 14 & $-$58 54 51 & 7×2 & 0.74 $\pm$ 0.04 & & Y & 25\% & $^a$6\\
        G320.6$-$0.9 & 15 15 04 & $-$58 47 00 &	4 & 0.5 $\pm$ 0.1 & $-$0.5 $\pm$ 0.10 & N & & 6\\
        G321.3$-$3.9 & 15 32 14 & $-$60 51 54 & 109×64 & N/A & & Y & $>$70\% & $^a$2,3\\
    \hline
        \multicolumn{8}{l}{Probable SNRs / Strong candidates} \\
    \hline
        G281.2$-$0.1 & 10 06 31 & $-$55 47 35 & 68×46 & 8 $\pm$ 1 & & N & & 1,4\\
        G286.5+1.0 & 10 44 24 & $-$57 48 10 & 24×20 & 0.4 $\pm$ 0.2 & & N & & 1\\
        G289.2$-$0.8 & 10 55 52 & $-$60 35 33 & 39×29 & $<$8 & & N & & 1\\
        G296.6$-$0.4 & 11 55 51 & $-$62 34 27 & 14$\times$10 & 0.81 $\pm$ 0.07 & & N & & $^a$3\\
        G299.7$-$0.0 & 12 23 03 & $-$62 42 35 & 25×22 & 3.0 $\pm$ 0.4 & $-$0.57 $\pm$ 0.08 & N & & 1\\
        G308.7+1.4 & 13 39 28 & $-$60 56 25 & 22$\times$18 & 0.37 $\pm$ 0.02 & & N & & 7\\
        G310.9$-$0.3 & 13 59 55 & $-$62 03 47 & 17$\times$11 & 2.6 $\pm$ 0.4 & & N & & $^a$3\\
        G311.0$-$0.6 & 14 01 27 & $-$62 19 54 & 5×4 & 0.16 $\pm$ 0.01 & \(-0.55 \pm 0.10\) & N & & 1\\
        G317.6+0.9 & 14 47 23 & $-$58 36 52 & 34×29 & 2.0 $\pm$ 0.3 & & N & & 5,6\\
        G320.8$-$0.3 & 15 13 36 & $-$58 12 57 & 10×9 & 0.36 $\pm$ 0.09 & & N & & 1\\
        G321.3$-$0.9 & 15 19 13 & $-$58 21 30 & 6 & 0.05 $\pm$ 0.02* & & N & & 1\\
        G322.7$-$0.6 & 15 26 53 & $-$57 22 32 & 42×32 & N/A & & N & & 1\\
    \hline
        \multicolumn{8}{l}{Possible SNRs / Weak candidates} \\
    \hline
        G278.8$-$1.2 & 09 48 12 & $-$55 12 41 & 64×56 & N/A & & N & & 1\\
        G280.2+0.2 & 10 01 49 & $-$54 59 21 & 18 & $<$0.2 & & N & & 1\\
        G280.4$-$0.9 & 09 58 22 & $-$55 58 41 & 66 & N/A & & N & & 4\\
        G283.9$-$1.4 & 10 17 17 & $-$58 27 40 & 78×41 & N/A & & N & & 1\\
        G286.8+0.0 & 10 42 09 & $-$58 43 24 & 34 & 18 $\pm$ 2 & & N & & 1\\
        G288.9$-$1.4 & 10 51 27 & $-$60 56 09 & 48×35 & N/A & & N & & 1\\
        G289.4$-$2.5 & 10 51 06 & $-$62 10 07 & 28×23 & 1.1 $\pm$ 0.1* & & N & & 4\\
        G290.6+0.5 & 11 10 23 & $-$59 58 00 & 25 & N/A & & N & & 1\\
        G297.0$-$1.0 & 11 58 24 & $-$63 12 30 & 55×52 & N/A & & N & & 1\\
        G298.5+2.1 & 12 15 39 & $-$60 29 50 & 50 & N/A & & N & & 4\\
        G304.2$-$0.5 & 13 03 09 & $-$63 20 30 & 12 & 0.49 $\pm$ 0.04 & & N & & 1\\
        G304.4$-$0.2 & 13 04 42 & $-$63 03 30 & 15×13 & 0.31 $\pm$ 0.07* & & N & & 8\\
        G306.2$-$0.8 & 13 21 03 & $-$63 30 49 & 32×30 & N/A & & N & & 1\\
        G307.1$-$0.7 & 13 28 23 & $-$63 16 16 & 54 & N/A & & N & & 1\\
        G307.5$-$1.0 & 13 32 45 & $-$63 27 17 & 22×20 & N/A & & N & & 1\\
        G307.9+0.1 & 13 34 48 & $-$62 19 13 & 30×22 & N/A & & N & & 1\\
        G308.3$-$0.2 & 13 38 25 & $-$62 33 40 & 11×8 & N/A & & N & & 1\\
        G308.8$-$0.5 & 13 43 19 & $-$62 48 26 & 29 & N/A & & N & & 1\\
        G309.2$-$0.1 & 13 45 52 & $-$62 19 00 & 13×10 & N/A & & N & & 1,5\\
        G310.5$-$0.6 & 13 57 25 & $-$62 29 25 & 26×24 & N/A & & N & & 1\\
        G311.3+1.1 & 14 00 14 & $-$60 37 33 & 10 & N/A & & Y? & & 9\\
        G312.7+2.9 & 14 06 55 & $-$58 34 22 & 10 & 0.15 $\pm$ 0.02 & & N & & 10\\
        G316.3$-$0.4 & 14 43 07 & $-$60 17 40 & 21 & $<$5 & & N & & 1\\
        G316.7+0.4 & 14 43 09 & $-$59 26 55	& 13 & N/A & & N & & 1\\
        G318.1$-$0.4 & 14 55 35 & $-$59 33 47 & 8 & 1.3 $\pm$ 0.2 & & N & & 1\\
        G319.3$-$0.7 & 15 05 12 & $-$59 12 09 & 34×20 & N/A & & N & & 1\\
    \hline
    \end{tabular}
    \caption{Previously identified SNR candidates. $^a$Appears in SNRcat \citep{Ferrand2012}. References: (1) \cite{Anderson2024} (2) \cite{Duncan1997} (3) \cite{Green2014}  (4) \cite{Duncan1995} (5) \cite{Mantovanini2025} (6) \cite{Whiteoak1996} (7) \cite{Lazarevic2024} (8) \cite{Sushch2017} (9) \cite{Gaensler2000} (10) \cite{Smeaton2024}.}
    \label{tab:known_cands}
\end{table}

\subsubsection{The SMGPS SNR Candidates}

Recently, \cite{Anderson2024} published a list of 237 Galactic SNR candidates in the longitude range \(251^\circ \leq \ell \leq358^\circ\),  \(2^\circ\leq \ell \leq 61^\circ\) and latitude range \(|b| \leq 1.5^\circ\) using the SMGPS. Of the 57 candidates they identified that fall within the region described in this work, 32 are included in our catalogue. However, we list these sources as 30 SNRs/SNR candidates because they split a couple of our candidates into two separate sources. The sources that appear in both lists can be found in Table~\ref{tab:known_cands}, marked by the reference number. Of these, we can confirm G292.3+0.6 and G318.9$-$0.5 as true SNRs using polarization and spectral indices. We also classify G299.7$-$0.0 and G311.0$-$0.6 as likely SNRs based on spectral indices. Descriptions of the sources that are included in our catalogue can be found in Section~\ref{sec:new_snrs_cands}.

The SMGPS candidates not included in our catalogue are briefly discussed in this section. These sources are omitted from our catalogue because (1) we find a possible MIR counterpart, (2) we did not find a morphology consistent with expectations for an SNR in the ASKAP images, or (3) they are very small and/or faint and are not clearly visible in the ASKAP data, which in general has slightly poorer resolution and sensitivity when compared to the SMGPS data. In Table~\ref{tab:MK_cands}, we list the SMGPS candidates that have been omitted from our catalogue and the reason why they were not included.

For SMGPS SNR candidates of small angular size ($<$10'), we attempt to calculate spectral indices using 943~MHz data from ASKAP and the 1.36~GHz data from the SMGPS. These results should be viewed with skepticism for the reasons discussed in Section~\ref{sec:known_indices}. As discussed, we find that the indices that we calculate with only these two fluxes are generally steeper than expected. Thus, steep indices indicate that a source may be an SNR, but we cannot make definitive conclusions. Errors for the indices are based on flux uncertainties and in many cases are likely underestimated. In Table~\ref{tab:MK_cands} we provide the results of these calculations for sources that were not included in our catalogue. 

We found that G288.824$-$00.155, G310.702$-$00.573, G315.501$-$00.584, and G316.456$-$00.095 have indices comparable to the values determined for known SNRs in Section~\ref{sec:known_indices}. We do not include them in our catalogue, however, as they appear to have MIR counterparts. We found G277.162+00.396 and G295.855$-$01.388 to have flat indices, consistent with the expected value for \ion{H}{2} regions. G304.071+01.227, G310.702$-$00.573, and G318.033$-$00.935 have intermediate indices, and we cannot make conclusions either way. For G318.033$-$00.935, we also note that the brightest edge of the source has a positive index and that there is a small MIR source that overlaps with this candidate. We find that we cannot properly determine a flux with the ASKAP data for G306.195$-$00.520 as it is too faint to properly distinguish above the background noise. This may also be the case (to a lesser extent) for G292.472+00.167 and G295.855$-$01.388.

\begin{table}[]
    \centering
    \begin{tabular}{cccccc}
    \hline
        SNR Candidate & Size & Flux Density & Flux Density & Spectral  & Reason \\
         & [$'$] & [Jy at 943~MHz] & [Jy at 1360~MHz] & Index & Omitted\\
        \hline
        G275.986$-$01.077 & 9.6 & 0.08 $\pm$ 0.02 & 0.09 $\pm$ 0.02 & 0.3 $\pm$ 0.3 & Faint, unclear\\
        G277.162+00.396 & 9.6 & 0.030 $\pm$ 0.002 & 0.029 $\pm$ 0.002 & $-$0.09 $\pm$ 0.09 & 22$\mu$m emission, faint\\
        G282.674$-$00.737 & 20.8 & & & & Possible MIR\\
        G283.234$-$00.945 & 12.8 & & & & Possible MIR\\
        G283.629$-$00.943 & 20.4 & & & & Possible MIR\\
        G288.824$-$00.155 & 5.4 & 0.34 $\pm$ 0.06 & 0.24 $\pm$ 0.03 & $-$0.9 $\pm$ 0.2 & Possible MIR\\
        G288.863$-$00.046 & 41.2 & & & & Possible MIR\\
        G290.003$-$01.510 & 8.6 & & & & Possible MIR\\
        G292.472+00.167 & 1.8 & 0.0070 $\pm$ 0.0003 & 0.0077 $\pm$ 0.0004 & 0.25 $\pm$ 0.07 & Possible MIR\\
        G295.855$-$01.388 & 1.8 & 0.0081 $\pm$ 0.0008 & 0.0077 $\pm$ 0.0006 & $-$0.1 $\pm$ 0.1 & Possible MIR\\
        G302.324$-$01.300 & 12.0 & & & & Unclear morphology\\
        G303.365$-$00.050 & 42.6 & & & & Unclear morphology\\
        G304.071+01.227 & 7 & 0.037 $\pm$ 0.008 & 0.030 $\pm$ 0.006 & $-$0.6 $\pm$ 0.3 & Faint, unclear\\
        G306.195$-$00.520 & 3.8 & & & & Not visible in ASKAP\\
        G310.275$-$00.171 & 10.4 & & & & Possible MIR\\
        G310.702$-$00.573 & 2.4 & 0.0330 $\pm$ 0.0003 & 0.0290 $\pm$ 0.0007 & $-$0.34 $\pm$ 0.03 & Possible MIR\\
        G315.501$-$00.584 & 3.2 & 0.014 $\pm$ 0.002 & 0.010 $\pm$ 0.002 & $-$1.0 $\pm$ 0.3 & 22$\mu$m emission, faint\\
        G315.715$-$00.227 & 1.8 & 0.017 $\pm$ 0.001 & 0.0100 $\pm$ 0.0004 & $-$1.4 $\pm$ 0.1 & Possible MIR\\
        G315.905$-$00.817 & 26.4 & & & & Unclear morphology\\
        G316.456$-$00.095 & 2.8 & 0.25 $\pm$ 0.04 & 0.18 $\pm$ 0.03 & $-$0.8 $\pm$ 0.2 & Possible MIR\\
        G316.587$-$00.097 & 9.8 & & & & Possible MIR\\
        G318.033$-$00.935 & 3.0 & 0.075 $\pm$ 0.001 & 0.065 $\pm$ 0.002 & $-$0.38 $\pm$ 0.03 & Flatter spectrum\\
        G318.602$-$00.884 & 18.4 & & & & Unclear morphology\\
        G320.944$-$01.289 & 14.8 & & & & Part of known SNR?\\
        \hline
    \end{tabular}
    \caption{SNR candidates from the SMGPS \citep{Anderson2024} that are not included in our catalogue. Spectral indices are determined for sources $<$10' using 943~MHz flux values from ASKAP and 1360~MHz flux values from MeerKAT. In the last column, we indicate why they were omitted from our candidate catalogue.}
    \label{tab:MK_cands}
\end{table}

\subsubsection{SNR Candidates from X-rays and Optical} \label{sec:xray_optical}

\begin{table}[]
    \centering
    \begin{tabular}{cccccccc}
    \hline
       Name & RA & Dec & Size & Flux Density & Pol & Peak \% & Ref\\
        & (J2000) & (J2000) & [$'$] & [Jy at 943 MHz] & & Pol & \\
        \hline
        \multicolumn{6}{l}{X-ray SNR Candidates} \\
        \hline
        G284.0$-$1.8 & 10 16 21 & $-$58 57 10 & 6$\times$3 & 0.21 $\pm$ 0.04 & N & &$^a$1\\
        G284.2$-$0.4 & 10 23 16 & $-$57 46 00 & 2$\times$1 & N/A & N& &$^a$2\\
        G304.1$-$0.2 & 13 01 46 & $-$63 05 20 & 4$\times$3 & 0.04 $\pm$ 0.01 & N & &$^a$3\\
        G317.9$-$1.8 & 14 59 31 & $-$60 53 10 & 34$\times$20 & $<$1 & N & &$^a$5\\
        \hline
        \multicolumn{6}{l}{Optical SNR Candidates} \\
        \hline
        G283.8$-$4.0 & 10 05 06 & $-$60 32 25 & 54×22 & N/A & Y & $>$70\% &6\\
        G304.6$-$3.0 & 13 08 07 & $-$65 49 13 & 16 & N/A & N & &7\\
        G310.6+0.8 & 13 55 31 & $-$61 04 31 & 27×25 & $<$1 & N & &8\\
    \hline
    \end{tabular}
    \caption{SNR candidates that have been previously identified in X-rays or optical. $^a$Appears in SNRcat \citep{Ferrand2012}. References: (1) \cite{Camilo2001,Camilo2004} (2) \cite{Karagaltsev2013} (3) \cite{HESS2012} (4) \cite{Reynolds2012} (5) \cite{Abdo2009} (6) \cite{Stupar2008} (7) \cite{Stupar2010} (8) \cite{Stupar2011}.}
    \label{tab:xray_opt_cands}
\end{table}

Of the X-ray SNRs and SNR candidates from \cite{Ferrand2012} that fall within our surveyed region, we find that one has a clear radio counterpart (G284.0$-$1.8) and three have possible radio counterparts in EMU/POSSUM data. We also note that three of our radio SNR candidates have been previously identified as SNR candidates in the optical with the SuperCOSMOS H$\alpha$ Survey (SHS) \citep{Parker2005}. We list these sources in Table~\ref{tab:xray_opt_cands} and provide descriptions of each candidate below.

\begin{figure}
    \centering
    \subfigure[G284.0$-$1.8]{\includegraphics[width=0.35\textwidth]{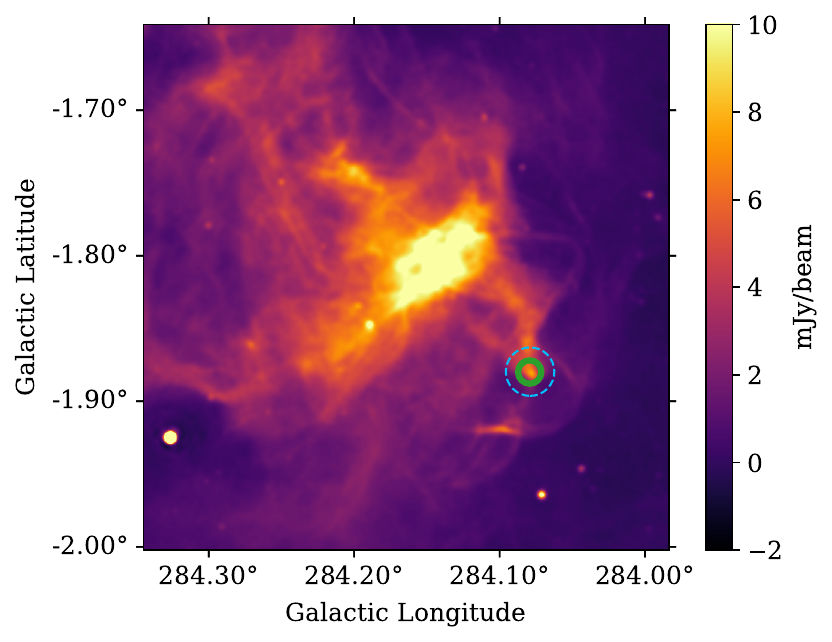}\label{fig:g284.0-1.8}}
    \subfigure[G284.2$-$0.4]{\includegraphics[width=0.35\textwidth]{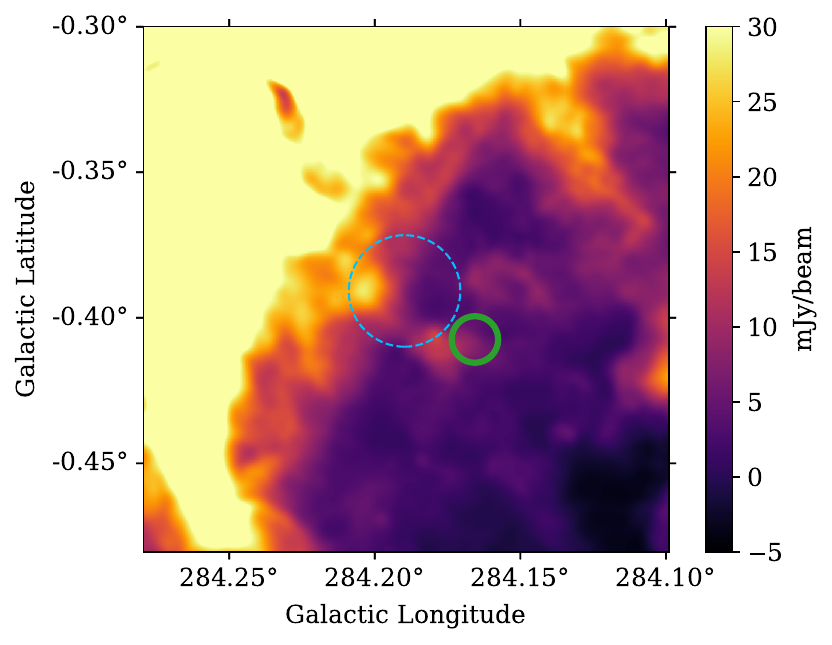}\label{fig:g284.2-0.4}}\\
    \subfigure[G304.1$-$0.2]{\includegraphics[width=0.35\textwidth]{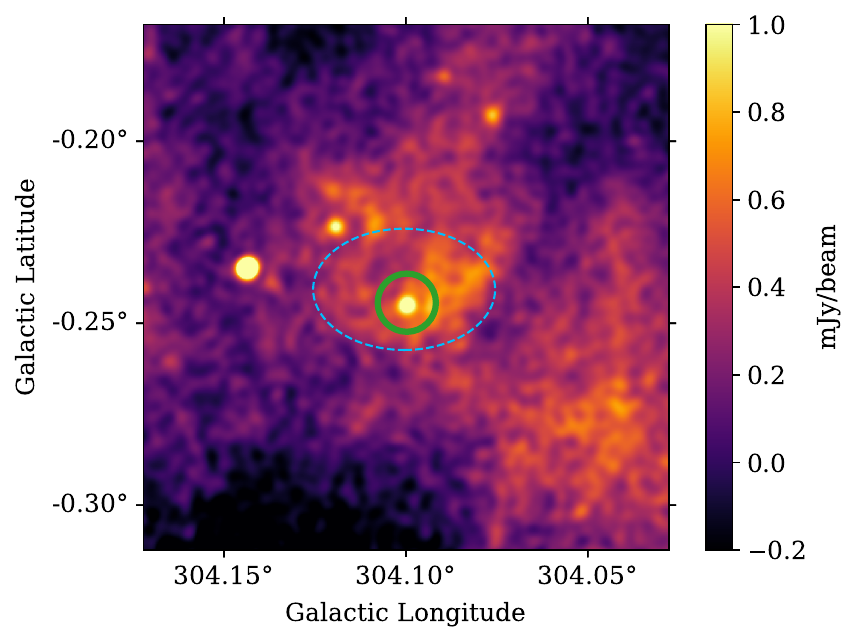}\label{fig:g304.1-0.2}}
    \subfigure[G317.9$-$1.8]{\includegraphics[width=0.35\textwidth]{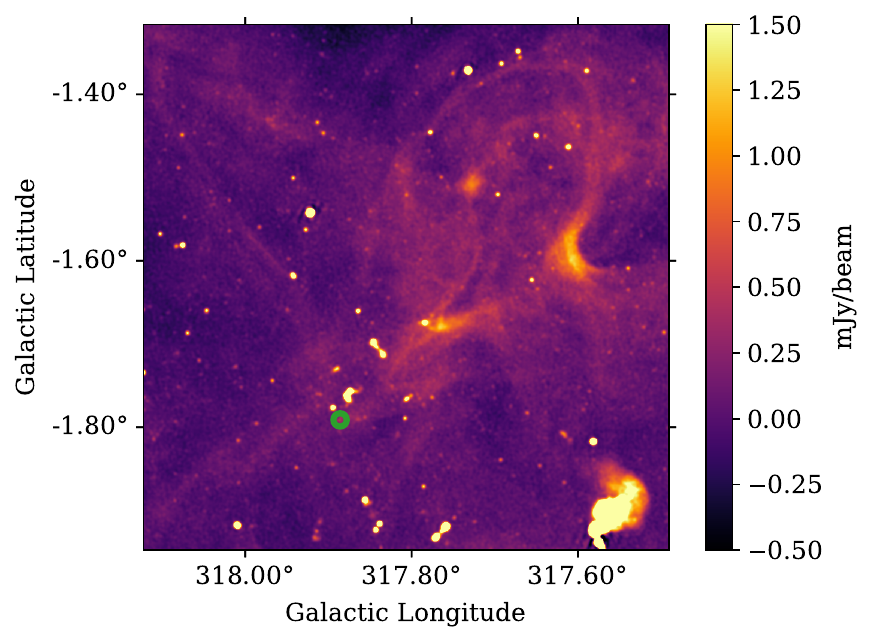}\label{fig:g317.9-1.8}}
    \caption{ASKAP 943 MHz images of X-ray SNRs/SNR candidates that show possible signs of a radio counterpart. Green circles indicate the locations of young pulsars. Blue dashed circles indicate the extent of the X-ray emission.}
    \label{fig:xray_snrs}
\end{figure}

\paragraph{G284.0$-$1.8 (Figure~\ref{fig:g284.0-1.8})} A known X-ray PWN around pulsar J1016$-$5857 (characteristic age 21 kyrs). The source is likely related to the known SNR G284.3$-$1.8 \citep{Camilo2001,Camilo2004}. We see a tail of radio emission that connects the pulsar to G284.3$-$1.8, indicating that it is likely its host SNR. The pulsar and host SNR are clearly visible in polarization but the PWN is not.

\paragraph{G284.2$-$0.4 (Figure~\ref{fig:g284.2-0.4})} X-ray PWN candidate around the very young (characteristic age 4.6 kyrs) pulsar J1023$-$5746 \citep{Karagaltsev2013}. We find no previous radio references for this source.  Because the source is embedded in a large, bright \ion{H}{2} region, it is unclear if there is a radio counterpart. 

\paragraph{G304.1$-$0.2 (Figure~\ref{fig:g304.1-0.2})} X-ray PWN \citep{HESS2012} around a young (characteristic age 11 kyrs) pulsar J1301$-$6305 \citep{Manchester2001}. The source was studied in the radio by \cite{Sushch2017} with ATCA, but no radio counterpart was found. We find very faint radio emission around the pulsar that could be the radio counterpart, but this is not definitive. Only the pulsar is clearly polarized. \cite{Sushch2017} suggest that the source could be related to the SNR candidate G304.4$-$0.2.

\paragraph{G317.9$-$1.8 (Figure~\ref{fig:g317.9-1.8})} X-ray PWN candidate around pulsar J1459$-$6053 (characteristic age 64.7 kyrs) \citep{Abdo2009}. There is no known radio counterpart. We detect some faint shells to the northwest in the radio that could be the host SNR. There is a faint tail of radio emission tracing from the pulsar back towards these shells, supporting this interpretation.

\begin{figure}
    \centering
    \subfigure[G283.8$-$4.0]{\includegraphics[width=0.34\textwidth]{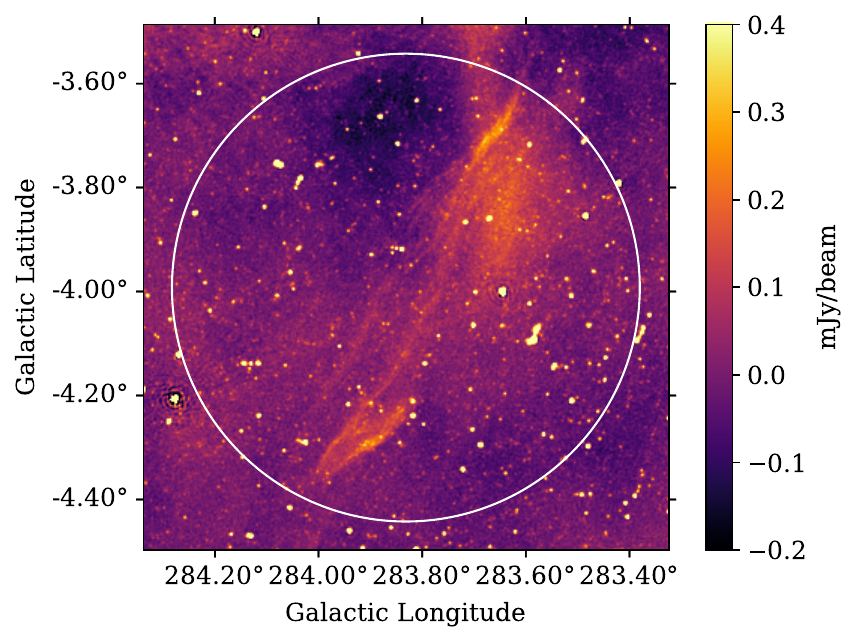}\label{fig:g283.8-4.0}}
    \subfigure[G283.8$-$4.0 PI]{\includegraphics[width=0.34\textwidth]{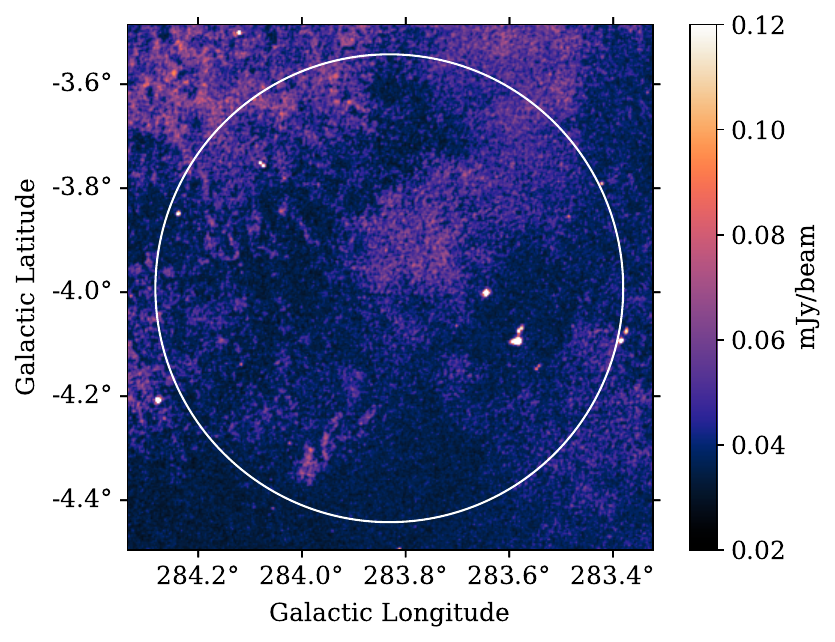}\label{fig:g283.8-4.0_pi}}
    \subfigure[G283.8$-$4.0 H$\alpha$]{\includegraphics[width=0.28\textwidth]{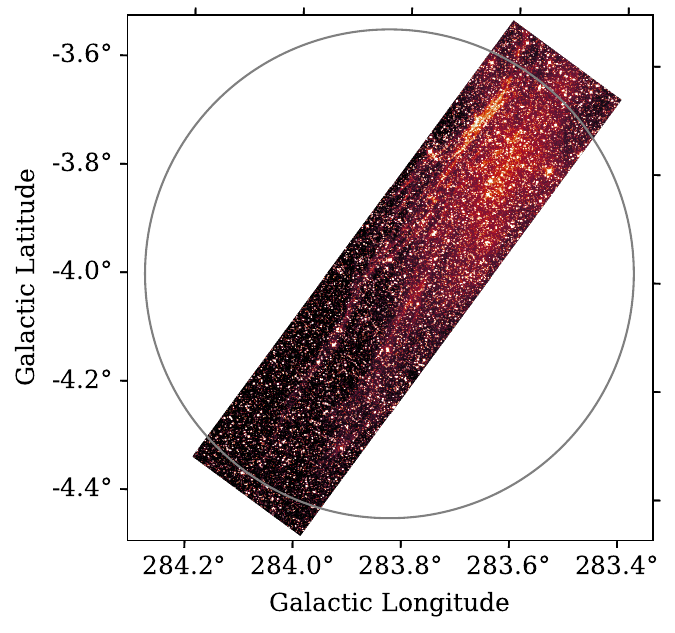}\label{fig:g283.8-4.0_ha}}
    \subfigure[G304.6$-$3.0]{\includegraphics[width=0.37\textwidth]{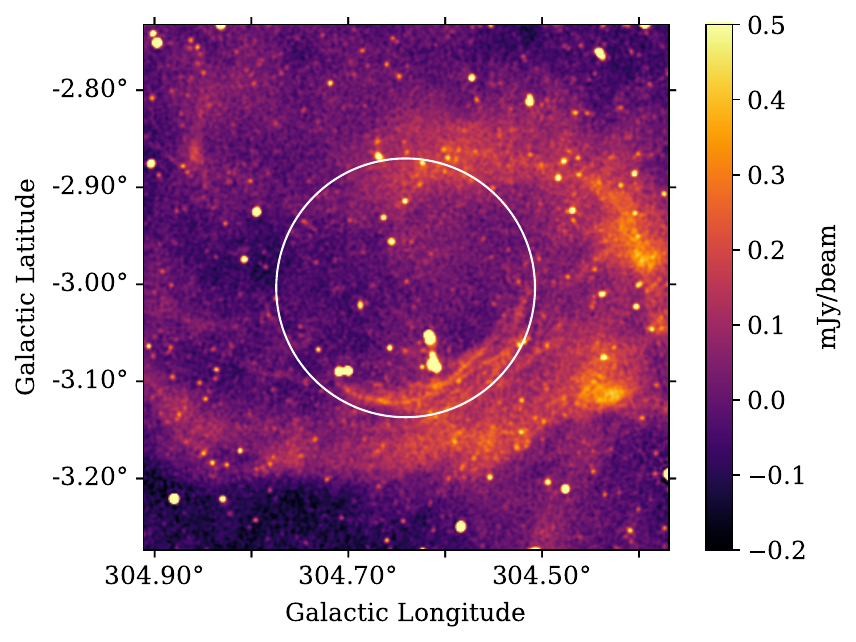}\label{fig:g304.6-3.0}}
    \subfigure[G304.6$-$3.0 H$\alpha$]{\includegraphics[width=0.3\textwidth]{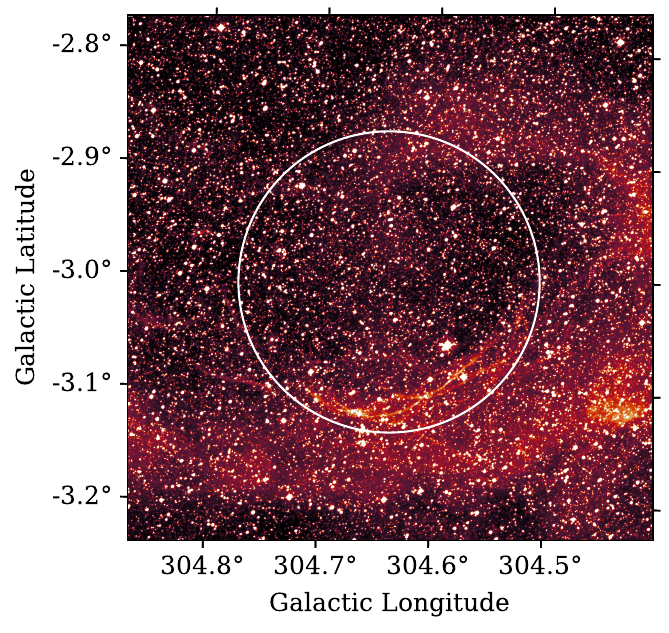}\label{fig:g304.6-3.0_ha}}
    \subfigure[G310.6+0.8]{\includegraphics[width=0.37\textwidth]{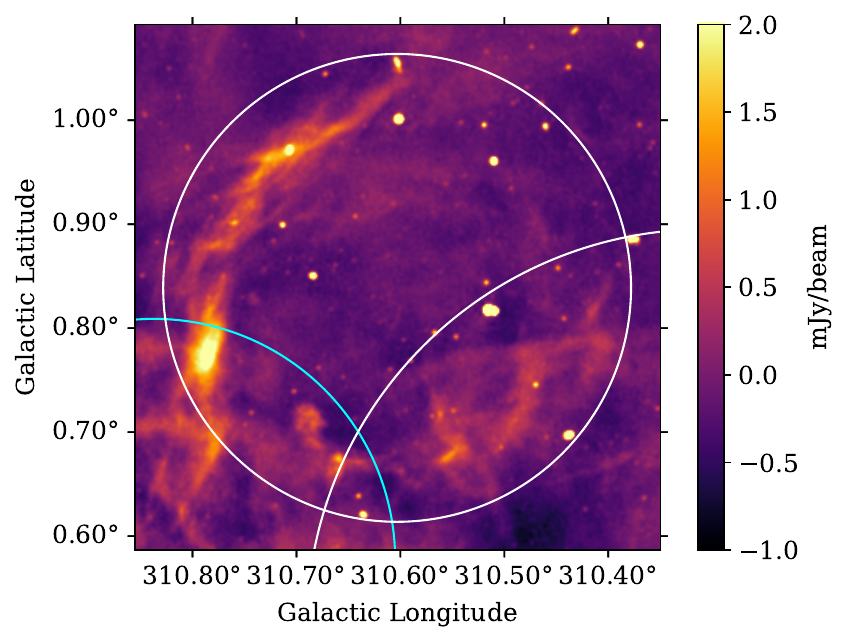}\label{fig:g310.6+0.8}}
    \subfigure[G310.6+0.8 H$\alpha$]{\includegraphics[width=0.3\textwidth]{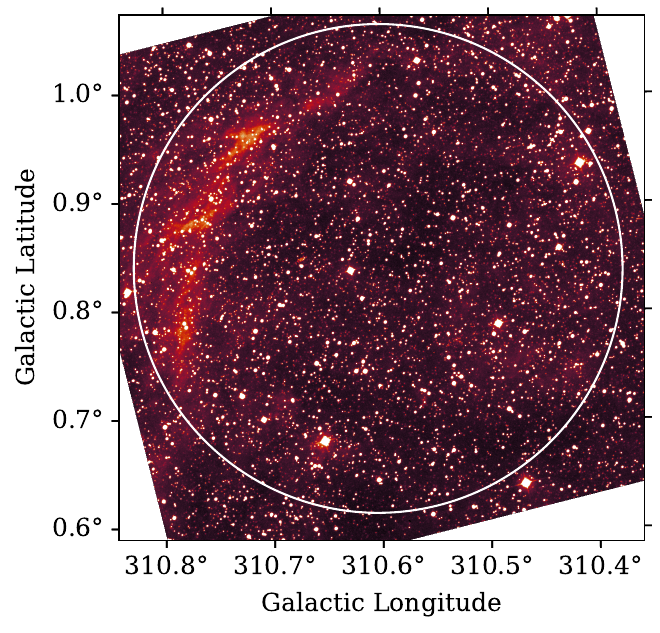}\label{fig:g310.6+0.8_ha}}
    \caption{Radio SNR candidates that have been previously detected in the optical. 943 MHz radio continuum images (left) and PI (centre) are from ASKAP. H$\alpha$ data (right) is from the SHS \citep{Parker2005}.}
    \label{fig:op_snrs}
\end{figure}

\paragraph{G283.8$-$4.0 (Figure~\ref{fig:g283.8-4.0}, \ref{fig:g283.8-4.0_pi}, \ref{fig:g283.8-4.0_ha})} A long filament, identified by \cite{Stupar2008} as an SNR candidate (G283.7$-$3.8) based on a high [SII]/H$\alpha$ ratio. The source has not previously been detected in the radio. The northern end of the filament is clearly visible in H$\alpha$ while the southernmost end is visible in polarization. The source is further discussed in Section~\ref{sec:new_snrs_cands} due to its proximity and possible relationship to another SNR candidate. 

\paragraph{G304.6$-$3.0 (Figure~\ref{fig:g304.6-3.0}, \ref{fig:g304.6-3.0_ha})} Multiple thin filaments located in a complex region near a Wolf-Rayet star and several possible \ion{H}{2} regions. The source was proposed as an optical SNR candidate (G304.4$-$3.1) by \cite{Stupar2010} as the thin filaments are clearly visible in H$\alpha$. The source was also detected in the radio at 4850~MHz in the PMN survey and SUMSS at 843~MHz, but with limited detail due to poor resolution so that a definitive relationship could not be established \citep{Stupar2010}. 

\paragraph{G310.6+0.8 (Figure~\ref{fig:g310.6+0.8}, \ref{fig:g310.6+0.8_ha})} Thin shell-like feature seen to the northeast with fainter filaments to the southwest. There are also two polarized point sources located within the arc. The source was proposed as an optical SNR candidate (G310.5+0.8) by \cite{Stupar2011} based on optical spectroscopy. It was also detected in the radio at 4850~MHz in the PMN survey and SUMSS at 843~MHz \citep{Stupar2011}. The brighter northeastern shell can be clearly seen in H$\alpha$.

\subsection{New SNRs and SNR Candidates} \label{sec:new_snrs_cands}

\begin{table}[]
    \centering
    \begin{tabular}{cccccccc}
    \hline
       Name & RA & Dec & Size & Flux Density & Spectral & Pol & Peak \%\\
        & (J2000) & (J2000) & [$'$] & [Jy at 943 MHz] & Index & & Pol\\
    \hline
        \multicolumn{7}{l}{New SNRs} \\
    \hline
        G282.1$-$4.1 & 09 54 06 & $-$59 39 12 & 71×48 & N/A & & Y & $>$70\%\\
        G283.1$-$0.6 & 10 15 38	& $-$57 20 07 & 3×2 & 0.08 $\pm$ 0.01 & \(-0.55 \pm 0.13\) & N\\
        G285.0$-$3.2 & 10 16 48 & $-$60 32 39 & 38×25 & $<$1 & & Y & $>$70\%\\
        G289.6+5.8 & 11 18 22 & $-$54 37 03 & 38×32 & 1.4 $\pm$ 0.2 & & Y & $>$70\%\\
        G295.6+0.5 & 11 49 09 & $-$61 29 25 & 11 & 0.23 $\pm$ 0.01 & & Y & 67\%\\
        G300.1$-$1.6 & 12 25 10 & $-$64 20 38 & 48×31 & 3.2 $\pm$ 0.2* & & Y & $>$70\%\\
    \hline
        \multicolumn{7}{l}{Probable SNRs / Strong candidates} \\
    \hline
        G291.2$-$0.6 & 11 12 02 & $-$61 09 35 & 14×12 & N/A & & N\\
        G298.0$-$0.2 & 12 08 19 & $-$62 40 38 & 10×9 & N/A & \(-0.43 \pm 0.08\) & N\\
        G305.4$-$2.2 & 13 15 02 & $-$64 57 45 & 22 & 0.31 $\pm$ 0.04* & & N\\
        G311.1+1.4 & 13 58 36 & $-$60 21 15 & 18×16 & N/A & & N\\
        G317.4$-$1.1 & 14 53 15 & $-$60 31 37 & 6×5 & 0.19 $\pm$ 0.02 & \(-0.46 \pm 0.16\) & N \\
        G320.9$-$0.7 & 15 16 00 & $-$58 27 46 & 16×13 & N/A & & N\\
    \hline
        \multicolumn{7}{l}{Possible SNRs / Weak candidates} \\
    \hline
        G276.1+3.2 & 09 52 58 & $-$50 07 28 & 113×110 & N/A & & N\\
        G280.7+3.2 & 10 16 28 & $-$52 48 51 & 110×96 & N/A & & N\\
        G281.1$-$2.8 & 09 54 00 & $-$57 56 06 & 94×68 & N/A & & N\\
        G282.3$-$1.9 & 10 05 14 & $-$57 58 09 & 28 & $<$5 & & N\\
        G283.9+0.5 & 10 25 12 & $-$56 49 30 & 16×14 & N/A & & N\\
        G284.2$-$0.9 & 10 21 46 & $-$58 10 47 & 24×20 & N/A & & N\\
        G285.5$-$0.1 & 10 32 58 & $-$58 09 40 & 37×26 & $<$7 & & N\\
        G287.1$-$2.4 & 10 34 57 & $-$60 57 38 & 38 & N/A & & N\\
        G288.0+1.1 & 10 53 46 & $-$58 16 30 & 5×4 & 0.05 $\pm$ 0.01 & & N\\
        G290.1+0.5 & 11 06 40 & $-$59 42 07 & 56 & N/A & & N\\
        G290.6$-$1.5 & 11 04 12 & $-$61 46 32 & 38×36 & $<$10 & & N\\
        G291.1$-$0.4 & 11 11 41 & $-$60 58 37 & 12×8 & 2.5 $\pm$ 0.3 & & N\\
        G291.3+4.5 & 11 26 13 & $-$56 23 29 & 36×24 & $<$0.8 & & N\\
        G294.5$-$0.6 & 11 37 57 & $-$62 14 02 & 16 & N/A & & N\\
        G295.2$-$0.5 & 11 44 10 & $-$62 21 19 & 9 & 0.8 $\pm$ 0.2 & & N\\
        G296.4$-$2.8 & 11 49 22 & $-$64 52 28 & 91 & N/A & & N\\
        G298.0+0.3 & 12 08 51 & $-$62 12 27 & 44×29 & N/A & & Y?\\
        G298.2$-$1.9 & 12 07 15 & $-$64 19 53 & 75×65 & N/A & & N\\
        G299.2+0.3 & 12 18 48 & $-$62 21 38 & 30 & N/A & & N\\
        G300.0$-$1.6 & 12 21 20 & $-$64 15 55 & 40 & N/A & & N\\
        G301.8$-$2.1 & 12 40 26 & $-$64 55 38 & 90×40 & $<$3 & & N\\
        G306.4+2.6 & 13 19 22 & $-$60 06 02 & 100 & N/A & & N\\
        G308.5+0.4 & 13 39 25 & $-$61 54 57 & 44×30 & $<$3 & & N\\
        G309.9$-$0.4 & 13 52 11 & $-$62 29 25 & 29×22 & N/A & & N\\
        G310.3+0.5 & 13 53 47 & $-$61 27 49 & 46×34 & N/A & & N\\
        G311.5$-$0.1 & 14 04 49 & $-$61 44 23 & 12×8 & N/A & & N\\
        G311.9+0.9 & 14 05 31 & $-$60 39 28 & 16 & $<$1 & & N\\
        G319.4+0.2 & 15 02 50 & $-$58 27 00 & 8 & N/A & & N\\
        G320.0$-$1.7 & 15 13 53 & $-$59 47 52 & 52 & N/A & & N\\
        G320.6$-$0.8 & 15 14 04 & $-$58 41 00 & 17×16 & N/A & & N\\
        G320.9$-$0.3 & 15 13 49 & $-$58 04 46 & 29×21 & N/A & & N\\    
    \hline
    \end{tabular}
    \caption{New SNRs and SNR candidates discovered with the EMU/POSSUM surveys.}
    \label{tab:newcands}
\end{table}

Here, we describe the newly confirmed SNRs and SNR candidates identified with the EMU and POSSUM surveys. In Table~\ref{tab:newcands}, we list SNRs and SNR candidates that, to our knowledge, have not been previously identified. In total, we found six new SNRs, six strong candidates, and 31 weak candidates that have not been previously identified. In the following sections, we provide descriptions of the new SNRs and SNR candidates, including those that have been identified previously (see Table~\ref{tab:known_cands}). In total, we find 14 newly confirmed SNRs, 18 strong SNR candidates, and 57 weak candidates. 

\subsubsection{New SNRs}

Here, we list 14 sources that we believe can be confirmed as SNRs using the EMU and POSSUM surveys. Of these, eight have previously been identified as SNR candidates. Eleven of the sources show evidence of polarization and we determine nonthermal spectral indices for four.

\begin{figure}
    \centering
    \subfigure[G282.1$-$4.1]{\includegraphics[width=0.32\textwidth]{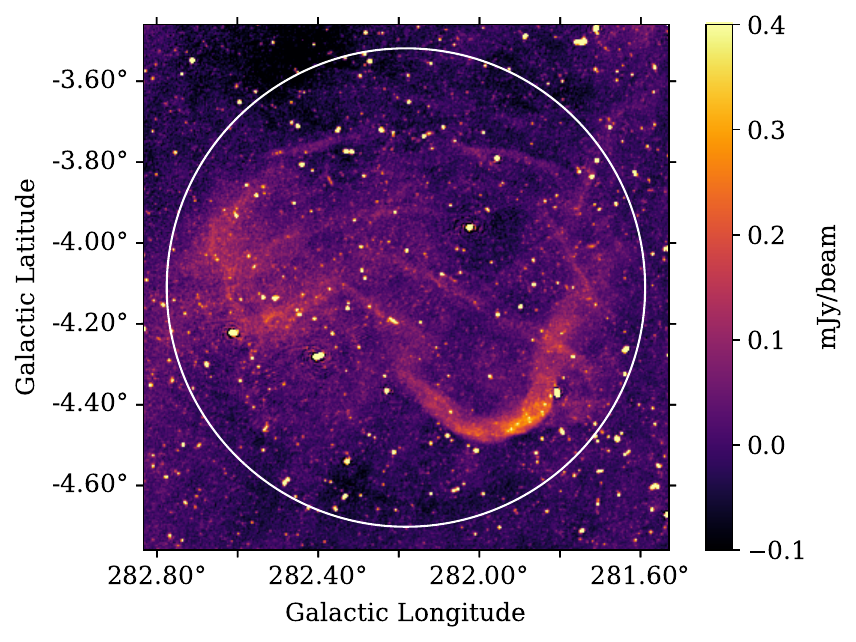}\label{fig:g282.1-4.1}}
    \subfigure[G289.6+5.8]{\includegraphics[width=0.32\textwidth]{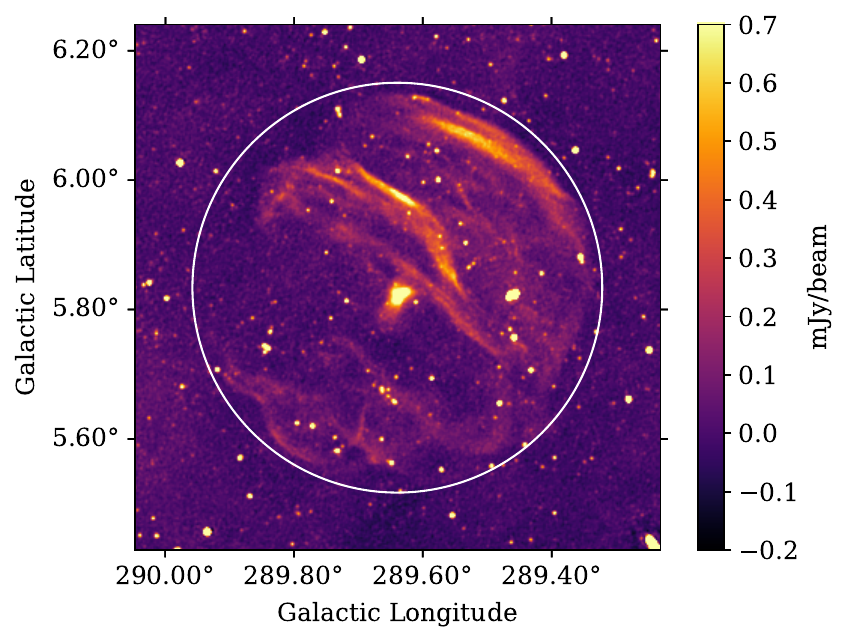}\label{fig:g289.6+5.8}}
    \subfigure[G292.3+0.6]{\includegraphics[width=0.32\textwidth]{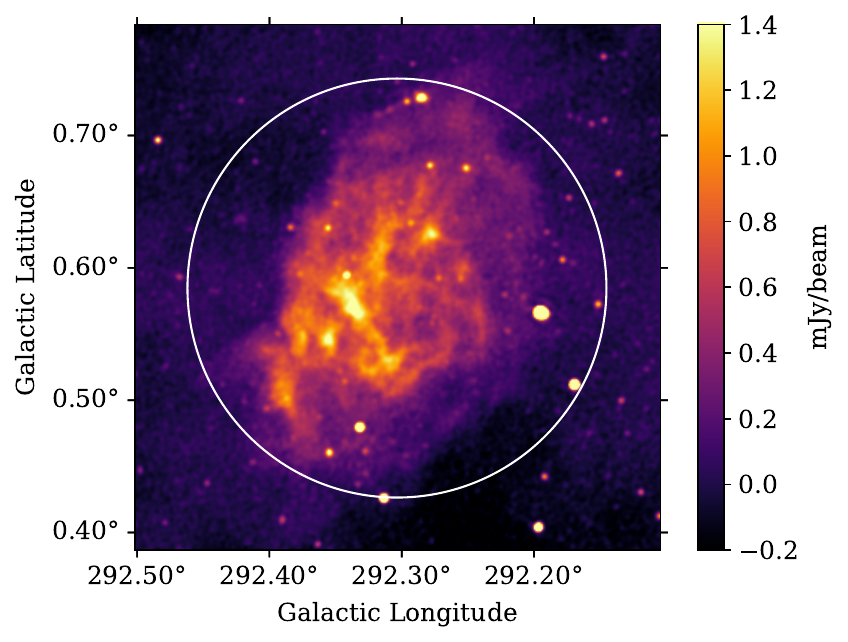}\label{fig:g292.3+0.6}}
    \subfigure[G282.1$-$4.1 PI]{\includegraphics[width=0.32\textwidth]{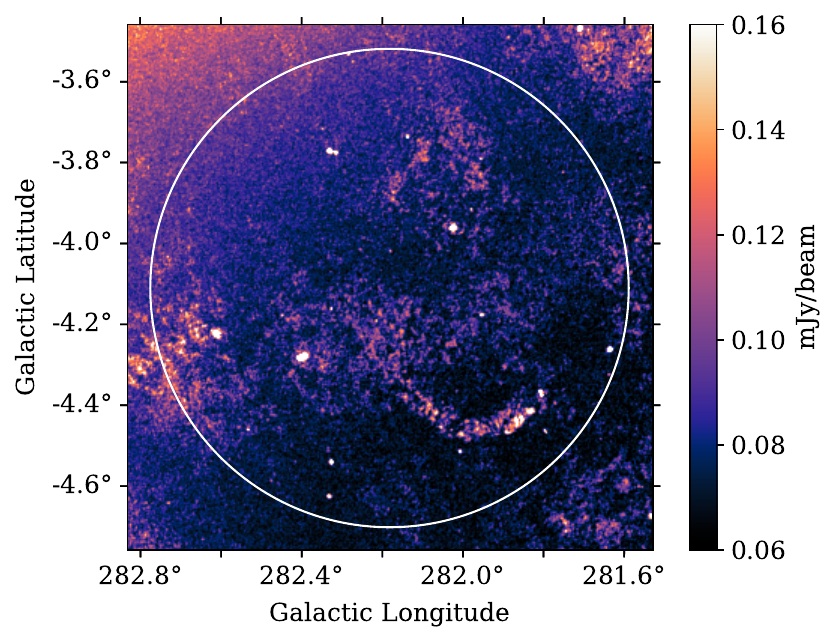}\label{fig:g282.1-4.1_pi}}
    \subfigure[G289.6+5.8 PI]{\includegraphics[width=0.32\textwidth]{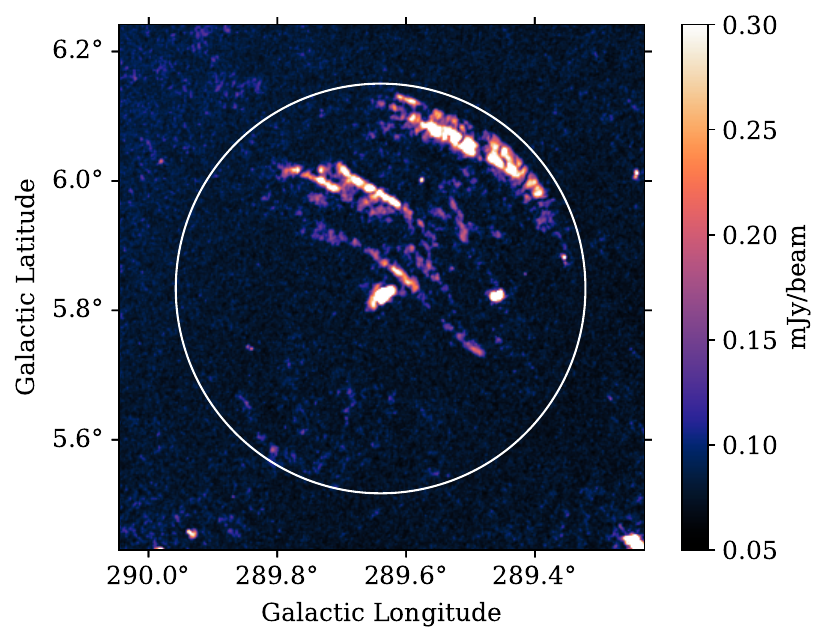}\label{fig:g289.6+5.8_pi}}
    \subfigure[G292.3+0.6 PI]{\includegraphics[width=0.32\textwidth]{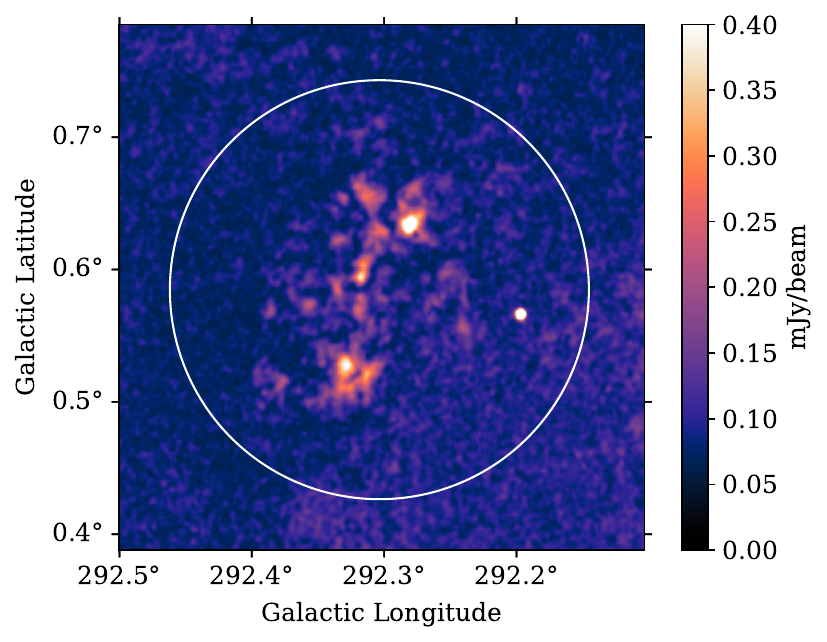}\label{fig:g292.3+0.6_pi}}
    \subfigure[G295.6+0.5]{\includegraphics[width=0.32\textwidth]{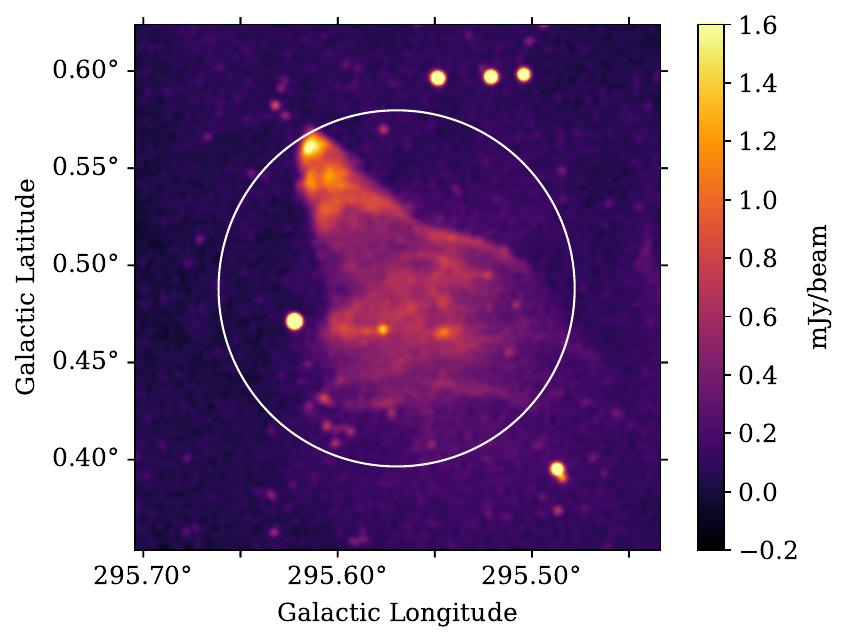}\label{fig:g295.6+0.5}}
    \subfigure[G300.1$-$1.6]{\includegraphics[width=0.32\textwidth]{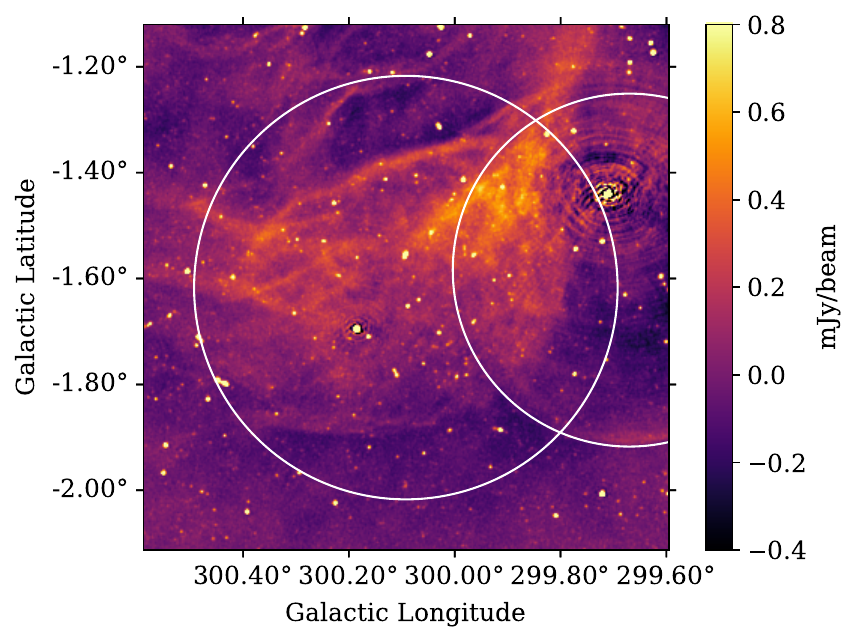}\label{fig:g300.1-1.6}} \\
    \subfigure[G295.6+0.5 PI]{\includegraphics[width=0.32\textwidth]{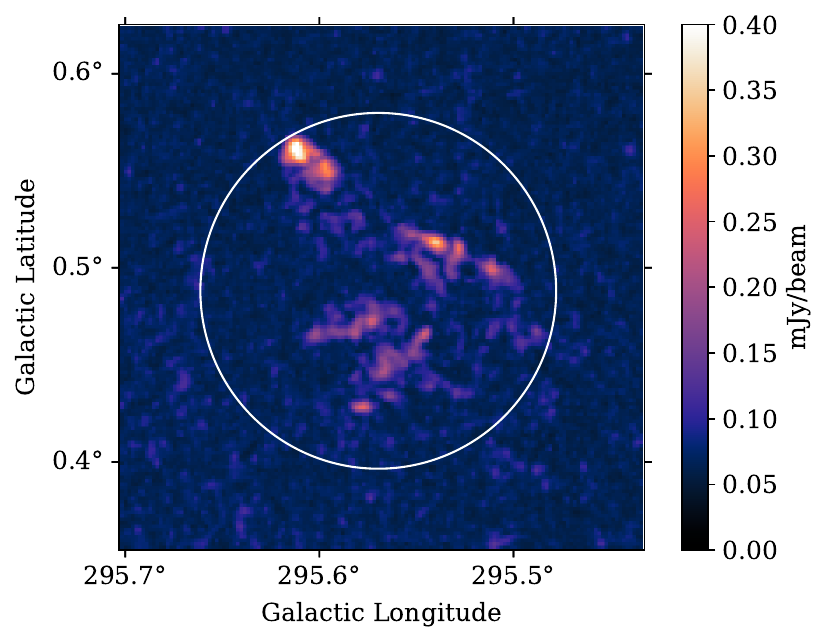}\label{fig:g295.6+0.5_pi}}
    \subfigure[G300.1$-$1.6 PI]{\includegraphics[width=0.32\textwidth]{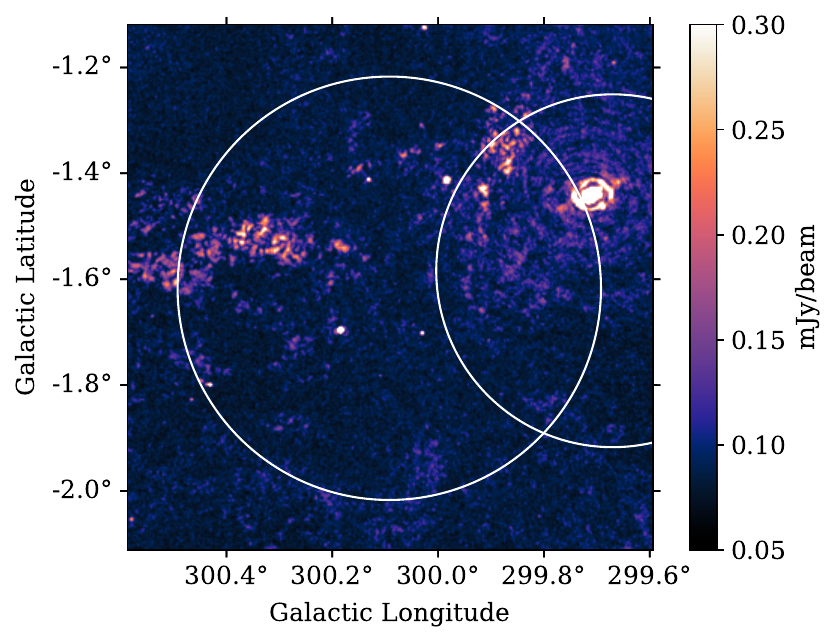}\label{fig:g300.1-1.6_pi}}
    \caption{ASKAP 943 MHz images of new polarized SNRs in total power and PI.} 
    \label{fig:new_snrs_pol}
\end{figure}

\paragraph{G282.1$-$4.1 (Figure~\ref{fig:g282.1-4.1}, \ref{fig:g282.1-4.1_pi})} This source has a unique morphology with one well-defined shell structure to the south and another to the east with several fainter filaments stretching between them. The southern shell is the brightest part of the SNR and is very clearly polarized. Follow-up work studying the source in greater depth utilizing EMU and POSSUM is underway (W. Jing et al., in preparation).

\begin{figure}
    \centering
    \subfigure[G283.1$-$0.6]{\includegraphics[width=0.33\textwidth]{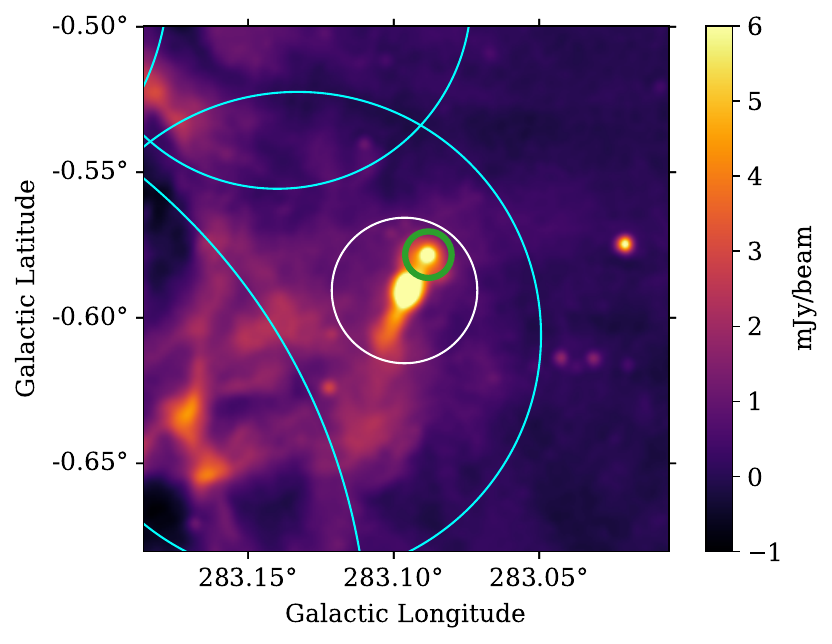}}
    \subfigure[Region surrounding G283.1$-$0.6]{\includegraphics[width=0.33\textwidth]{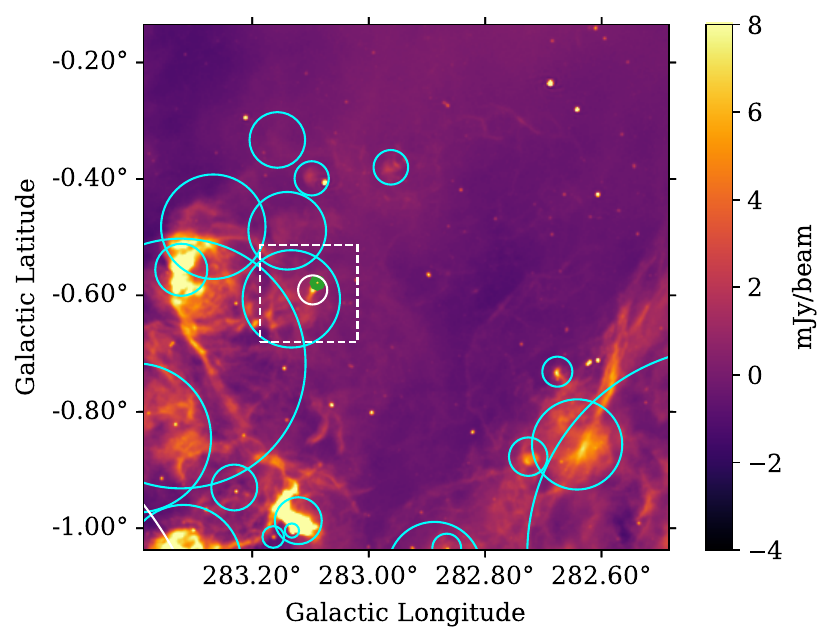}}
    \subfigure[Surrounding region with 943~MHz (red), 12~$\mu$m (blue), 22~$\mu$m (green)]{\includegraphics[width=0.28\textwidth]{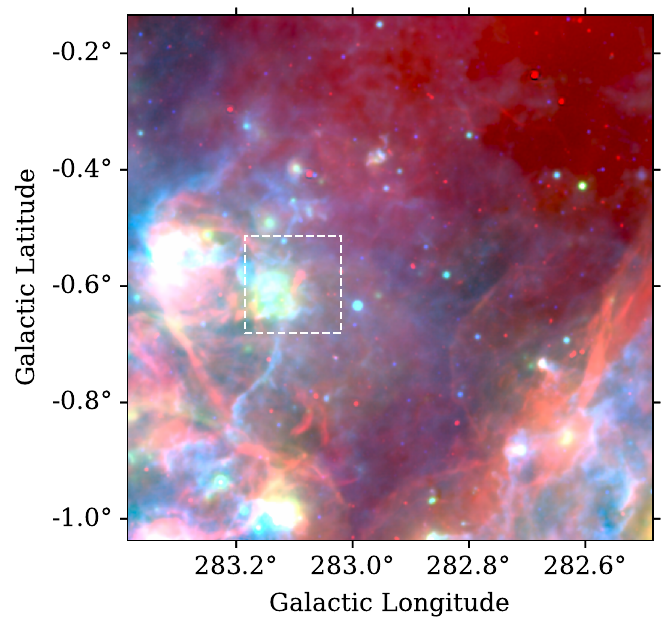}}
    \caption{ASKAP 943 MHz images of PWN candidate G283.1$-$0.6 and possible host SNR. The white circles indicate the position of G283.1$-$0.6. The green circles indicate the location of the pulsar. The cyan circles indicate the positions of \ion{H}{2} regions. The dashed white square is the region shown in panel (a).}
    \label{fig:G283.1-0.6}
\end{figure}

\paragraph{G283.1$-$0.6 (Figure~\ref{fig:G283.1-0.6})}  We find a tail of radio emission extending south from a young (38.6 kyrs) pulsar, J1015$-$5719 \citep{Kramer2003}. This morphology is characteristic of a bow shock PWN. East of this source there is a faint circular ring of emission that does not have an MIR counterpart, which could be the host SNR. There is also a longer filament extending to the southwest from this ring that is not visible in MIR. Many thin faint filaments can be seen in the region to the southwest of the PWN candidate. However, these features are located in a complex area that overlaps with multiple \ion{H}{2} regions, so the association is unclear, and lack of an MIR counterpart is difficult to establish. Only the pulsar is visible in polarization. While the host SNR is uncertain, we believe that the bright tail extending from the pulsar can be classified as a PWN. The values provided in Table~\ref{tab:newcands} are for the PWN only. The source is also visible in the SMGPS \citep{Goedhart2024}. At 1360~MHz, we calculate a flux density of 0.065 $\pm$ 0.003 Jy, producing a spectral index of $-$0.55 $\pm$ 0.13, steep for a PWN. As described in Section~\ref{sec:known_indices}, we expect the indices we calculate using only MeerKAT and ASKAP fluxes to be steeper than the true value.

\begin{figure}
    \centering
    \subfigure[G285.0$-$3.2]{\includegraphics[width=0.32\textwidth]{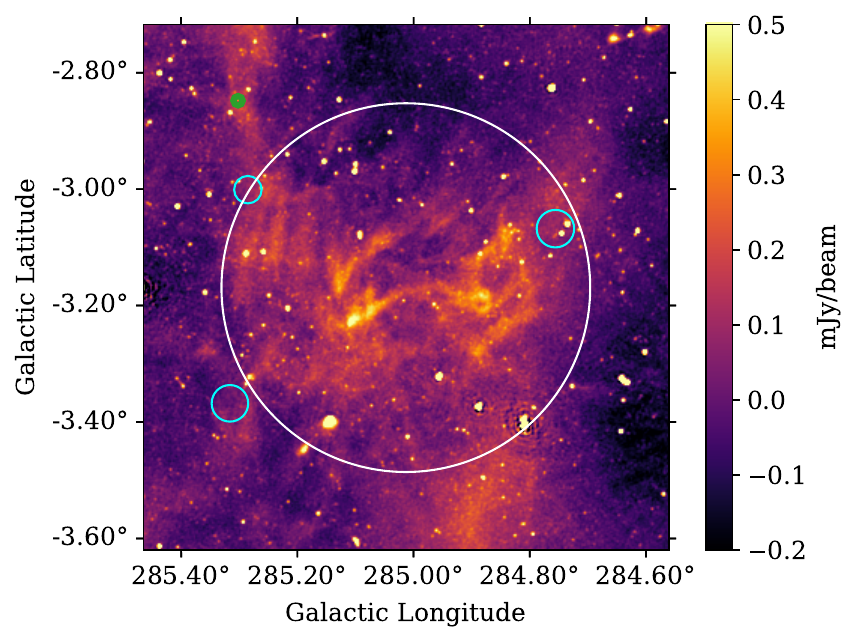}}
    \subfigure[G285.0$-$3.2 and G283.8$-$4.0]{\includegraphics[width=0.32\textwidth]{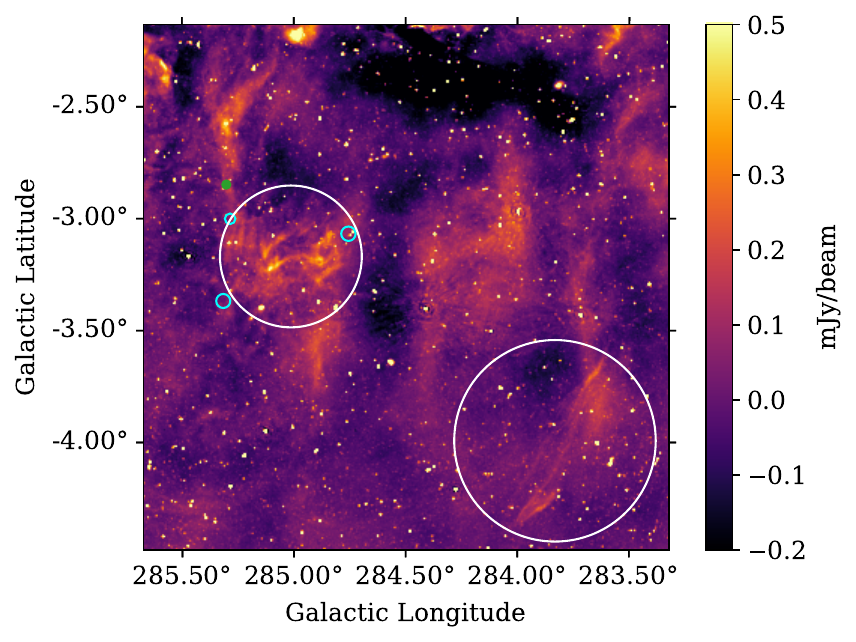}}
    \subfigure[G283.8$-$4.0]{\includegraphics[width=0.32\textwidth]{new_snrs/G283.8-4.0.pdf}}
    \subfigure[G285.0$-$3.2 PI]{\includegraphics[width=0.32\textwidth]{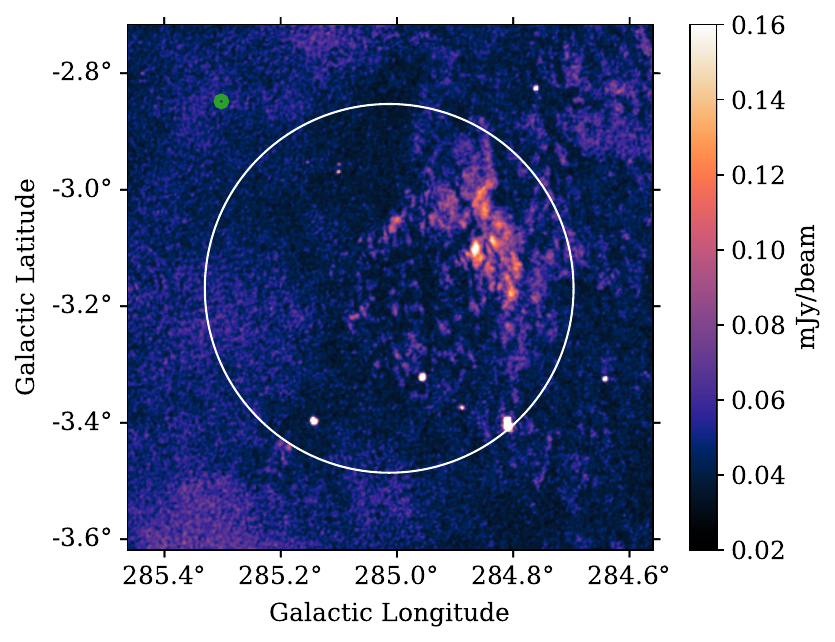}}
    \subfigure[G285.0$-$3.2 and G283.8$-$4.0 PI]{\includegraphics[width=0.32\textwidth]{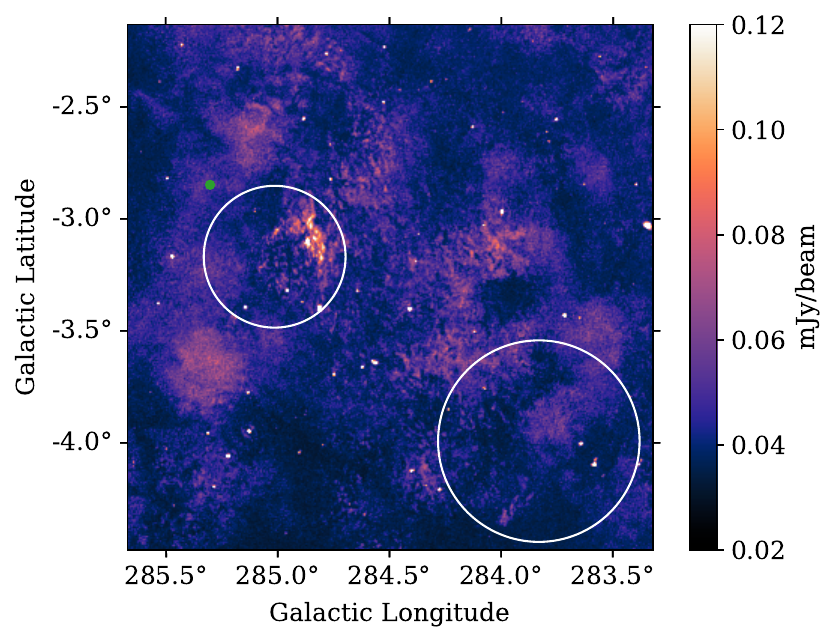}}
    \subfigure[G283.8$-$4.0 PI]{\includegraphics[width=0.32\textwidth]{pol/G283.8-4.0_pi.pdf}}
    \caption{ASKAP 943 MHz images of G285.9$-$3.2 and a possibly associated filament G283.8$-$4.0. The blue dashed circle marks the location of the X-ray source. The cyan circles indicate the positions of \ion{H}{2} regions.}
    \label{fig:g285.0-3.2}
\end{figure}

\paragraph{G285.0$-$3.2 (Figure~\ref{fig:g285.0-3.2})} This source has a unique radio morphology with multiple shell-like structures. The source is coincident with a candidate X-ray PWN, indicated by the blue circle in Figure~\ref{fig:g285.0-3.2}, proposed by \cite{Hare2019}. At this position, we also see a small ring-like structure in the radio with a larger non-circular shell around it. There is also a radio tail extending east from the location of the X-ray source toward the eastern filamentary emission. polarization can be seen around the western filaments that surround the X-ray source, though the polarized emission has a different morphology than the Stokes I. There is a long filament (labeled G283.8$-$4.0) to the southwest of the source, which is slightly polarized at its southernmost end. We also describe this filament in Section~\ref{sec:xray_optical} as it has been previously identified as a candidate in the optical \citep{Stupar2008}. This filament could be associated with G285.0$-$3.2, though it is at a considerable distance. If this is the case, its counterpart could be the thin shell of emission extending to the north of G285.0$-$3.2. However, if the PWN is young, as \cite{Hare2019} speculate, it is unlikely that the SNR could expand to this size in such a short period of time, in which case these larger filaments may comprise a separate SNR.

\paragraph{G289.6+5.8 (Figure~\ref{fig:g289.6+5.8}, \ref{fig:g289.6+5.8_pi})} A high-latitude source composed of two bright northwestern shells and a fainter shell that runs along the south. There is also a small bright nebula located near the centre of the source. This nebula is coincident with an X-ray source that has been identified as a possible low-mass X-ray binary \citep{Coleiro2013}. The northwestern shells are clearly polarized as is the nebula located near the centre, indicating that it may be a PWN. This source will be further described by Lazarevi\'c et al. (in preparation).

\begin{figure}
    \centering
    \subfigure{\includegraphics[width=0.4\textwidth]{new_snrs/G292.3+0.6.pdf}}
    \subfigure{\includegraphics[width=0.42\textwidth]{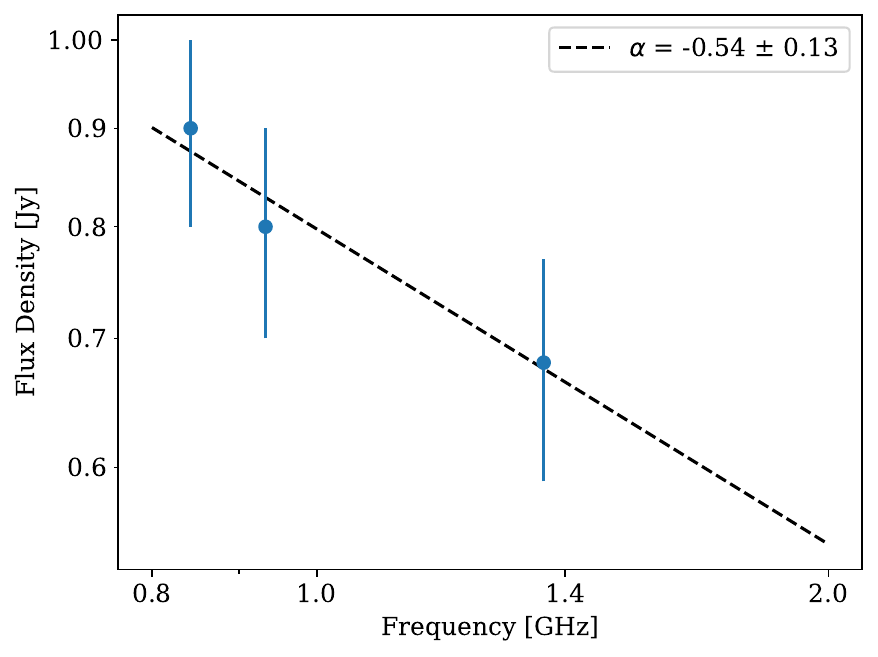}}
    \caption{ASKAP 943 MHz image of G292.3+0.6 and its spectrum. The 843~MHz value is from SUMSS, the 943~MHz value is from EMU, and the 1360~MHz value is from the SMGPS.}
    \label{fig:index_G292.3+0.6}
\end{figure}

\paragraph{G292.3+0.6 (Figure~\ref{fig:g292.3+0.6}, \ref{fig:g292.3+0.6_pi})} Roughly elliptical source with several internal filaments. Some of these internal filaments form a roughly circular shape. This source does not exhibit the typical SNR morphology as the edges are less well defined than is characteristic of a shock front. However, there is no MIR counterpart, and clear polarization can be seen coming from the centre of the source, extending in an arc along the north-south axis. In polarization, we see a bright circular source at the northern end of this arc, which is slightly offset from the point source we see in Stokes I. This circular region has a low negative RM (approx. $-$10 rad/m$^2$), whereas the rest of the source has a higher positive RM (approx. 40$-$50 rad/m$^2$). Perhaps this could be a PWN with the point source as its pulsar. The source is also visible in SUMSS and the SMGPS. Calculating flux densities at 843~MHz and 1360~MHz, we find a spectral index of \(-0.54 \pm 0.13\) (Figure~\ref{fig:index_G292.3+0.6}). This source has also been listed as an SNR candidate in the SMGPS \citep{Anderson2024}.

\paragraph{G295.6+0.5 (Figure~\ref{fig:g295.6+0.5}, \ref{fig:g295.6+0.5_pi})} A point source with a radio tail that extends back toward several shell-like features. This source exhibits the typical morphology of a bow shock PWN. There is clear polarization coming from the point source, likely a pulsar, and from the tail. The pulsar does not appear to be known as it is not included in the ATNF catalogue \citep{Manchester2005}. We believe that this source should be classified as a PWN based on its morphology and strong linear polarization. 

\begin{figure}
    \centering
    \subfigure[G299.3$-$1.5]{\includegraphics[width=0.32\textwidth]{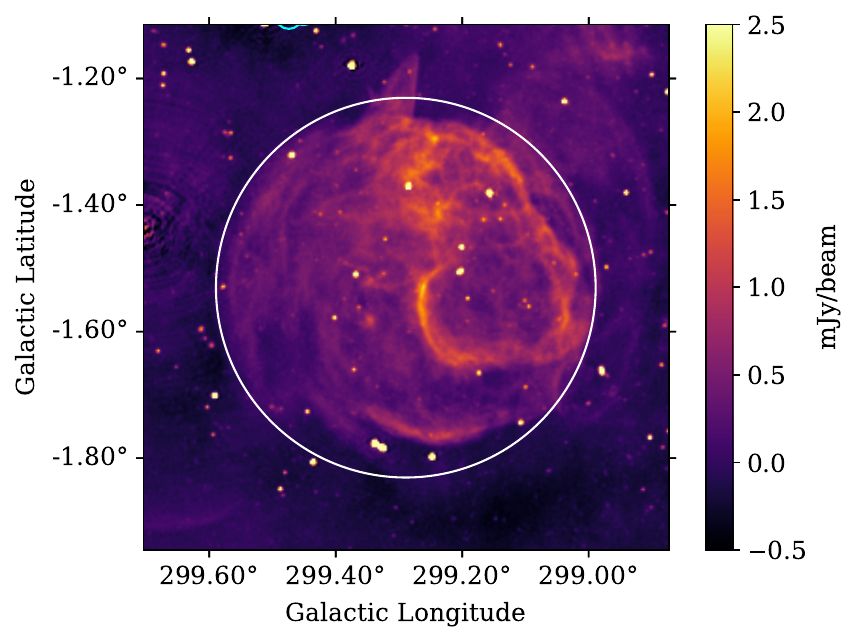}\label{fig:g299.3-1.5}}
    \subfigure[G309.8$-$2.6]{\includegraphics[width=0.32\textwidth]{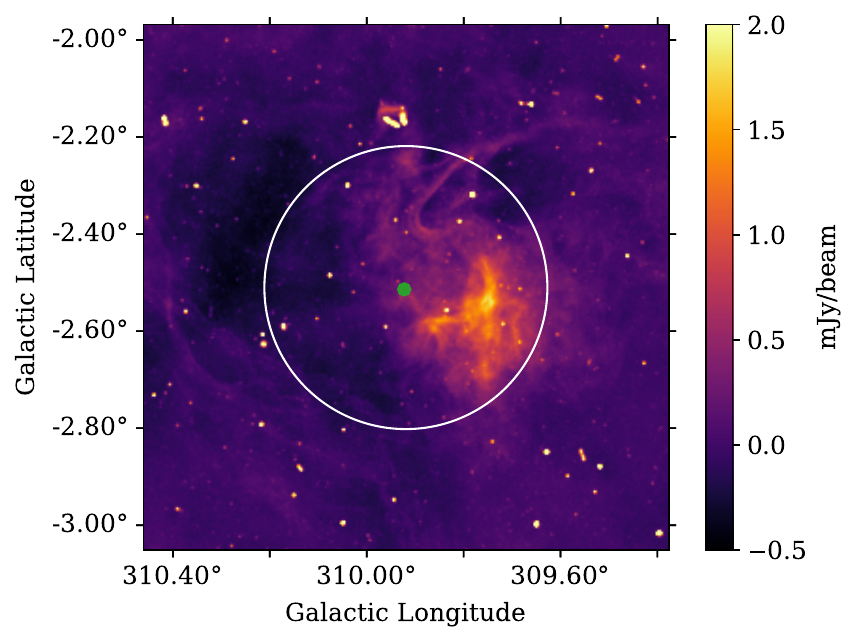}\label{fig:g309.8-2.6}}
    \subfigure[G310.7$-$5.4]{\includegraphics[width=0.32\textwidth]{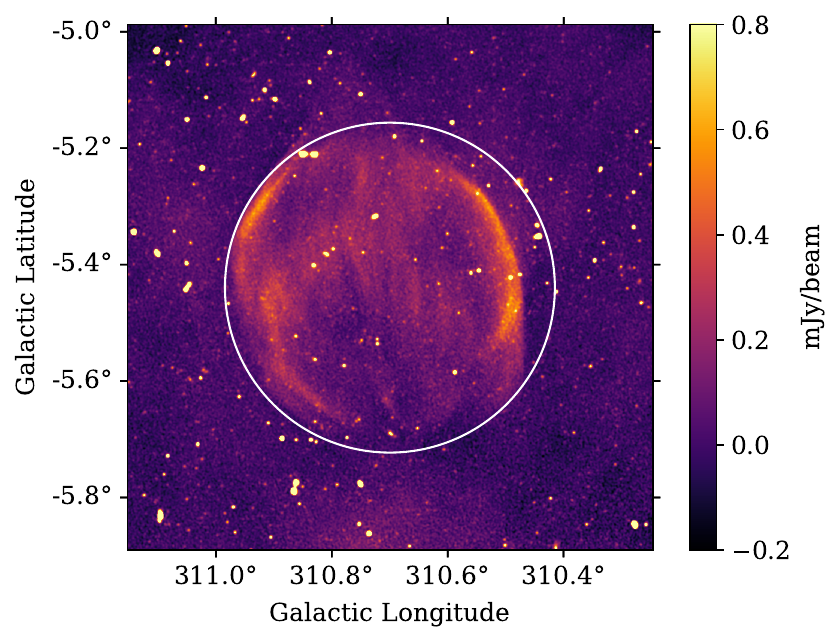}\label{fig:g310.7-5.4}}
    \subfigure[G299.3$-$1.5 PI]{\includegraphics[width=0.32\textwidth]{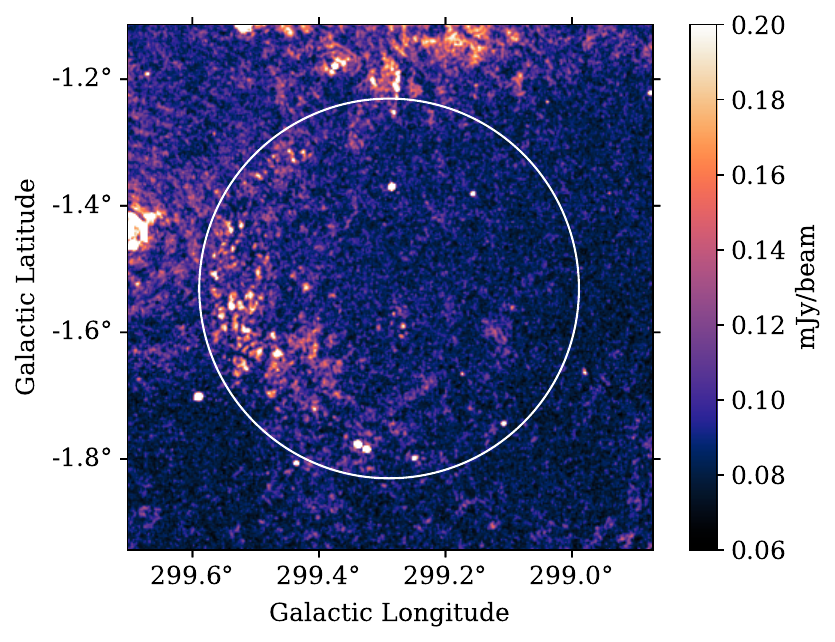}\label{fig:g299.3-1.5_pi}}
    \subfigure[G309.8$-$2.6 PI]{\includegraphics[width=0.32\textwidth]{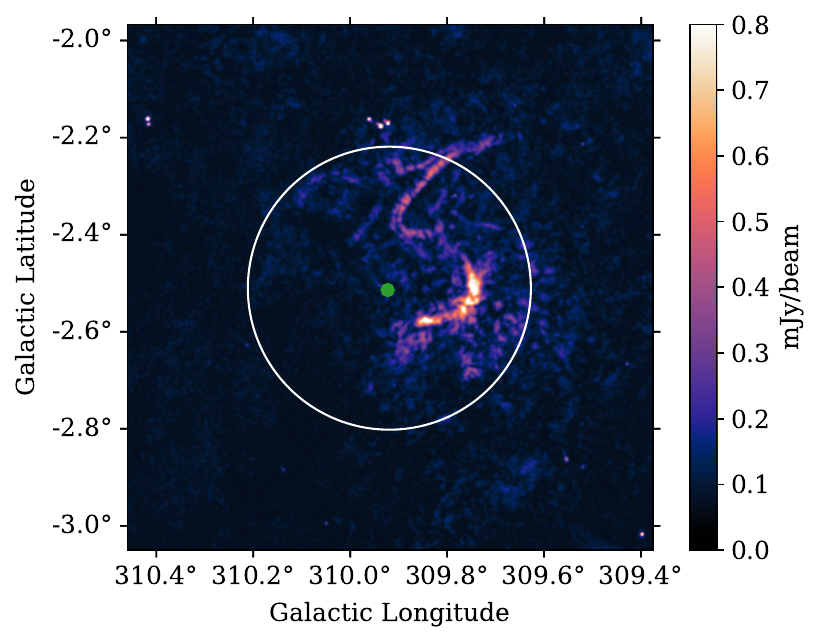}\label{fig:g309.8-2.6_pi}}
    \subfigure[G310.7$-$5.4 PI]{\includegraphics[width=0.32\textwidth]{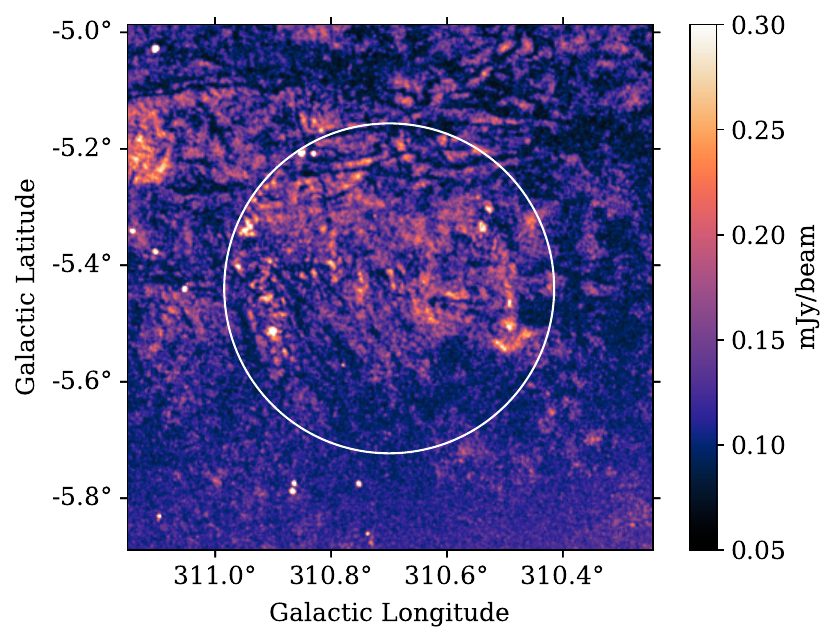}\label{fig:g310.7-5.4_pi}}
    \subfigure[G319.9$-$0.7]{\includegraphics[width=0.32\textwidth]{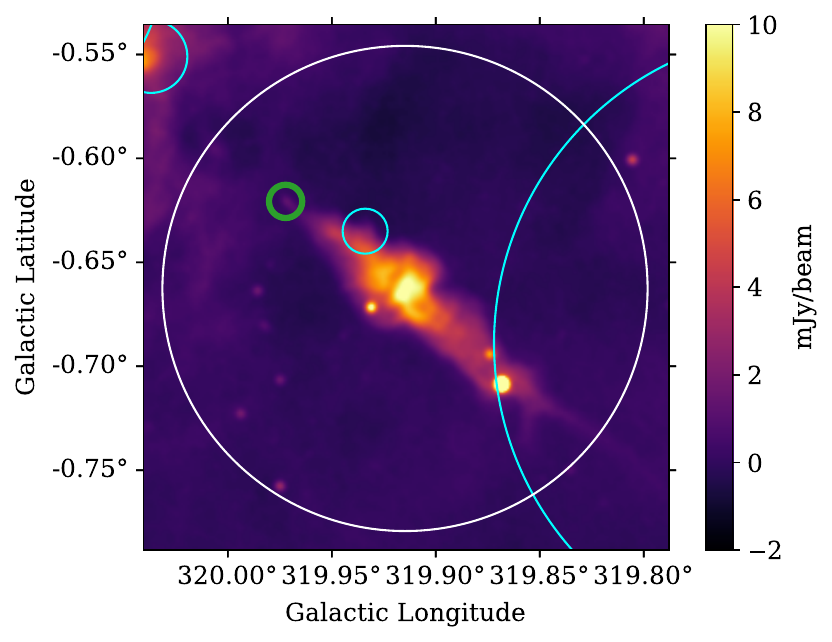}\label{fig:g319.9-0.7}}
    \subfigure[G321.3$-$3.9]{\includegraphics[width=0.32\textwidth]{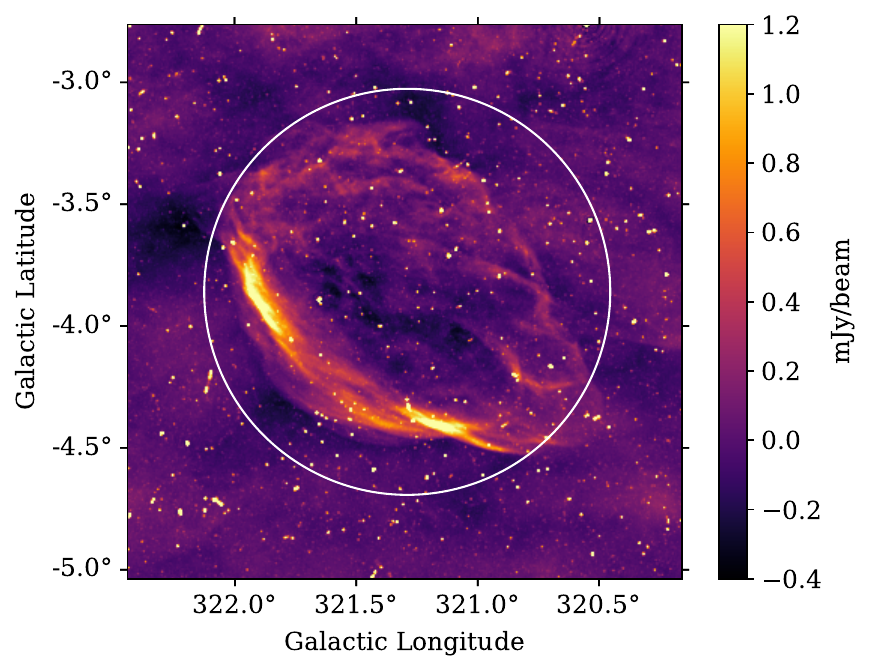}\label{fig:g321.3-3.9}} \\
    \subfigure[G319.9$-$0.7 PI]{\includegraphics[width=0.32\textwidth]{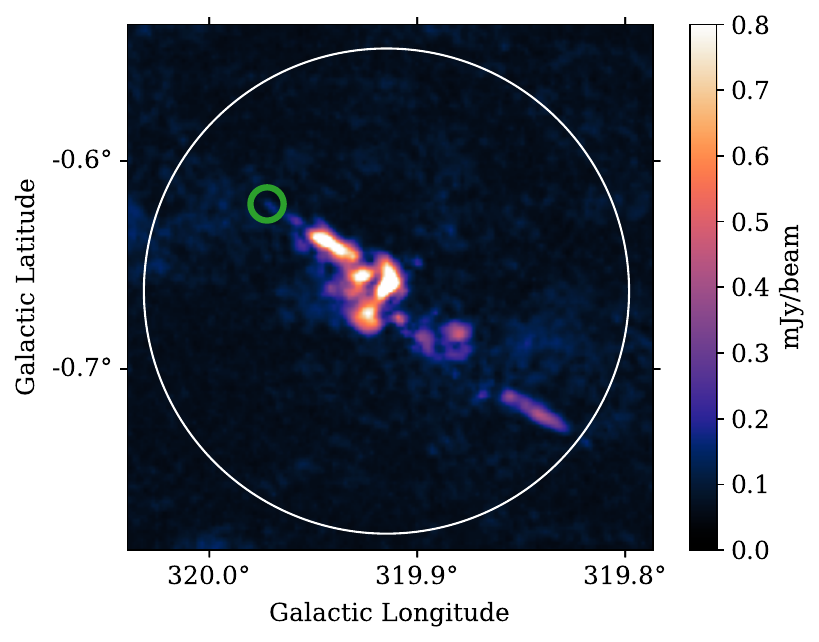}\label{fig:g319.9-0.7_pi}} 
    \subfigure[G321.3$-$3.9 PI]{\includegraphics[width=0.32\textwidth]{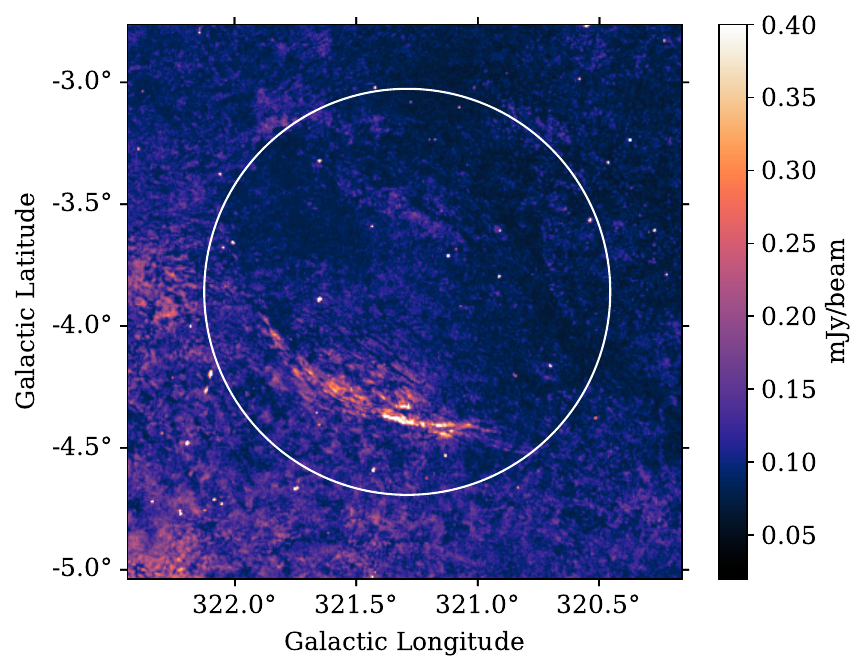}\label{fig:g321.3-3.9_pi}}
    \caption{ASKAP 943 MHz images of known SNR Candidates that appear in SNRcat \citep{Ferrand2012} and that we find to be polarized. The cyan circles indicate the positions of \ion{H}{2} regions. Green circles indicate the locations of pulsars.}
    \label{fig:known_cands_pol}
\end{figure}

\paragraph{G299.3$-$1.5 (Figure~\ref{fig:g299.3-1.5}, \ref{fig:g299.3-1.5_pi})} Proposed as an SNR candidate by \cite{Duncan1997} and \cite{Green2014}. The source is roughly circular with the brightest shells extending along the northwestern and southern edges. The outer edges are well defined and there are additional filamentary structures within the source. There is clear evidence of polarized emission coming from the eastern shell. There is also a faint tail of radio emission extending north from the source that is strongly polarized. Based on the morphology, this could be evidence of a relic PWN from an escaped pulsar, though the pulsar is not known. This source should be classified as an SNR based on polarization.

\paragraph{G300.1$-$1.6 (Figure~\ref{fig:g300.1-1.6}, \ref{fig:g300.1-1.6_pi})} A roughly elliptical source formed by a brighter northern shell and faint southern shell with several filamentary structures between them. There is also a filament that appears to run across the northern edge of the source, from the east to the northwest, in an upward arc. This filament, as well as the north-most shell, are polarized. 

\paragraph{G309.8$-$2.6 (Figure~\ref{fig:g309.8-2.6}, \ref{fig:g309.8-2.6_pi})} Proposed as an SNR candidate by \cite{Duncan1995,Duncan1997}. The source is associated with a young (7.3 kyrs) pulsar, J1357$-$6429 \citep{Camilo2004b}. A PWN has been detected in both radio and X-rays \citep{HESS2011,Chang2012}. ASKAP observations have revealed new features, in particular, a polarized ``S" shape. There is also a faint filament to the east of the source that may or may not be associated. The PWN sits near the geometric centre of the strongly polarized southwestern arc. These unique features require further study (W. Jing et al., in preparation) but the source should be definitively classified as an SNR.

\paragraph{G310.7$-$5.4 (Figure~\ref{fig:g310.7-5.4}, \ref{fig:g310.7-5.4_pi})} Proposed as an SNR candidate by \cite{Green2014}. The source is roughly circular with the brightest emission coming from the eastern and western shells. There are also some fainter filaments running through the centre, primarily in the north-south direction. The outer shells are both clearly polarized and thus the source should be classified as an SNR. This source will be studied in greater depth by \cite{BurgerScheidlin_inprep}.

\begin{figure}
    \centering
    \subfigure{\includegraphics[width=0.4\textwidth]{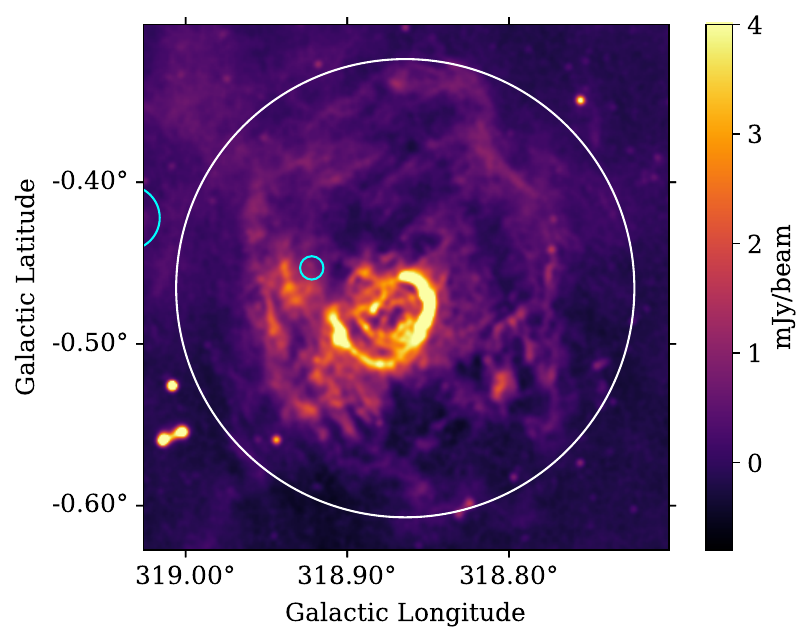}}
    \subfigure{\includegraphics[width=0.42\textwidth]{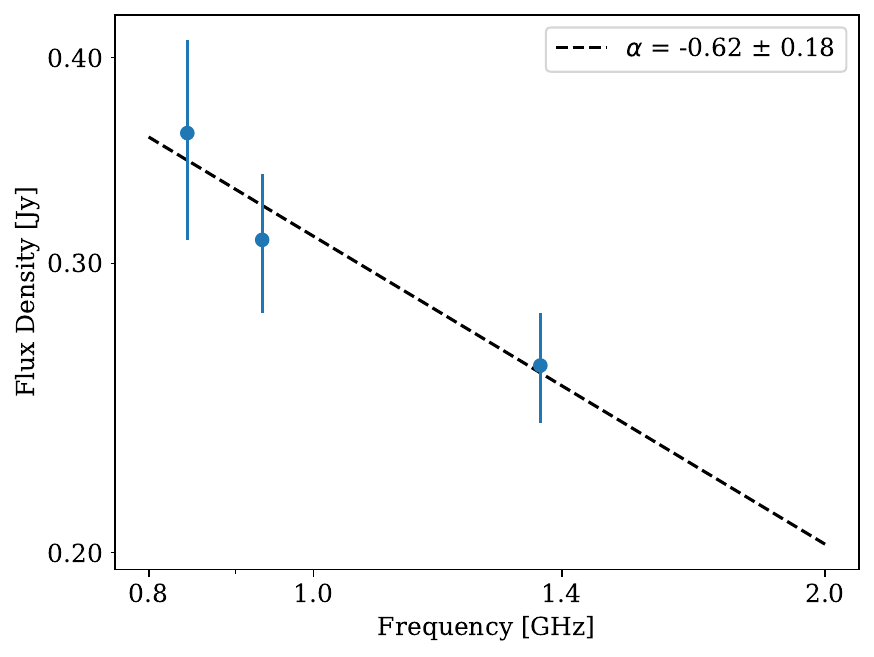}}
    \caption{ASKAP 943 MHz image of G318.9$-$0.5 and its spectrum. Flux values represent only the bright circular shell. The 843~MHz value is from SUMSS, the 943~MHz value is from EMU, and the 1360~MHz value is from SMGPS.}
    \label{fig:index_G318.9-0.5}
\end{figure}

\paragraph{G318.9$-$0.5 (Figure~\ref{fig:index_G318.9-0.5})} Bright, roughly circular thin shell with several fainter filaments surrounding it. The bright shell is also visible in SUMSS and the SMGPS so we are able to obtain a spectrum for the bright shell. We find a spectral index of $-$0.62 $\pm$ 0.18, consistent with expectations for an SNR. It is unclear whether or not the fainter filaments are associated. This source has also been listed as an SNR candidate in the SMGPS \citep{Anderson2024}.

\paragraph{G319.9$-$0.7 (Figure~\ref{fig:g319.9-0.7}, \ref{fig:g319.9-0.7_pi})} A bow shock PWN around pulsar J1509$-$5850 (154 kyrs) \citep{Kramer2003}. It was first observed in the radio by \cite{Whiteoak1996} and further studied by \cite{Hui2007}. \cite{Ng2010} report on its radio polarization and argue that this source should be confirmed as a bow shock PWN. Our findings support this conclusion as we also observe a high degree of linear polarization. 

\begin{figure}
    \centering
    \subfigure{\includegraphics[width=0.4\textwidth]{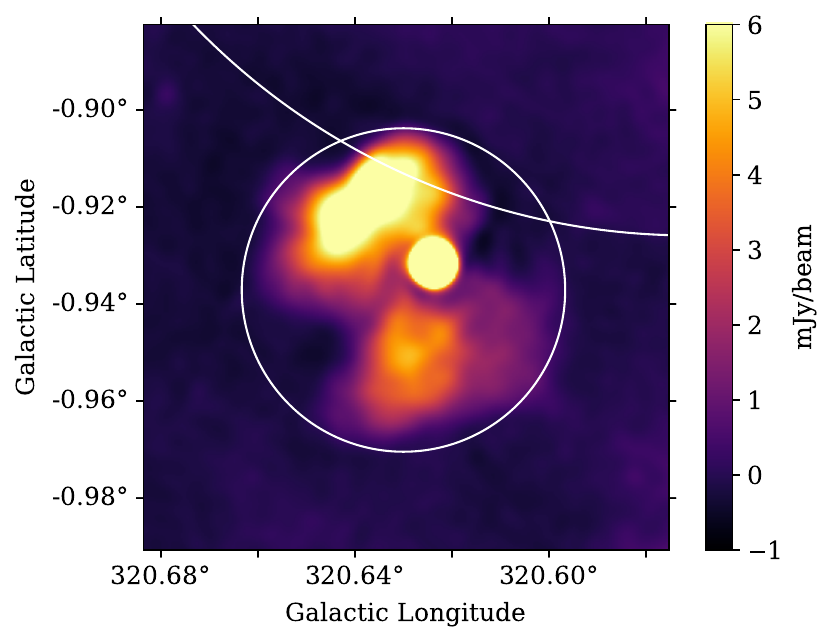}}
    \subfigure{\includegraphics[width=0.42\textwidth]{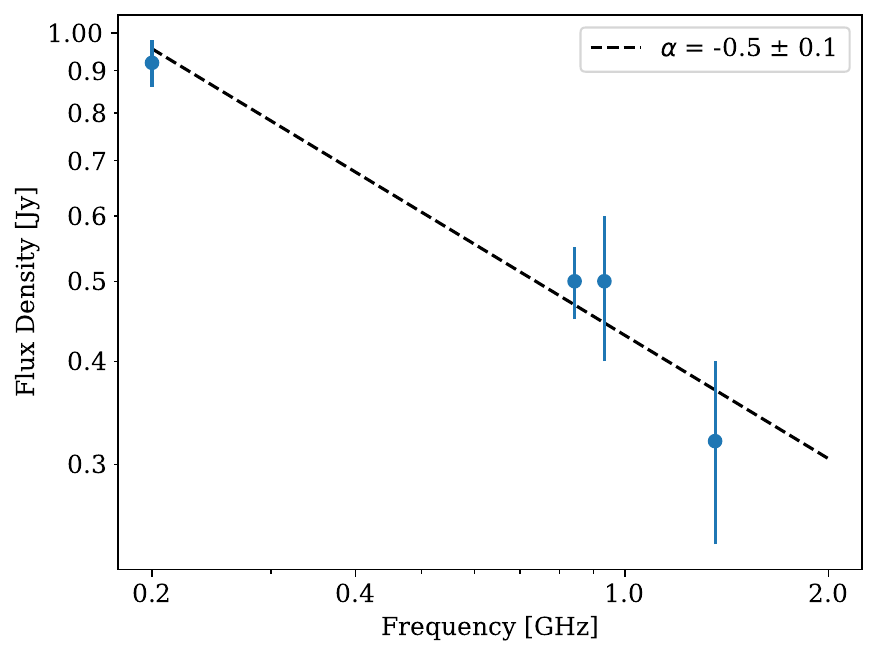}}
    \caption{ASKAP 943 MHz image of G320.6$-$0.9 and its spectrum. The 200~MHz value is from GLEAM, the 843~MHz value is from SUMSS, the 943~MHz value is from EMU, and the 1360~MHz value is from SMGPS.}
    \label{fig:index_G320.6-0.9}
\end{figure}

\paragraph{G320.6$-$0.9 (Figure~\ref{fig:index_G320.6-0.9})} Small, bright source first identified as an SNR candidate by \cite{Whiteoak1996}. We calculate flux densities for the source at four frequencies using data from GLEAM, SUMSS, EMU, and the SMGPS and find a spectral index of $-$0.5 $\pm$ 0.1, consistent with the expected value for a supernova remnant.

\paragraph{G321.3$-$3.9 (Figure~\ref{fig:g321.3-3.9}, \ref{fig:g321.3-3.9_pi})} A large high-latitude source proposed as an SNR candidate by \cite{Duncan1997} and again by \cite{Green2014}. The southern shell is clearly polarized and thus, the source should be classified as an SNR. Our findings support the recent work by \cite{Mantovanini2024}, which finds the source is a polarized SNR. \rev{This classification is further supported by observations in the optical published by \cite{Fesen2024}.}

\subsubsection{Probable SNRs / Strong candidates}

Here, we list 18 SNR candidates that we deem to be strong candidates, 12 of which have been identified as SNR candidates in previous works. Most are sources that exhibit characteristic SNR morphologies, i.e. well-defined shell-like features, and lack MIR counterparts. We classify five of these sources as strong candidates based on spatial coincidence and a probable relationship with a young pulsar. For three of the sources, we provide spectral indices determined using the ASKAP and MeerKAT data. The indices determined are consistent with expected SNR values. However, we do not include them in the new SNRs section because of the lack of reliability with these indices, as described in Section~\ref{sec:known_indices}.

\begin{figure}
    \centering
    \subfigure[G281.2$-$0.1]{\includegraphics[width=0.32\textwidth]{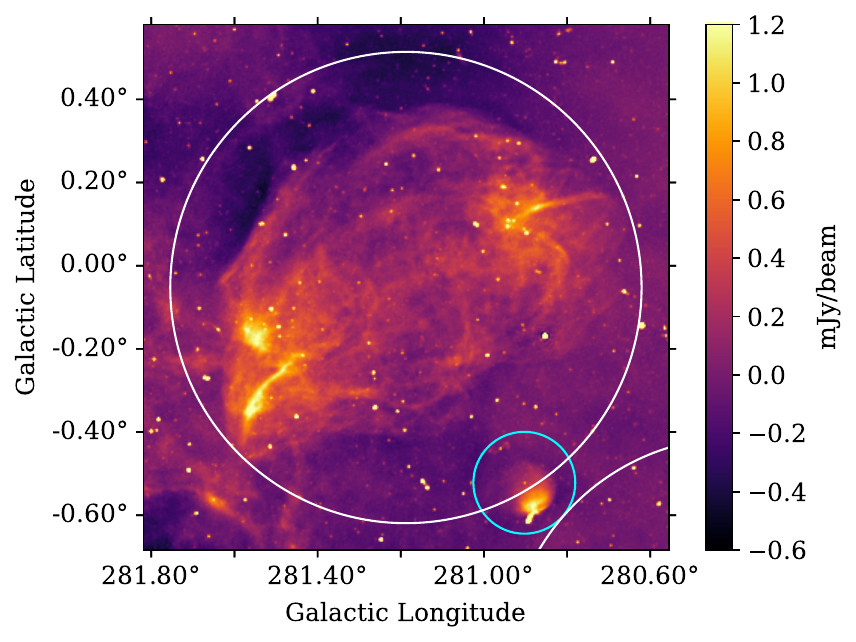}\label{fig:g281.2-0.1}}
     \subfigure[G286.5+1.0]{\includegraphics[width=0.32\textwidth]{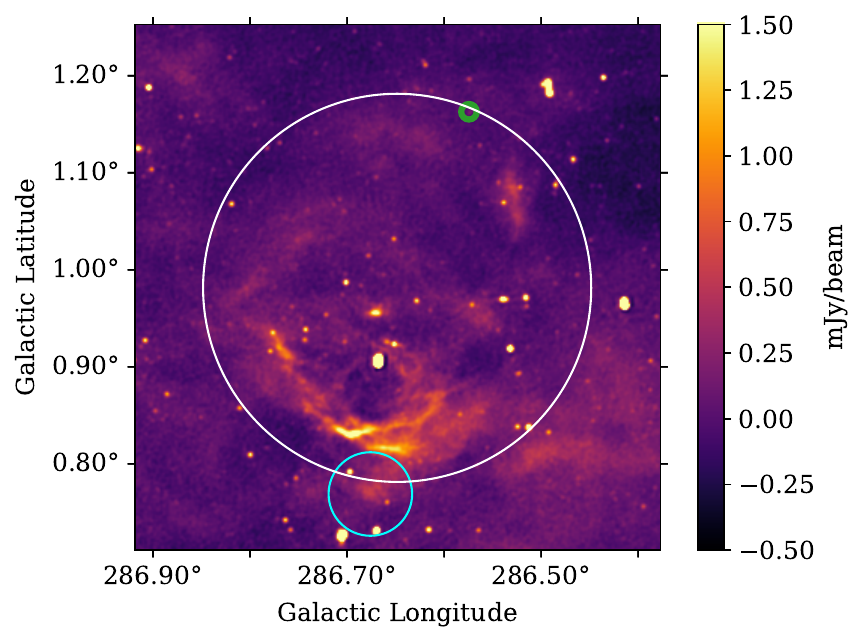}\label{fig:g286.5+1.0}}
    \subfigure[G289.2$-$0.8]{\includegraphics[width=0.32\textwidth]{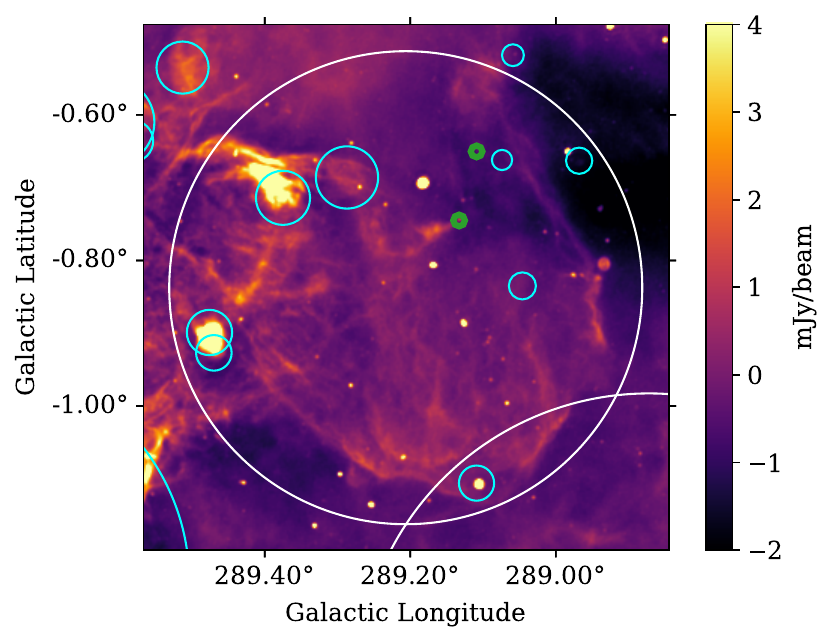}\label{fig:g289.2-0.8}}
    \subfigure[G291.2$-$0.6]{\includegraphics[width=0.32\textwidth]{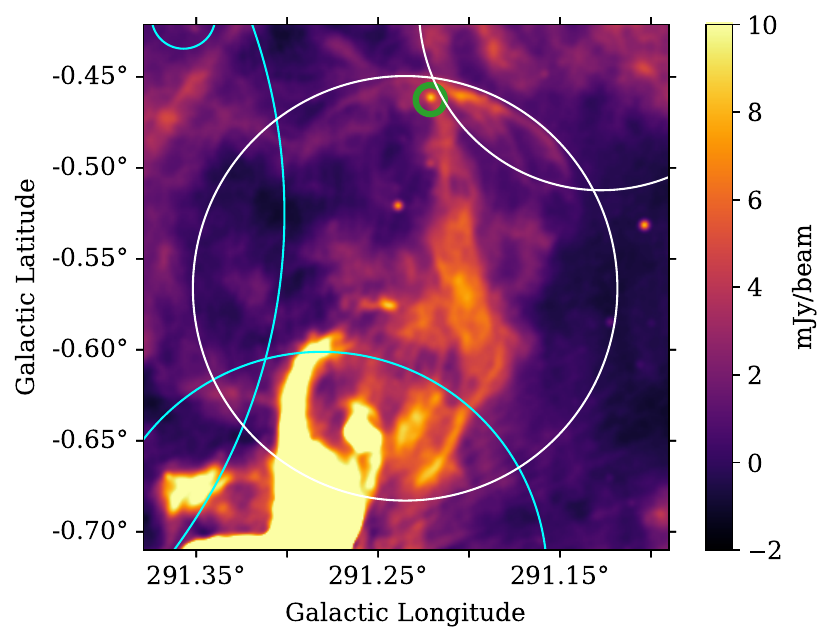}\label{fig:g291.2-0.6}}
    \subfigure[G296.6$-$0.4]{\includegraphics[width=0.32\textwidth]{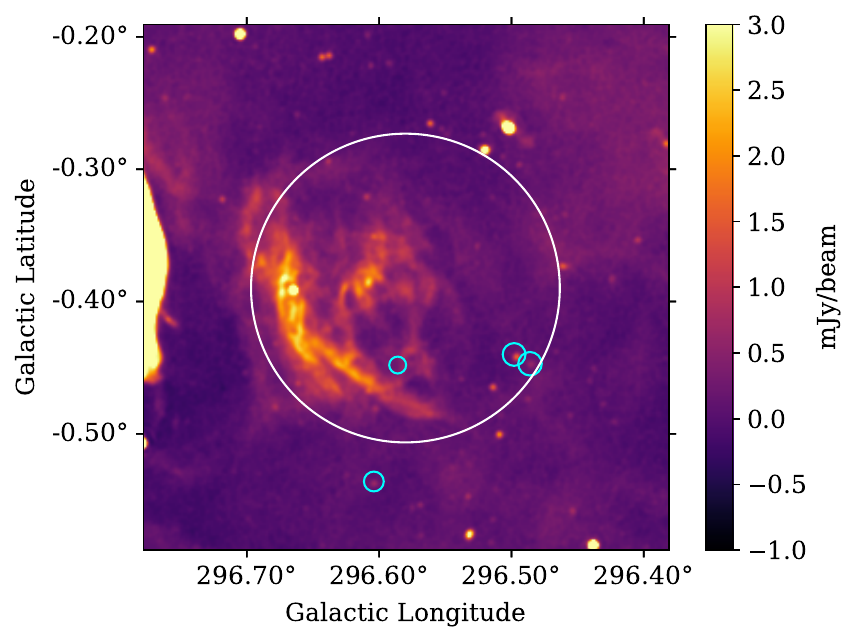}\label{fig:g296.6-0.4}}
    \subfigure[G298.0$-$0.2]{\includegraphics[width=0.32\textwidth]{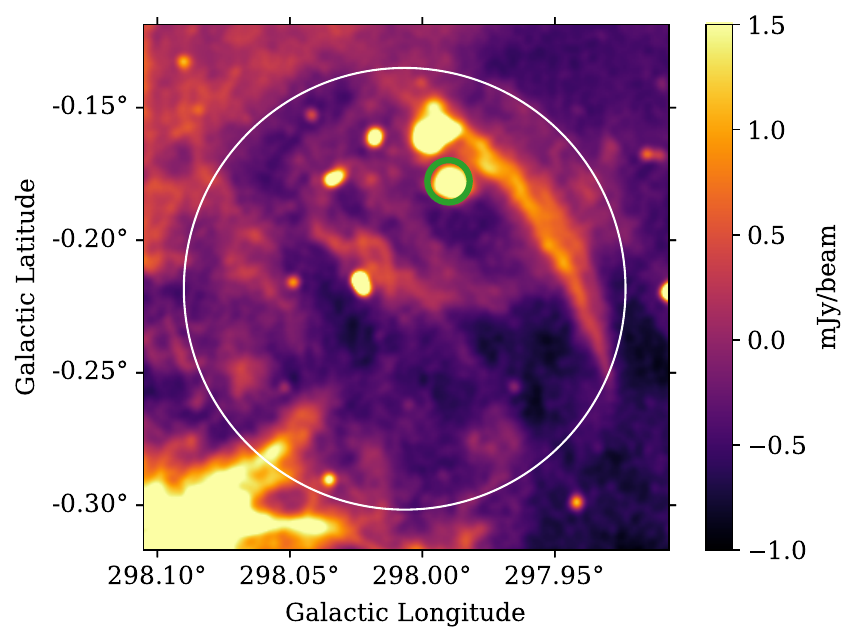}\label{fig:g298.0-0.2}}
    \subfigure[G299.7$-$0.0]{\includegraphics[width=0.32\textwidth]{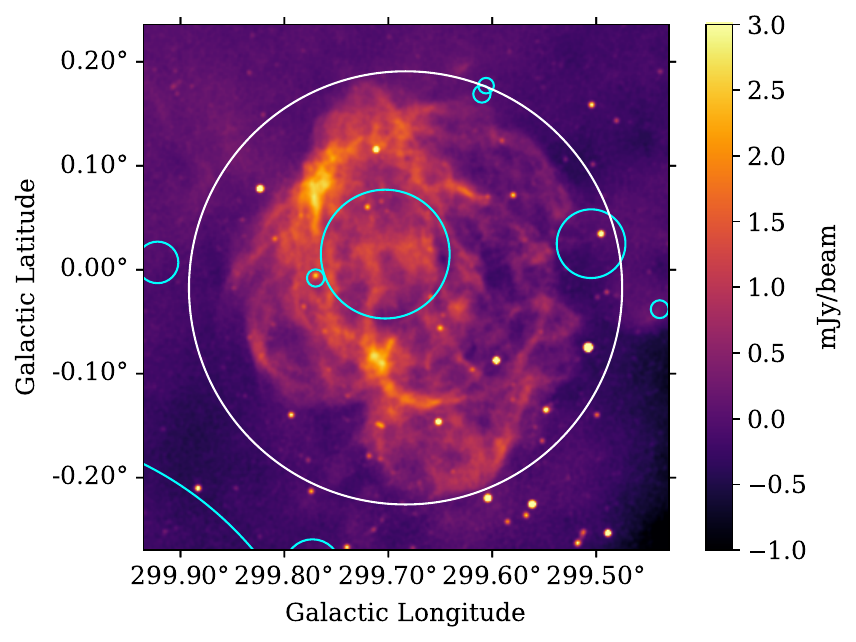}\label{fig:g299.7-0.0}}
    \subfigure[G305.4$-$2.2]{\includegraphics[width=0.32\textwidth]{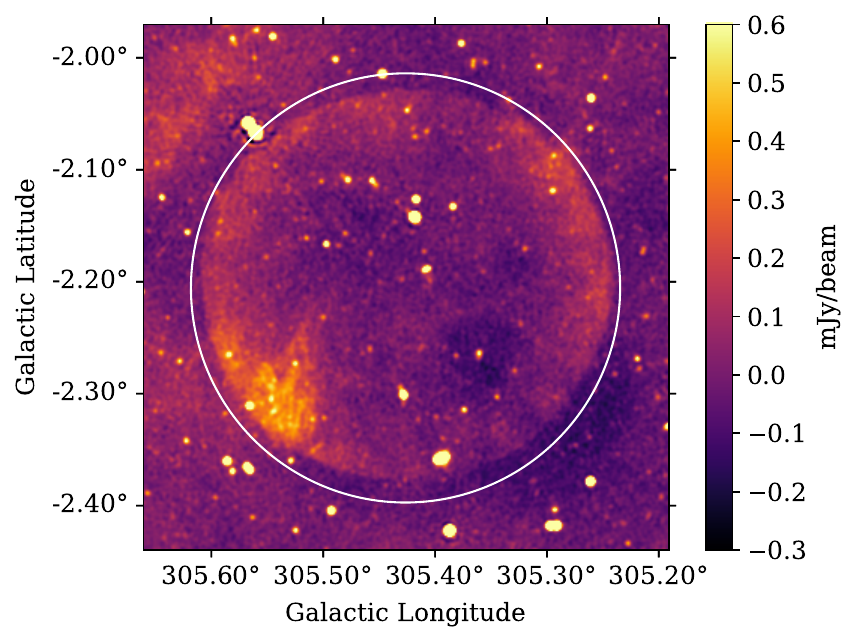}\label{fig:g305.4-2.2}}
    \subfigure[G308.7+1.4]{\includegraphics[width=0.32\textwidth]{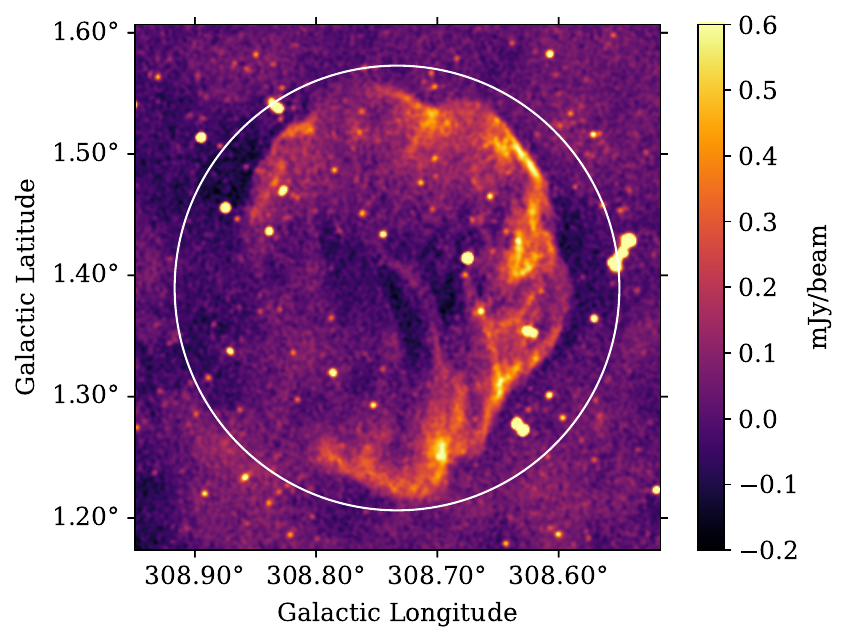}\label{fig:g308.7+1.4}}
    \subfigure[G310.9$-$0.3]{\includegraphics[width=0.32\textwidth]{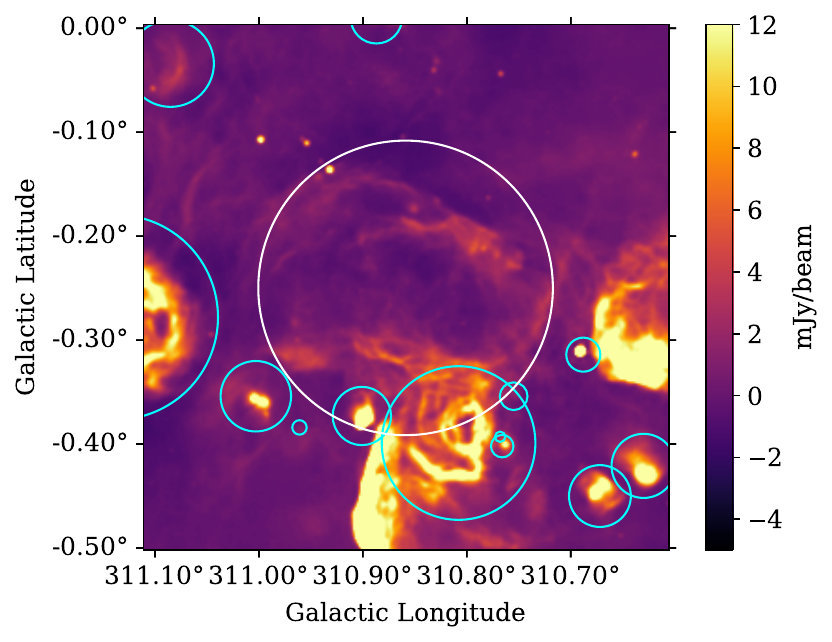}\label{fig:g310.9-0.3}}
    \subfigure[G311.0$-$0.6]{\includegraphics[width=0.32\textwidth]{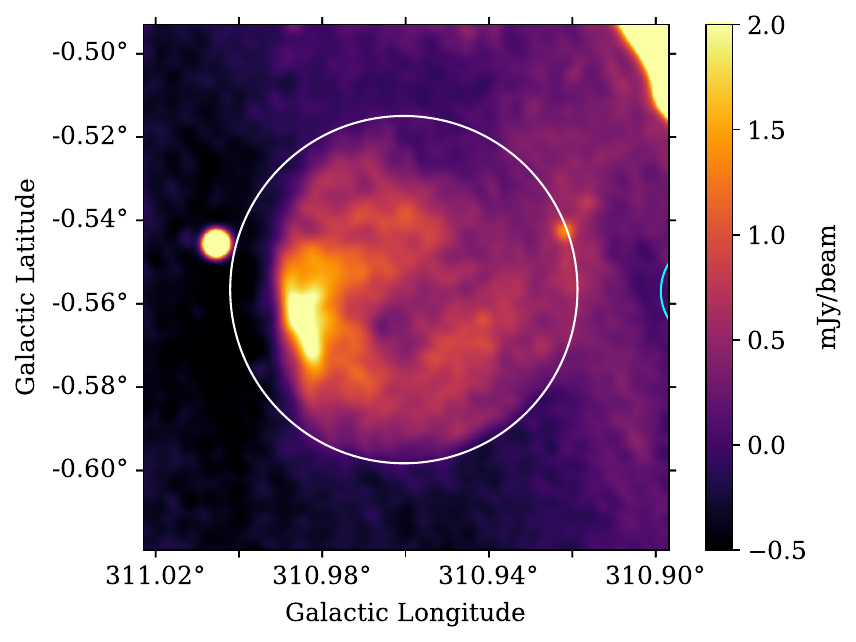}\label{fig:g311.0-0.6}}
    \subfigure[G311.1+1.4]{\includegraphics[width=0.32\textwidth]{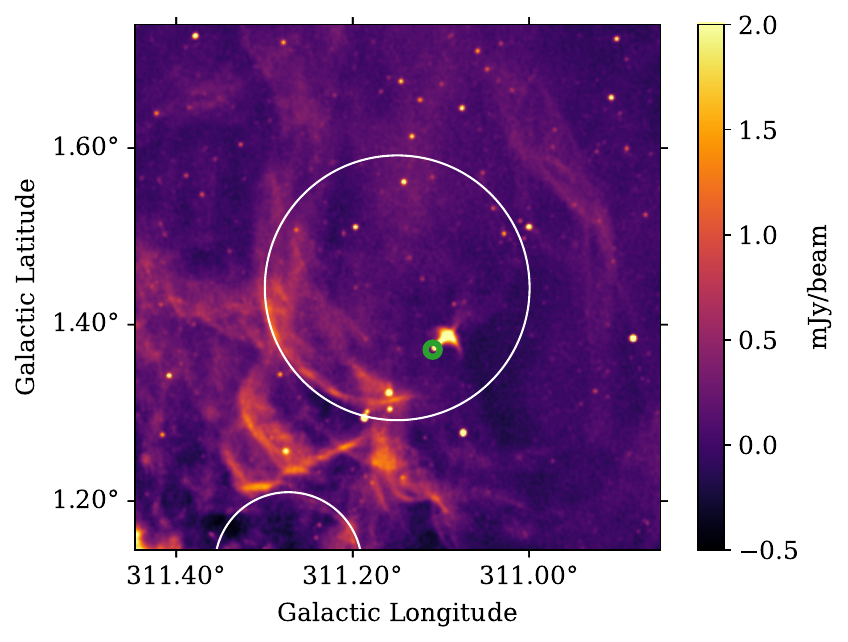}\label{fig:g311.1+1.4}}
    \caption{ASKAP 943 MHz images of Probable SNRs / Strong candidates. The green circles indicate the locations of young pulsars. The cyan circles indicate the positions of \ion{H}{2} regions.}
    \label{fig:good_cands}
\end{figure}

\paragraph{G281.2$-$0.1 (Figure~\ref{fig:g281.2-0.1})} First identified as an SNR candidate by \cite{Duncan1995}. The source has well-defined edges with many internal filamentary structures. There is a particularly bright filament running roughly through the centre of the source, which is brightest at the southeast and northwest ends.

\paragraph{G286.5+1.0 (Figure~\ref{fig:g286.5+1.0})} Shell-like structure to the south with fainter emission extending north. North of this shell, we find a young (40.3 kyrs) pulsar, J1044$-$5737 \citep{Saz2010}. This source has also been listed as an SNR candidate in the SMGPS \citep{Anderson2024}.

\paragraph{G289.2$-$0.8 (Figure~\ref{fig:g289.2-0.8})} We find a radio tail extending from the known young (53.5 kyrs) pulsar J1055$-$6028 \citep{Keith2009}, exhibiting the typical morphology of a bow shock PWN. There are also some shell-like structures to the south and west of the PWN candidate that may be the host SNR, but this association is not certain. This source has also been listed as an SNR candidate in the SMGPS \citep{Anderson2024}.

\paragraph{G291.2$-$0.6 (Figure~\ref{fig:g291.2-0.6})} Shell-like feature to the north that is spatially coincident with the young pulsar J1112$-$6103 (32.7 kyrs) \citep{Manchester2001}. There is also a tail of radio emission extending south from the pulsar, possibly evidence of a bow shock PWN.

\paragraph{G296.6$-$0.4 (Figure~\ref{fig:g296.6-0.4})} An SNR candidate proposed by \cite{Green2014} that appears in SNRcat \citep{Ferrand2012}. The source is roughly elliptical, with its most distinctive feature being a bright southeastern shell. Some additional fainter filamentary structures can be seen to the west of this shell.

\paragraph{G298.0$-$0.2 (Figure~\ref{fig:g298.0-0.2})} Shell-like feature near a very young (2.7 kyrs) pulsar J1208$-$6238 \citep{Clark2016}. \cite{Bamba2020} detected a possible X-ray PWN around the pulsar. The source is also visible in the SMGPS and discussed by \cite{Goedhart2024} as a possible PWN. They find an in-band index of $-$0.33 for the candidate PWN. We calculate a 943~MHz flux density of 0.020 $\pm$ 0.001 Jy and a 1360~MHz flux density of 0.017 $\pm$ 0.001 Jy, resulting in a spectral index of $-$0.43 $\pm$ 0.08. It is not clear if the shell and PWN candidate are related. 

\paragraph{G299.7$-$0.0 (Figure~\ref{fig:g299.7-0.0})} Well-defined outer edges with many internal filaments. There are a few overlapping \ion{H}{2} regions. The large \ion{H}{2} region near the centre has a clear MIR counterpart, but the rest of the source does not. This source has also been listed as an SNR candidate in the SMGPS \citep{Anderson2024}. We calculate a spectral index using a 1360~MHz flux density from MeerKAT. The resulting value, $-$0.57 $\pm$ 0.08, is consistent with the expected value for SNRs. However, we note that this value should be taken with caution as we see evidence of missing flux, particularly in the SMGPS data, which may be steepening the index. Conversely, we see evidence that the overlapping \ion{H}{2} region has a flattening effect on the index.

\paragraph{G305.4$-$2.2 (Figure~\ref{fig:g305.4-2.2})} Almost perfectly circular emission with a sharp, shell-like edge. There is some bright extended emission along the southeastern edge. The source is studied in greater depth in \cite{Filipovic2025}.

\paragraph{G308.7+1.4 (Figure~\ref{fig:g308.7+1.4})} Well-defined partial outer shell that is brightest along its western half. Some internal filaments can also be seen. This source was published by \cite{Lazarevic2024}, who nicknamed it ``Raspberry" and identify it as an SNR candidate using EMU.

\paragraph{G310.9$-$0.3 (Figure~\ref{fig:g310.9-0.3})} This source was proposed as an SNR candidate by \cite{Green2014} and appears in SNRcat \citep{Ferrand2012}. A northern and southern shell form a roughly elliptical shape, elongated along the east-west axis. The source is located in a complex area with many surrounding \ion{H}{2} regions. 

\begin{figure}
    \centering
    \subfigure[G317.4$-$1.1]{\includegraphics[width=0.32\textwidth]{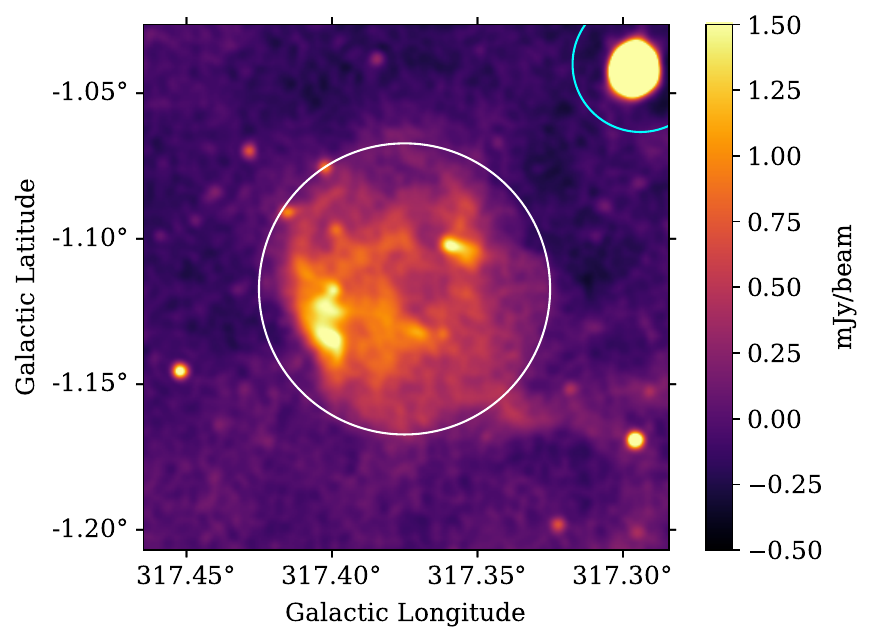}\label{fig:g317.4-1.1}}
    \subfigure[G317.6+0.9]{\includegraphics[width=0.32\textwidth]{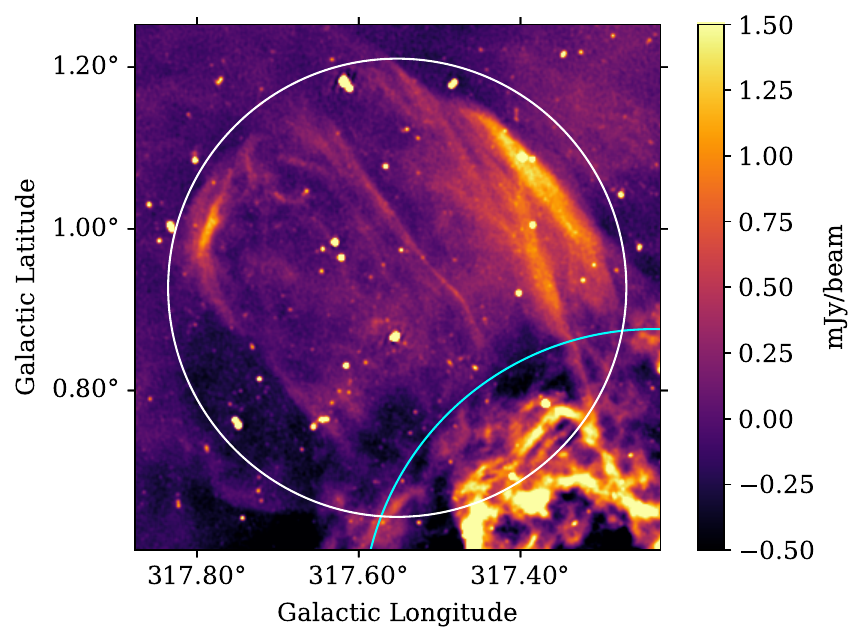}\label{fig:g317.6+0.9}}
    \subfigure[G320.8$-$0.3]{\includegraphics[width=0.32\textwidth]{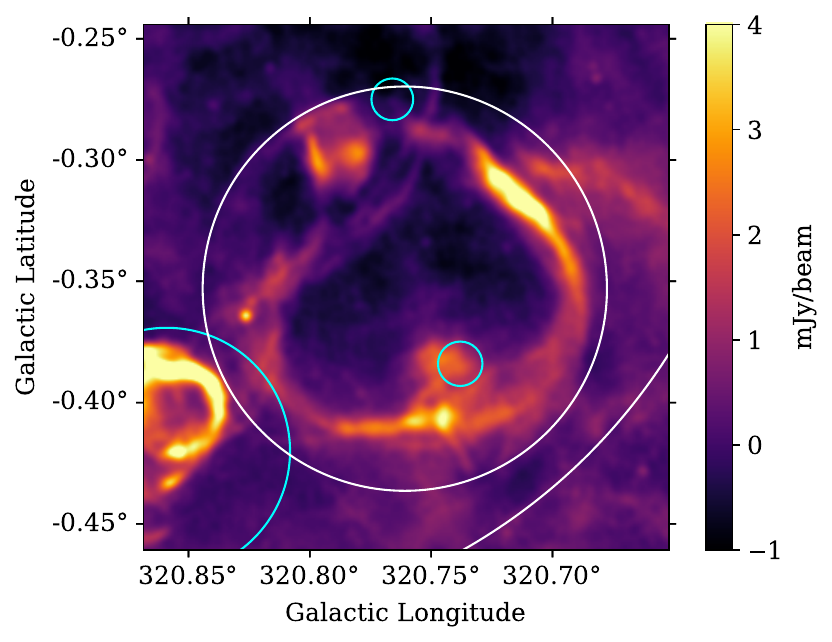}\label{fig:g320.8-0.3}}
    \subfigure[G320.9$-$0.7]{\includegraphics[width=0.32\textwidth]{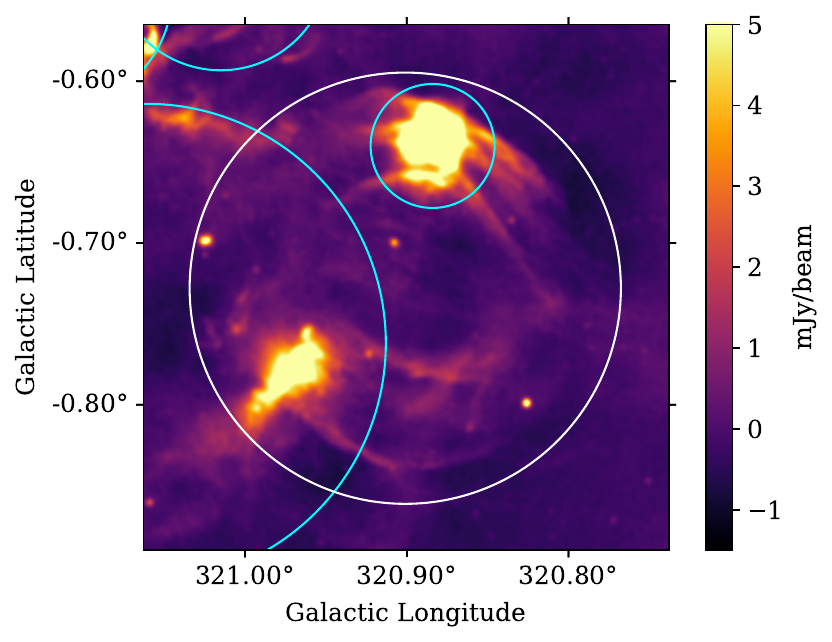}\label{fig:g320.9-0.7}}
    \subfigure[G321.3$-$0.9]{\includegraphics[width=0.32\textwidth]{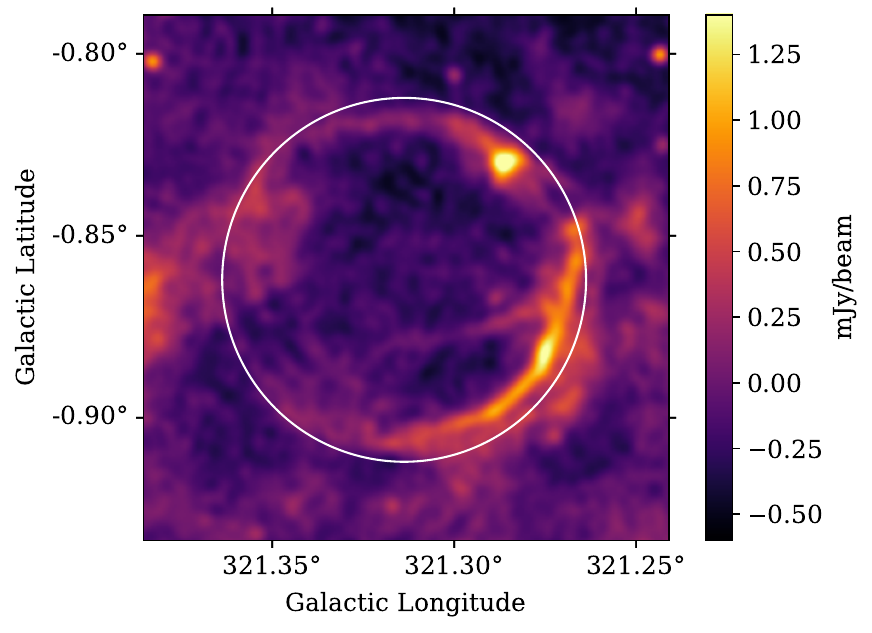}\label{fig:g321.3-0.9}}
    \subfigure[G322.7$-$0.6]{\includegraphics[width=0.32\textwidth]{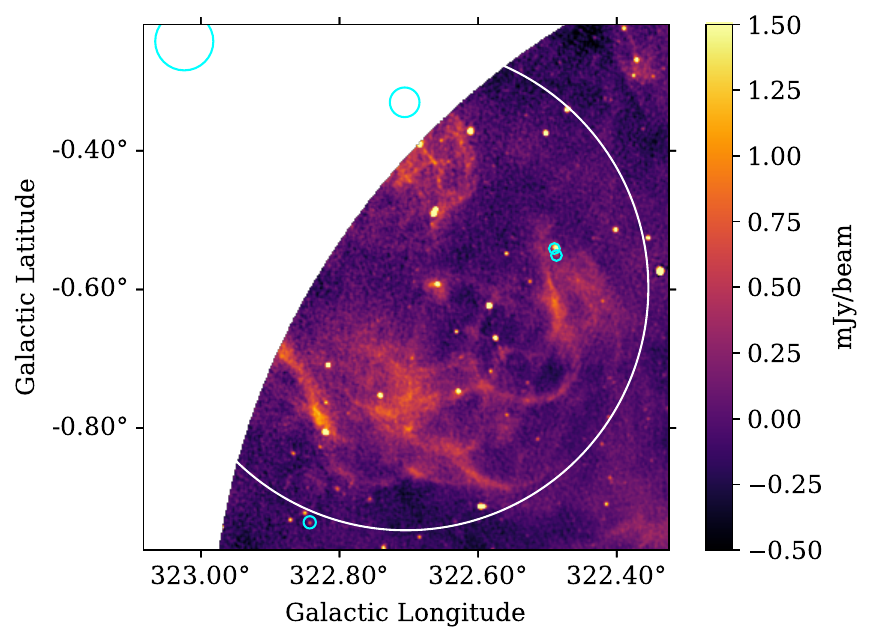}\label{fig:g322.7-0.6}}
    \caption{ASKAP 943 MHz images of Probable SNRs / Strong candidates. The cyan circles indicate the positions of \ion{H}{2} regions.}
    \label{fig:good_cands2}
\end{figure}

\paragraph{G311.0$-$0.6 (Figure~\ref{fig:g311.0-0.6})} Small roughly circular nebula, brightest along the eastern edge. We calculate a 1360~MHz flux density, using the SMGPS \citep{Goedhart2024}, of 0.13 $\pm$ 0.01 Jy, resulting in a spectral index of $-$0.55 $\pm$ 0.10. This is consistent with the expected values for an SNR. This source has also been listed as an SNR candidate in the SMGPS \citep{Anderson2024}.

\paragraph{G311.1+1.4 (Figure~\ref{fig:g311.1+1.4})} A tail of radio emission extending from the pulsar J1358$-$6025 (318 kyrs) \citep{Abdollahi2022} that could be a bow shock PWN. We also see several shell-like features to the southeast of the PWN candidate that could be related. This source was also found in the SMGPS and discussed as a possible PWN \citep{Goedhart2024}.

\paragraph{G317.4$-$1.1 (Figure~\ref{fig:g317.4-1.1})} Small roughly circular nebula, brightest along the eastern edge. We calculate a 1360~MHz flux density, using the SMGPS \citep{Goedhart2024}, of 0.16 $\pm$ 0.02 Jy, resulting in a spectral index of $-$0.46 $\pm$ 0.16. This is consistent with the expected values for an SNR.

\paragraph{G317.6+0.9 (Figure~\ref{fig:g317.6+0.9})} Clear shell-like feature to the northwest with well-defined edges. Several internal filaments are visible as well, running parallel to the bright outer shell. The source was listed as a possible SNR (G317.5+0.9) by \cite{Whiteoak1996} but no detailed description or image was provided.

\paragraph{G320.8$-$0.3 (Figure~\ref{fig:g320.8-0.3})} Bright, thin, roughly circular shell of emission. This shell is brightest on the northwestern and southern edges. The source is missing emission in its centre, caused by artifacts from nearby bright sources. There are three small overlapping \ion{H}{2} regions. This source has also been listed as an SNR candidate in the SMGPS \citep{Anderson2024}.

\paragraph{G320.9$-$0.7 (Figure~\ref{fig:g320.9-0.7})} Multiple thin partial shells, brightest in the north and south. Bright extended emission can be seen from overlapping \ion{H}{2} regions. The \ion{H}{2} regions have MIR counterparts but the shell features do not.

\paragraph{G321.3$-$0.9 (Figure~\ref{fig:g321.3-0.9})} Bright, thin, roughly circular shell of emission. The source is brightest along the western edge and the centre is missing emission, resulting from artifacts from a nearby bright point source. This source has also been listed as an SNR candidate in the SMGPS \citep{Anderson2024}.

\paragraph{G322.7$-$0.6 (Figure~\ref{fig:g322.7-0.6})} Multiple thin rounded filaments. The source is not yet fully imaged in EMU/POSSUM surveys. This source has also been listed as an SNR candidate in the SMGPS, though they separate the source into two distinct candidates \citep{Anderson2024}.

\subsubsection{Possible SNRs / Weak candidates}

Here, we list 57 SNR candidates that we consider to be weak candidates. Twenty-six of these sources have been previously identified as SNR candidates in other works. Many of the sources consist of very faint partial shells. Others are made up of wider filaments than would typically be expected for SNR shock fronts. Many of these candidates are located in complex areas with multiple overlapping \ion{H}{2} regions. Often, these bright \ion{H}{2} regions can create image artifacts that interfere with our view of the SNR candidate. The overlapping \ion{H}{2} regions can also make it difficult to tell whether or not the candidate has an MIR counterpart. Because of these uncertainties, we classify these candidates as ``weak". Images of these sources can be found in Appendix~\ref{sec:appA} in Figures~\ref{fig:weak_cands}--\ref{fig:weak_cands5}.

\paragraph{G276.1+3.2} Very faint, roughly circular thin shell that can be seen most prominently along the northern and southern edges. This field has a high background noise level, making it more difficult to distinguish these faint shells.

\paragraph{G278.8$-$1.2} Bright partial eastern shell with no detectable western counterpart. This source has also been listed as an SNR candidate in the SMGPS \citep{Anderson2024}.

\paragraph{G280.2+0.2} Two very faint curved filaments can be seen in the southeast. One of these filaments is spatially coincident with a pulsar, J1001$-$5507 (B0959$-$54, characteristic age 441 kyrs) \citep{Wielebinski1969}. This source has also been listed as an SNR candidate in the SMGPS \citep{Anderson2024}.

\paragraph{G280.4$-$0.9} A faint thin partial shell is visible in the south. An even fainter possible northern counterpart is also visible. There is MIR emission that runs through the centre of the source, but the shell itself has no MIR counterpart. It was previously identified as an SNR candidate (G280.3$-$0.9) by \cite{Duncan1995} who described the source as a faint ring with an angular size of around 1$^\circ$.

\paragraph{G280.7+3.2} Very faint, roughly circular filament. This shell can be seen most prominently along the eastern edge. There is another small shell to the east of the source that could be associated with this source or part of a separate candidate.

\paragraph{G281.1$-$2.8} Multiple thin shell-like structures can be seen to the northwest with a thicker filament to the northeast. There is a bright circular source in the centre, identified as a Wolf-Rayet star (WR-16) \citep{vanderhucht2001}.

\paragraph{G282.3$-$1.9} A thick, roughly circular filament. We see clear MIR emission from the two overlapping \ion{H}{2} regions, but the outer shell does not appear to have an MIR counterpart.

\paragraph{G283.9+0.5} Very faint roughly circular filament, brightest along the eastern and northwestern edges. 

\paragraph{G283.9$-$1.4} Multiple filaments form a roughly elliptical source with a smaller elliptical source near the centre. The most well-defined outer shells can be seen to the southwest and northeast. The region is very complex, with many nearby \ion{H}{2} regions and a bright SNR to the south, so the source is heavily affected by image artifacts. The filaments themselves do not appear to have MIR counterparts. This source has also been listed as an SNR candidate in the SMGPS \citep{Anderson2024}, though they divide the source into two separate candidates, with the central ellipse as one (G283.849$-$01.431) and the western part of the outer shell as the other (G283.429$-$01.414).

\paragraph{G284.2$-$0.9} Roughly circular shell of emission surrounding a pulsar, J1022$-$5813 (characteristic age 179 kyrs) \citep{Kramer2003}. It is unclear if there is an MIR counterpart due to the presence of overlapping \ion{H}{2} regions.

\paragraph{G285.5$-$0.1} Several shell-like structures elongated in the north-south direction. There are multiple overlapping \ion{H}{2} regions, which have MIR counterparts. The brightest emission, coming from the centre, also has an MIR counterpart. There does not appear to be an MIR counterpart for the shells, but it is difficult to state this definitively due to the complexity of the region.

\paragraph{G286.8+0.0} Relatively bright source composed of multiple thin filaments, reflecting a typical SNR morphology. However, we consider this source to be a weak candidate because it is located in a complex region, and there may be an MIR counterpart. In particular, there is MIR emission along the southern and eastern edges of the source. This source was also listed as a candidate by \cite{Anderson2024}.

\paragraph{G287.1$-$2.4} Faint, roughly circular filament. There is a brighter filament that runs along the top of the source that has an MIR counterpart but the fainter circular component does not appear to.

\paragraph{G288.0+1.1} Faint, roughly elliptical shell. There is an MIR source to the east that corresponds to the smaller circular shell. The rest of the source does not have an MIR counterpart. 

\paragraph{G288.9$-$1.4} Multiple thin filaments. There is some MIR emission running through the source, but the brighter southwestern shell has no MIR counterpart. This source has also been listed as an SNR candidate in the SMGPS \citep{Anderson2024}.

\paragraph{G289.4$-$2.5} Multiple shell-like structures, most elongated in the east-west direction, forming a roughly elliptical shape. There is some faint MIR emission, but it does not correspond to the radio. This source was also identified as an SNR candidate by \cite{Duncan1995}.

\paragraph{G290.1+0.5} Rounded shell-like feature, most prominent along the southeastern edge.

\paragraph{G290.6+0.5} Rounded shell-like structure, brightest along the southeastern edge. An MIR filament runs through the centre of the source from north to south, but the shell itself has no MIR counterpart. This source has also been listed as an SNR candidate in the SMGPS \citep{Anderson2024}.

\paragraph{G290.6$-$1.5} Roughly circular source with a faint well-defined edge. There is also brighter extended emission coming from the northern part of the source. There are some filaments running through the upper half of the source that have MIR counterparts. There is a young pulsar located outside of the source to the east.

\paragraph{G291.1$-$0.4} Elliptical source with well-defined edges and a bright shell-like edge to the northeast. There is some MIR emission running through the upper part of the source, extending from the \ion{H}{2} regions to the north. The bright shell does not appear to have an MIR counterpart.

\paragraph{G291.3+4.5} Two very faint partial shells, one to the north and one to the south. There may also be some fainter filaments in the centre running parallel to these shells. 

\paragraph{G294.5$-$0.6} Faint partial shell to the southeast. There is also a pulsar J1138$-$6207 (characteristic age 149 kyrs) \citep{Manchester2001} to the north of this shell.

\paragraph{G295.2$-$0.5} Two extended arcs, one to the northeast and a counterpart to the southwest. A pulsar, J1139$-$6247 (characteristic age 471 kyrs), is spatially coincident with the northern arc. The source lies near a very bright \ion{H}{2} region, but the arcs have no MIR counterpart. 

\paragraph{G296.4$-$2.8} Thin faint circular shell, observed most clearly along the northeastern and southwestern edges. There also appears to be a thin, roughly circular shell near the centre of the source. This source is being studied in greater depth by \cite{Becker_inprep}.

\paragraph{G297.0$-$1.0} Roughly circular thin shell, seen most clearly along the northern edge. The region is complex with multiple overlapping \ion{H}{2} regions. There is some MIR emission that runs through the source, but the shell itself does not appear to have an MIR counterpart. This source has also been listed as an SNR candidate in the SMGPS \citep{Anderson2024}.

\paragraph{G298.0+0.3} Faint extended emission with some filamentary features. Though this source does not exhibit a characteristic SNR morphology, we find evidence of linear polarization. Because of the ambiguous morphology and unclear association between the total power and polarized emission, we classify this source as a weak candidate. 

\paragraph{G298.2$-$1.9} Multiple thin shell-like features, most prominent to the southeast. 

\paragraph{G298.5+2.1} Multiple thin shell-like features. This source was previously identified as an SNR candidate by \cite{Duncan1995}.

\paragraph{G299.2+0.3} Faint filamentary features. The source is heavily affected by imaging artifacts from bright \ion{H}{2} regions.

\paragraph{G300.0$-$1.6} Very faint partial shell can be seen in the south. This sources lies between two brighter SNR candidates.

\paragraph{G301.8$-$2.1} Northern and southern filaments elongated in the east-west direction.

\paragraph{G304.2$-$0.5} Roughly circular extended emission with several internal filamentary features. We find a polarized point source located near the candidate's centre (RM of $-$420 rad/m$^2$). This source has also been listed as an SNR candidate in the SMGPS \citep{Anderson2024}.

\paragraph{G304.4$-$0.2} Thin circular shell that is brightest along the southwestern edge. This source, which we name ``Mavka", was first identified as an SNR candidate by \cite{Sushch2017} who suggested it could be related to the X-ray PWN G304.1$-$0.2.

\paragraph{G306.2$-$0.8} Roughly circular filament, most prominent along the northern and eastern edges. The candidate overlaps with a bright source, which is a known SNR. This source has also been listed as an SNR candidate in the SMGPS \citep{Anderson2024}.

\paragraph{G306.4+2.6} Long thin shell visible along the northern edge.

\paragraph{G307.1$-$0.7} Thin shell-like structure along the eastern edge. There are also several thin filamentary features to the west. This source has also been listed as an SNR candidate in the SMGPS \citep{Anderson2024}.

\paragraph{G307.5$-$1.0} Roughly circular shell that is brightest to the northwest. Some fainter shells can also be seen to the south of the candidate. This source has also been listed as an SNR candidate in the SMGPS \citep{Anderson2024}.

\paragraph{G307.9+0.1} Roughly circular shell-like feature, with the brightest emission coming from the western edge. This source has also been listed as an SNR candidate in the SMGPS \citep{Anderson2024}.

\paragraph{G308.3$-$0.2} Thin shell-like features. Two brighter shells can be seen along the western edge. This source has also been listed as an SNR candidate in the SMGPS \citep{Anderson2024}.

\paragraph{G308.5+0.4} Roughly elliptical filament that overlaps with several bright \ion{H}{2} regions. There is MIR emission coming from the eastern half of the source, particularly around where the radio emission is the brightest, but not from the western side. 

\paragraph{G308.8$-$0.5} Roughly circular shell-like feature. There is a filament running through the upper part of the source that appears to have an MIR counterpart, but the outer shell structure does not. This source has also been listed as an SNR candidate in the SMGPS \citep{Anderson2024}.

\paragraph{G309.2$-$0.1} Roughly circular source with a brightened shell-like edge, most prominent to the east. There are several bright overlapping \ion{H}{2} regions. This source has also been listed as an SNR candidate in the SMGPS \citep{Anderson2024}.

\paragraph{G309.9$-$0.4} Two thin rounded filaments, seen most prominently along the southwestern edge. There is a pulsar, J1350$-$6225 (characteristic age 246 kyrs) \citep{Clark2017}, located near the end of one of these filaments. The region is heavily affected by image artifacts. 

\paragraph{G310.3+0.5} Multiple faint filamentary features. The source is located in a complex area with multiple overlapping \ion{H}{2} regions and is affected by image artifacts. Thus, it is unclear if there is an MIR counterpart for the filaments. 

\paragraph{G310.5$-$0.6} Two thin shells that arc in the southwest direction. The candidate overlaps with a pulsar, J1356$-$6230 (B1353$-$62, no age estimate available) \citep{Komesaroff1973,Han2006}. This source has also been listed as an SNR candidate in the SMGPS \citep{Anderson2024}.

\paragraph{G311.3+1.1} Faint filamentary features around the pulsar B1356$-$60 (characteristic age 320 kyrs) \citep{Manchester1978}. The region was studied by \cite{Gaensler2000}, who identified the faint shell as an SNR candidate. Because of the complexity of the region, it is difficult to determine whether or not there is an MIR counterpart for the shell. There is extended polarized emission around the pulsar that is not seen in total power. 

\paragraph{G311.5$-$0.1} Thin bright shell that is pinched in the centre with a similar morphology to the bright shell in the known remnant G286.1$-$1.1 (see Figure~\ref{fig:known_snrs}). The bright shell also has a fainter western counterpart. 

\paragraph{G311.9+0.9} Several thin shells with a thin filament extending south from the brightest shell. There is a filament that runs through the source in the north-south direction that can be seen in the MIR but the shell itself has no MIR counterpart. 

\paragraph{G312.7+2.9} Very faint, roughly circular shell. This source was identified as an SNR candidate using EMU by \cite{Smeaton2024}.

\paragraph{G316.3$-$0.4} Several faint thin filaments surrounding a bright \ion{H}{2} region. A bright known SNR can be seen to the north of the candidate. The source is located in a complex region, so it is unclear if there is an MIR counterpart. This source was identified as an SNR candidate by \cite{Anderson2024}.

\paragraph{G316.7+0.4} Very faint, thin partial shell. The source is heavily affected by image artifacts from a nearby bright \ion{H}{2} region, so the southern half of the shell cannot be seen. The source was also identified as an SNR candidate by \cite{Anderson2024}.

\paragraph{G318.1$-$0.4} Bright shell-like feature arcs along the southern edge with a fainter thin shell running along the northern part of the source. The source was also identified as an SNR candidate by \cite{Anderson2024}.

\paragraph{G319.3$-$0.7} Multiple faint rounded filaments that form a roughly elliptical shape. The eastern side is composed of a thicker, brighter filament with the western half composed of multiple thin faint shells. The eastern part of the source has also been listed as an SNR candidate in the SMGPS \citep{Anderson2024}.

\paragraph{G319.4+0.2} Very faint, roughly circular thin shell. The source is spatially coincident with the pulsar J1502$-$5828 (characteristic age 291 kyrs) \citep{Kramer2003}. There is bright MIR emission to the south of the source, extending from the nearby \ion{H}{2} regions, but no MIR counterpart for the radio features within the source boundary. There is also a bright 12 $\mu$m source north of the pulsar.

\paragraph{G320.0$-$1.7} Faint, roughly circular shell with several internal filamentary features. The region is heavily impacted by imaging artifacts. There is some MIR emission running through the source, particularly in the south, but it is unclear whether or not it is associated.

\paragraph{G320.6$-$0.8} Faint, roughly circular shell of emission that is brightest along the northeastern edge. There is MIR emission along the western edge of the source from the overlapping \ion{H}{2} regions.

\paragraph{G320.9$-$0.3} Several faint filaments, most well defined to the southeast. There is a pulsar, J1512$-$5759 (characteristic age 298 kyrs) \citep{Johnston1992}, to the northwest of the filaments. The source overlaps with multiple \ion{H}{2} regions and another SNR candidate. There is also a bright point source with a ring around it near the centre that has been identified as a planetary nebula \citep{vandesteene1993}. The area is heavily impacted by image artifacts from nearby bright \ion{H}{2} regions. 

\section{Discussion} \label{sec:disc}

\subsection{The Distribution of SNRs and SNR Candidates}

\begin{figure}
    \centering
    \includegraphics[width=0.9\linewidth]{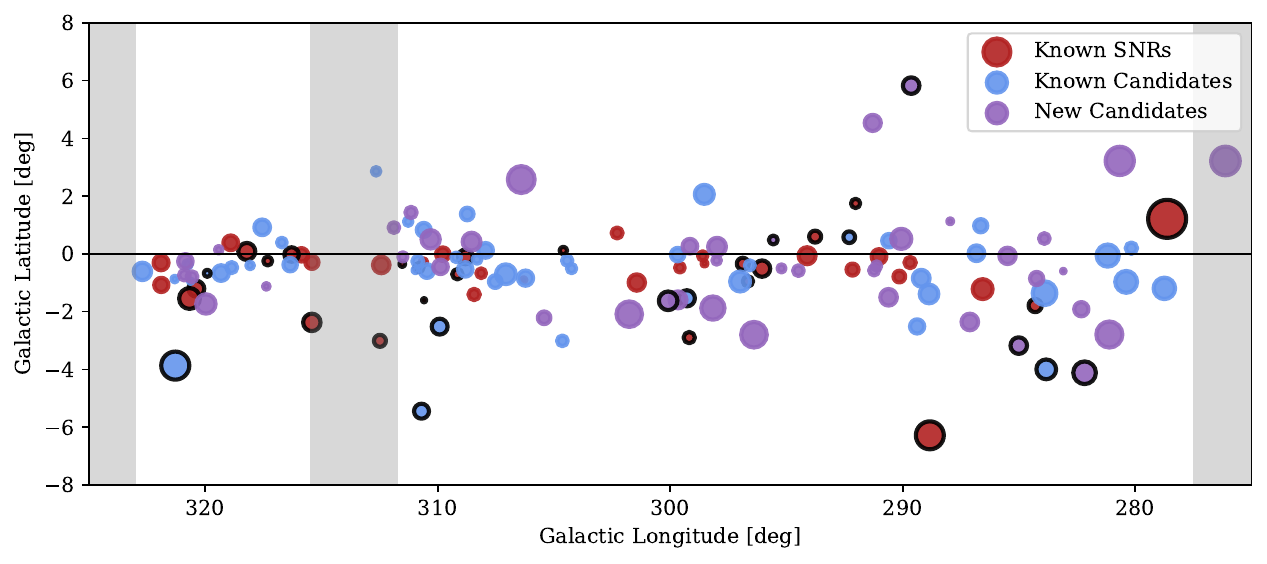}
    \caption{Distribution of known and new SNRs/SNR candidates. The sizes of the points are proportional to the sizes of the SNRs/SNR candidates. Black outlines indicate that we found the source to be polarized. The grey shaded regions indicate the approximate Galactic longitude limits of the region surveyed in this work.}
    \label{fig:dist}
\end{figure}

\begin{figure}
    \centering
    \subfigure{\includegraphics[width=0.49\textwidth]{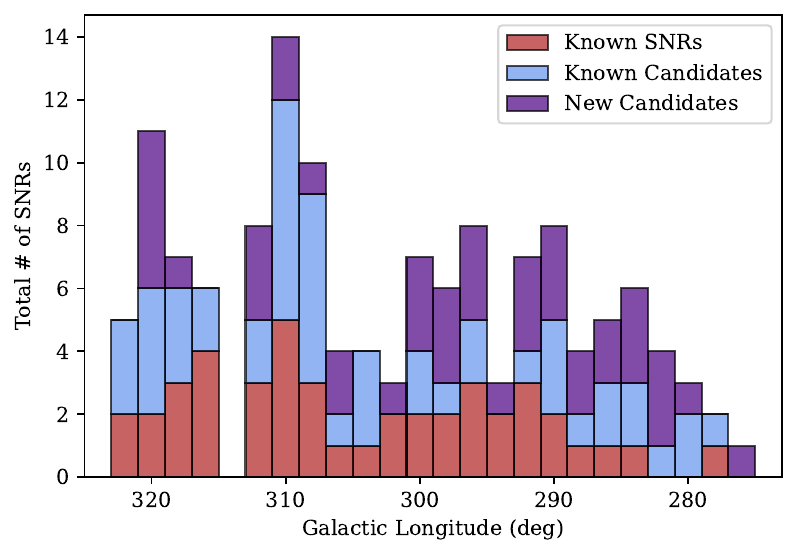}}
    \subfigure{\includegraphics[width=0.49\textwidth]{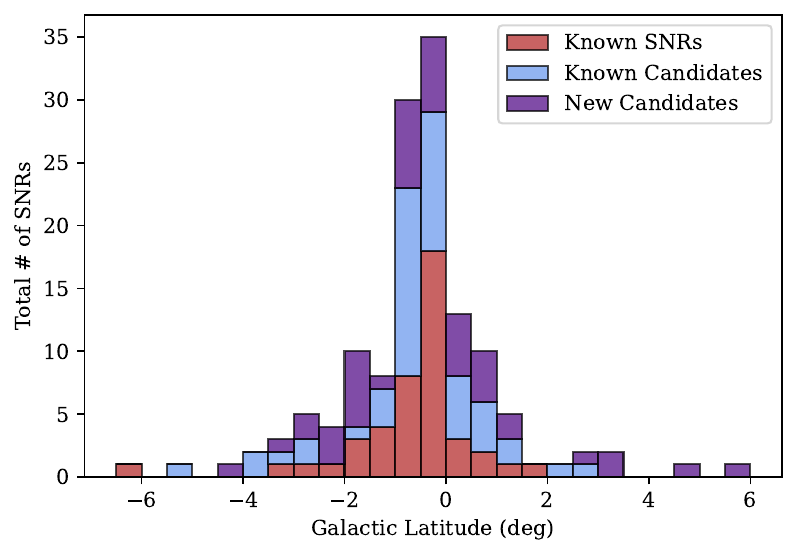}}
    \subfigure{\includegraphics[width=0.49\textwidth]{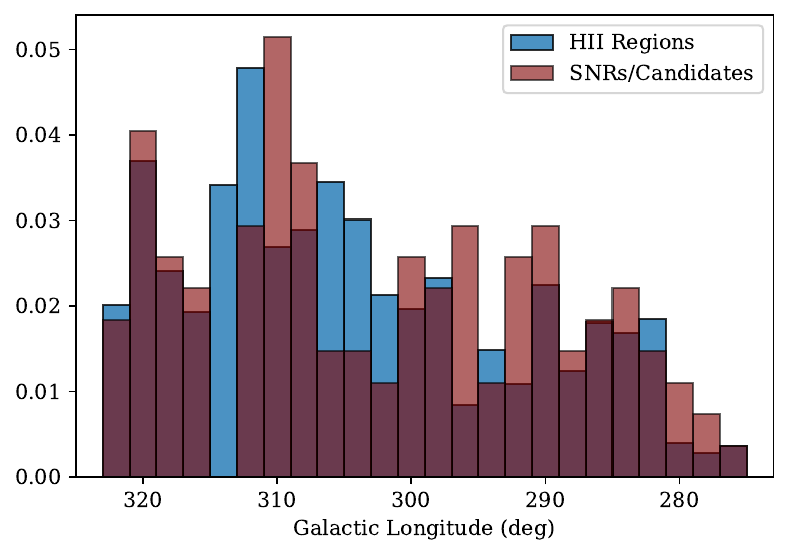}}
    \subfigure{\includegraphics[width=0.49\textwidth]{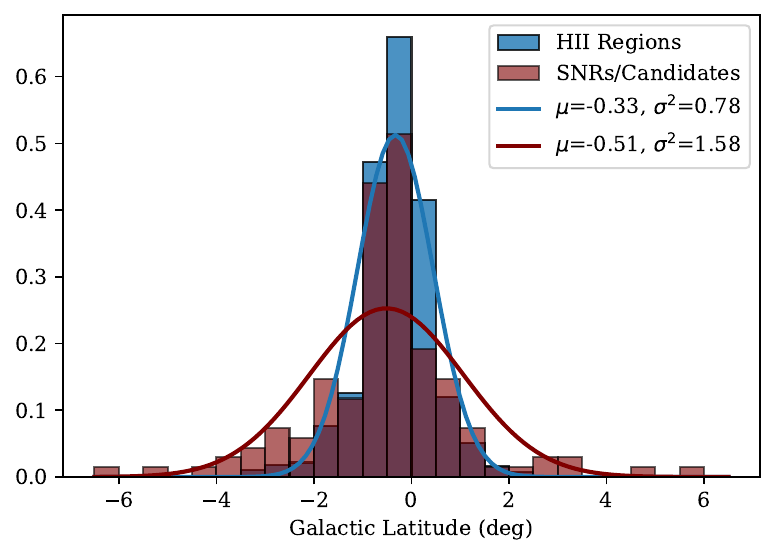}}
    \caption{The top figures show known and new SNRs/SNR candidates as a function of Galactic longitude and Galactic latitude. The bottom figures show the longitude and latitude distributions of SNRs/SNR candidates compared to the distributions of \ion{H}{2} regions covered in our surveyed area. The curves in the lower-right panel represent the PDFs for the \ion{H}{2} regions and SNRs/candidates, with mean $\mu$ and variance $\sigma^2$. These curves demonstrate the skew of both distributions to negative latitudes.}
    \label{fig:hist}
\end{figure}

Figure~\ref{fig:dist} shows the visual distribution of known SNRs (Table~\ref{tab:knownSNRs}), previously identified SNR candidates (Table~\ref{tab:known_cands} plus the optical candidates), and new SNRs/SNR candidates (Table~\ref{tab:newcands}) in our surveyed region. The points are proportional to the SNRs'/SNR candidates' sizes, and black outlines indicate that we found the source to be polarized. The top images in Figure~\ref{fig:hist} show the distribution of SNRs and SNR candidates as a function of Galactic longitude and Galactic latitude. The bottom images of Figure~\ref{fig:hist} compare the distributions of SNRs/SNR candidates to the distributions of \ion{H}{2} regions in our surveyed area, with the sample of \ion{H}{2} regions including all categories found in the WISE catalogue \citep{Anderson2014}. In the Galactic longitude distribution, we see a noticeable excess in SNRs/SNR candidates around 305$^\circ$-- 320$^\circ$. This is likely because within this region we are looking along a major spiral arm, Scutum-Centaurus \citep{Bland2016}. There is a gap in our distribution from approximately 312$^\circ$--315$^\circ$ because this field has not yet been covered by EMU/POSSUM. In the Galactic latitude distribution, we note that the SNR candidates are found at a larger range of latitudes than the known SNRs. The only known SNR found at a very high latitude in the survey area is G288.8$-$6.3, which was identified as an SNR using the EMU survey \citep{Filipovic2023}. The rest of the known SNRs are found within the latitude range \(-3.5^\circ \leq b \leq 2^\circ\), while the candidates extend up to latitudes of $\pm 6^\circ$. This demonstrates the effectiveness of EMU and POSSUM for uncovering and confirming high-latitude SNRs.

We also note that the latitude distribution skews toward negative values, as shown by the probability distribution function (PDF) in the lower-right panel of Figure~\ref{fig:hist}, with a mean value of $-0.51^\circ$. This bias can also clearly be seen looking at the distribution of points in Figure~\ref{fig:dist}, as the majority lie below the 0$^\circ$ line. We find a similar bias among \ion{H}{2} regions. While we found the total \ion{H}{2} region population from the WISE catalogue to exhibit a normal distribution centred at roughly 0$^\circ$ latitude, the \ion{H}{2} regions in our surveyed area also exhibit a negative bias, with a mean of $-0.33^\circ$. This bias primarily comes from sources in the longitude range 275$^\circ$-- 295$^\circ$. In Figure~\ref{fig:field}, we can see that the Galactic plane is clearly below the 0$^\circ$ latitude in this range while above 295$^\circ$, it is roughly centred around 0$^\circ$. This seems to be related to which spiral arm we are primarily looking along. From 300$^\circ$-- 320$^\circ$ we are looking along the Scutum-Centaurus arm, which is closer to the inner Galaxy. From 275$^\circ$-- 295$^\circ$ we are probably mostly looking at emission from the Sagittarius arm, closer to the outer edge where the Galactic disk is warped \citep{Djorgovski1989,Chrobakova2020}.

The longitude distribution of \ion{H}{2} regions is similar to the SNR distribution, though there are a few noticeable differences. As shown in Figure~\ref{fig:hist}, there is a discrepancy around approximately 305$^\circ$, where we see proportionally fewer SNRs than \ion{H}{2} regions, while in the 290$^\circ$-- 300$^\circ$ range there are proportionally more SNRs. Part of the explanation for this discrepancy may be that, in areas with a high density of \ion{H}{2} regions, it is harder to identify faint SNRs.

\subsection{Completeness of the Galactic SNR Population}

The continuous part of our survey, shown in Figure~\ref{fig:field} (within the longitude range \(277.5^\circ \leq \ell \leq 311.7^\circ\)), covers approximately 12$\%$ of the Galactic plane by surface area, or 55 kpc$^2$. We produce this estimate by approximating the Galaxy as a flat disk with a radius of 12.5 kpc, encompassing 95\% of the stellar mass \citep{Binney2023}. The line of sight through the Galaxy in this area ranges over 9 -- 16 kpc. Within this region, there are 33 known SNRs,  39 known SNR candidates, and 37 new SNR candidates (omitting the sources found in the non-contiguous field, EMU1505-60). This corresponds to a known SNR density of 0.6 SNRs/kpc$^2$, similar to the average known SNR density of the full Galaxy (0.63 SNRs/kpc$^2$), and a total SNR/SNR candidate density of 2.0 SNRs/kpc$^2$. If we only add our newly confirmed SNRs to the total SNR count, we find a density of 0.85 SNRs/kpc$^2$. To achieve the minimum estimate of 1000 Galactic SNRs, an average of 2.2 SNRs/kpc$^2$, we would need to find a total of 121 SNRs in this region. Our total of 109 SNRs and SNR candidates is very close to the expected value.

There are two major observational constraints to consider when analyzing the Galactic SNR population discrepancy: angular size and surface brightness relative to the surrounding region. We are able to detect sources at angular sizes down to around 2'. At the outer edge of the Galaxy at distances of 9 -- 16 kpc, this corresponds to physical sizes of around 5 -- 10 pc. Only very young remnants (\(\lesssim 1000\) years) should be smaller than this \citep{Ranasinghe2023}. Based on the Galactic supernova rate, there should only be around 20 -- 30 Galactic SNRs younger than this. Additionally, the young SNR population is believed to be mostly complete \citep{Leahy2020}, and thus angular resolution is likely not a significant factor in limiting the detection of new remnants in modern radio surveys. 

The more significant limitation is likely surface brightness, particularly in relation to the background brightness. At high latitudes, we are able to detect fainter sources than we can detect within the Galactic plane. Within the plane, we must deal with bright \ion{H}{2} regions and known SNRs as well as the image artifacts that bright sources create. These bright dense areas are also where we are more likely to find a higher density of SNRs, within spiral arms near star forming regions. Indeed, if we look at the longitude distribution of SNRs and SNR candidates, we find the highest concentration near the major spiral arm. To understand the scope of the Galactic SNR population discrepancy, we must account for these variations in detectability based on latitude and proximity to bright radio sources. 

\begin{figure}
    \centering
    \includegraphics[width=\linewidth]{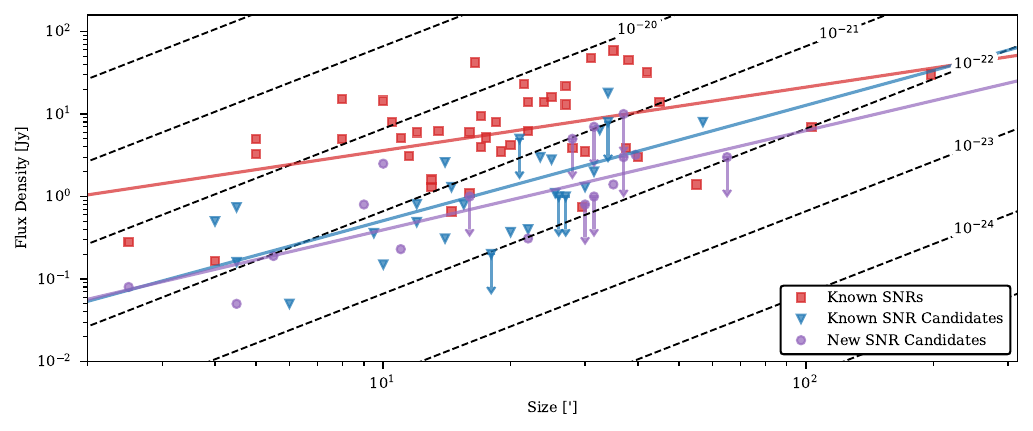}
    \caption{Size vs. flux density plot for known SNRs, known SNR candidates, and new SNR candidates in the surveyed region of the Galactic plane. Arrows indicate that the flux density is an upper limit. Dotted lines represent lines of constant surface brightness in units of  Wm$^{-2}$Hz$^{-1}$sr$^{-1}$. We show trend lines for each population with linear fits performed in the log space. The trend lines do not include the points for which we only have flux upper limits.}
    \label{fig:SB}
\end{figure}

As shown in Figure~\ref{fig:SB}, our SNR/SNR candidate sample includes sources with angular sizes $>$2' and with surface brightnesses above approximately 10$^{-22}$ Wm$^{-2}$Hz$^{-1}$sr$^{-1}$. As discussed, we do not believe that small angular size is a significant limiting factor to the completeness of our sample. The surface brightness limit is variable and depends on the brightness of the background surrounding the source. We can detect fainter sources at higher latitudes than we can detect near the large, bright clusters of \ion{H}{2} regions. In these regions, it is difficult to find faint SNR candidates. It is also more difficult to determine flux densities for fainter sources, especially large ones, so many of the faint candidates we identified are not included in this plot. 

Based on the trend lines shown in Figure~\ref{fig:SB}, we can see that the SNR candidates are generally fainter than the known SNRs, as expected. We also see that there is not a significant difference in the size and brightness distributions between the known SNR candidates and new SNR candidates. This is also not surprising, as many of the known candidates come from the recent SMGPS, which has a similar resolution and sensitivity to ASKAP. Two conclusions can be drawn from this: (1) low surface brightness sources are missing from the current SNR catalogues and are being uncovered by the new generation of radio surveys, and (2) confirming low surface brightness sources as true SNRs is challenging. Confirming faint sources as SNRs is difficult because flux densities often cannot be reliably determined, making it impossible to calculate spectral indices. This is particularly true for the large, faint candidates where we can often only see partial filaments and there is likely missing diffuse emission. Additionally, we find we are less likely to be able to detect polarization from these faint sources, especially when they are located near the Galactic plane. Low-frequency observations may play a role in helping to study and confirm faint SNR candidates as SNRs have much steeper indices than \ion{H}{2} regions and therefore should be comparatively brighter at low frequencies. WALLABY, a 1.4 GHz all-sky survey currently being conducted with ASKAP, may also be of use here \citep{Koribalski2020}. \rev{Finally, complementary surveys in the optical, X-rays, or gamma-rays may also be useful to supplement radio observations and potentially help to confirm radio SNR candidates.}

\subsection{Confirming SNR Candidates with Radio Polarization}

\begin{wrapfigure}{R}{0.5\textwidth}
    \centering
    \includegraphics[width=0.5\textwidth]{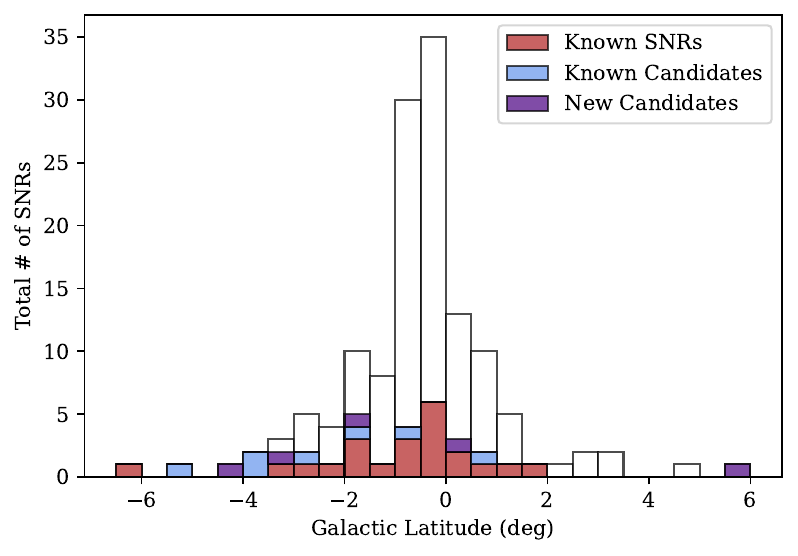}
    \caption{Known SNRs and SNR candidates for which we find evidence of linear polarization compared to the total sample of SNRs and SNR candidates in the surveyed region. We note that proportionally more of the high-latitude SNRs/SNR candidates were observed to be polarized.}
    \label{fig:hist_pol}
\end{wrapfigure}

We have shown in this work that radio polarization is an important and useful tool for confirming new SNR candidates. Of the 14 sources that we claim as new SNRs, 11 of these claims are made based on evidence of linear polarization. This may be especially true for high-latitude remnants, where we are more likely to be able to detect polarization. In total, we detected polarization for 50\% of known SNRs (22/44), 14\% of known SNR candidates (7/49), and 12\% (5/43) of new SNR candidates. Figure~\ref{fig:hist_pol} shows the polarized SNRs and SNR candidates as a function of latitude compared to the total numbers of SNRs/SNR candidates. At latitudes \(|b|>2^\circ\), we find 46\% of the SNRs/SNR candidates to be polarized (11/24) whereas at latitudes \(|b|<2^\circ\), we find only 21\% of the SNRs/SNR candidates to be polarized (23/112).

Confirming new candidates is a major challenge in modern studies of the Galactic SNR population. Many more SNR candidates have been proposed than what is currently included in the SNR catalogues. The confirmation of candidates can be very difficult, particularly for large faint sources where flux densities, and thus spectral indices, cannot be reliably determined. Better polarization data may be a useful tool for confirming new sources, particularly at higher frequencies, where the effects of Faraday rotation are less significant. Accordingly, future radio telescopes, particularly the Square Kilometre Array, may play an important role in this endeavour. \rev{The combination of improved sensitivity and higher resolution could make more SNR candidate detections and confirmations possible, particularly in polarization where beam depolarization effects may be a factor.} However, there are likely some candidates that we can never really confirm, sources that are radio observable but not ``radio confirmable". It remains unclear how these sources should be accounted for when analyzing the completeness of the Galactic SNR population.

\section{Conclusions} \label{sec:conc}
Using the EMU and POSSUM sky surveys, we have compiled a catalogue of SNRs and candidate SNRs within the region of \(277.5^\circ \leq \ell \leq 311.7^\circ\) Galactic longitude, \(|b| \leq 5.4^\circ\) Galactic latitude, and an additional region of the Galactic plane in the longitude range \(315.5^\circ \leq \ell \leq 323.0^\circ\). In the area studied, we found 44 known SNRs, 46 previously identified radio SNR candidates, and 43 new SNRs/SNR candidates. We also found possible radio counterparts for four known X-ray SNRs/SNR candidates and three optical SNR candidates. We argue for the confirmation of 14 sources as SNRs based on spectral indices and polarization, six of which had not been previously identified as SNR candidates. We list 16 strong SNR candidates, six of which are new, and 57 weak candidates, 31 of which are new.

ASKAP is proving to be a particularly effective tool for detecting high-latitude remnants and for confirming these sources as SNRs based on polarization. In particular, we note that at least 22 of our sources were not covered (or fully covered) in the SMGPS because they were found at higher latitudes. Issues of missing flux may present challenges to the detection and confident classification of large low surface brightness SNRs, particularly those located within the densest regions of the Galactic plane. We note that we are unlikely to detect polarization from SNRs within the Galactic plane unless they are very bright. 

The EMU and POSSUM surveys are currently ongoing and will eventually cover the Galactic plane over the longitude range of approximately 220$^\circ$--18$^\circ$, or around 62\% of the Galaxy by surface area, which is five times the area studied here. There are currently 186 known SNRs \citep{Green2024_cat} in this longitude range, an average of 0.62 SNRs/kpc$^2$. Extrapolating from our results, we would expect to find over 400 SNR candidates with the EMU/POSSUM surveys, 200 of which would be new (though we may not be equally successful in all fields, particularly when looking near the Galactic centre). If these candidates were to be confirmed, this would nearly resolve the Galactic SNR discrepancy in the surveyed region, if we assume the lower limit of 1000 Galactic SNRs. It is clear that EMU and POSSUM are proving to be powerful tools for identifying Galactic SNR candidates. As large numbers of new candidates are discovered, the most significant obstacle to resolving the Galactic SNR discrepancy may be confirming these candidates as true SNRs. High-quality polarization observations at higher frequencies, where depolarization effects are less pronounced, may be the most promising approach in this endeavour.

\begin{acknowledgments}
\rev{We are grateful to an anonymous referee whose comments improved the quality of the paper.} This scientific work uses data obtained from Inyarrimanha Ilgari Bundara / the Murchison Radio-astronomy Observatory. We acknowledge the Wajarri Yamaji People as the Traditional Owners and native title holders of the Observatory site. CSIRO’s ASKAP radio telescope is part of the Australia Telescope National Facility (https://ror.org/05qajvd42). Operation of ASKAP is funded by the Australian Government with support from the National Collaborative Research Infrastructure Strategy. ASKAP uses the resources of the Pawsey Supercomputing Research Centre. Establishment of ASKAP, Inyarrimanha Ilgari Bundara, the CSIRO Murchison Radio-astronomy Observatory and the Pawsey Supercomputing Research Centre are initiatives of the Australian Government, with support from the Government of Western Australia and the Science and Industry Endowment Fund. B.D.B., R.K., and E.R. acknowledge the support of the Natural Sciences and Engineering Research Council of Canada (NSERC), funding reference number RGPIN-2022-03499. M.D.F. and S.L. acknowledge Australian Research Council (ARC) funding through grant DP200100784. This research was funded in part by the National Science Centre, Poland (grant numbers 2018/30/E/ST9/00208 and 2023/49/B/ST9/00066).
\end{acknowledgments}

\appendix

\section{Weak Candidates} \label{sec:appA}

ASKAP 943 MHz images of the possible SNRs/weak candidates. Green circles indicate the positions of young pulsars. Cyan circles indicate the positions of \ion{H}{2} regions.

\begin{figure}
    \centering
    \subfigure[G276.1+3.2]{\includegraphics[width=0.32\textwidth]{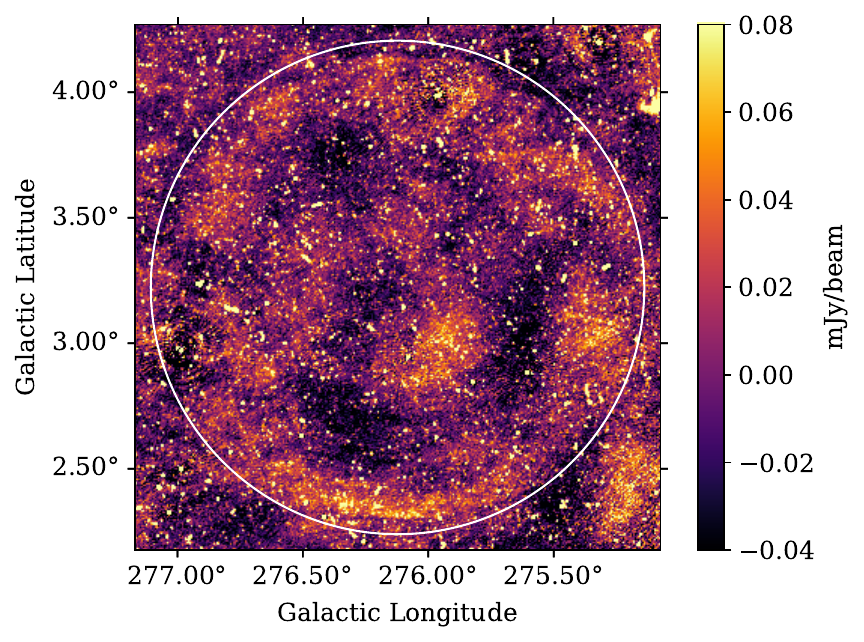}}
    \subfigure[G278.8$-$1.2]{\includegraphics[width=0.32\textwidth]{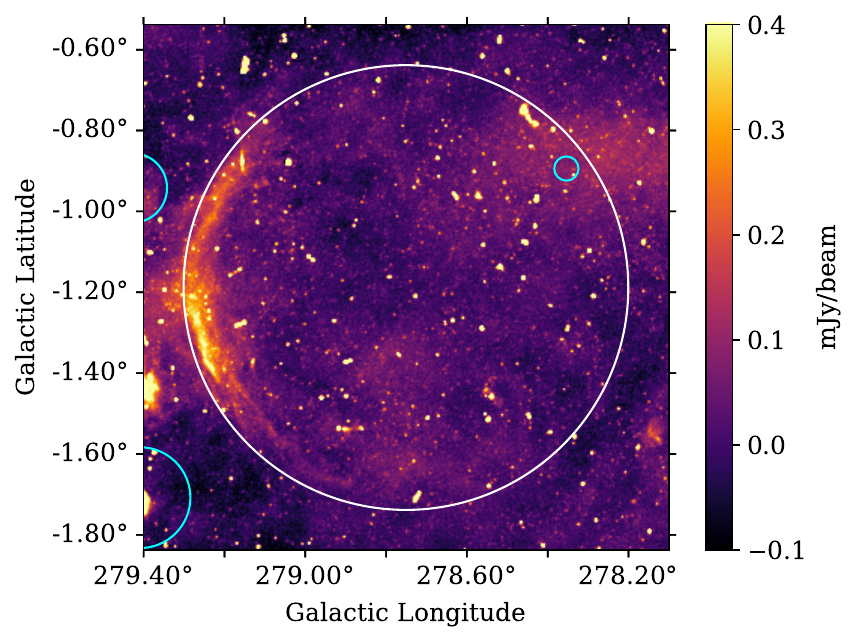}}
    \subfigure[G280.2+0.2]{\includegraphics[width=0.32\textwidth]{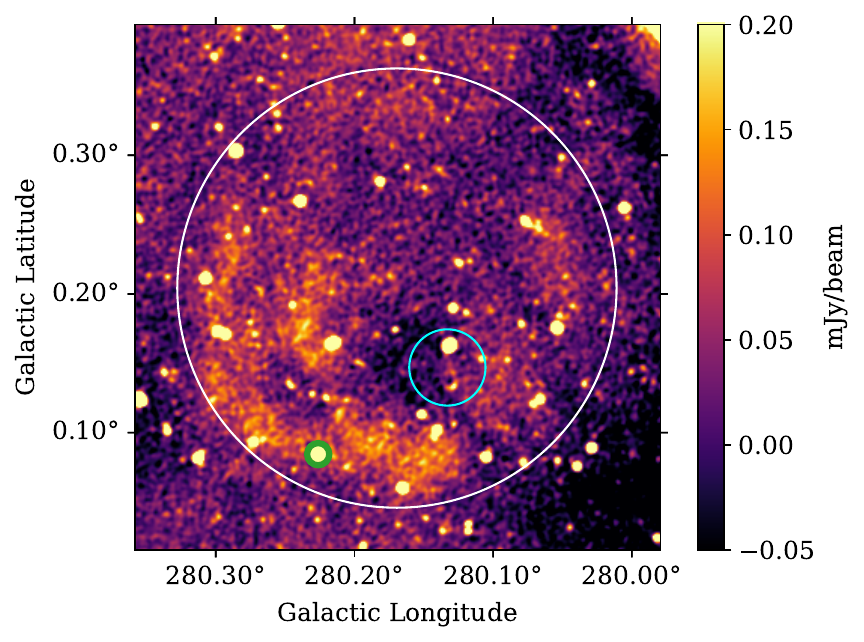}}
    \subfigure[G280.4$-$0.9]{\includegraphics[width=0.32\textwidth]{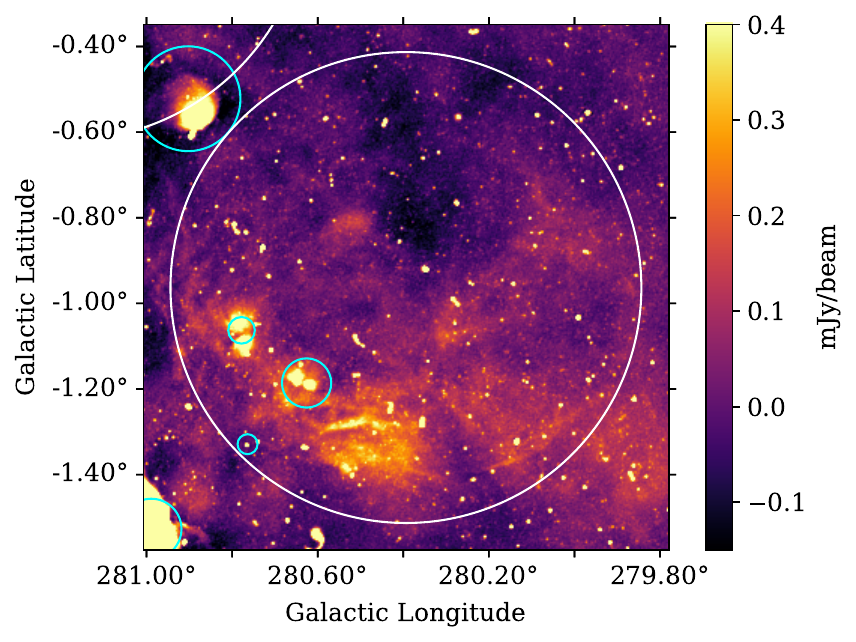}}
    \subfigure[G280.7+3.2]{\includegraphics[width=0.32\textwidth]{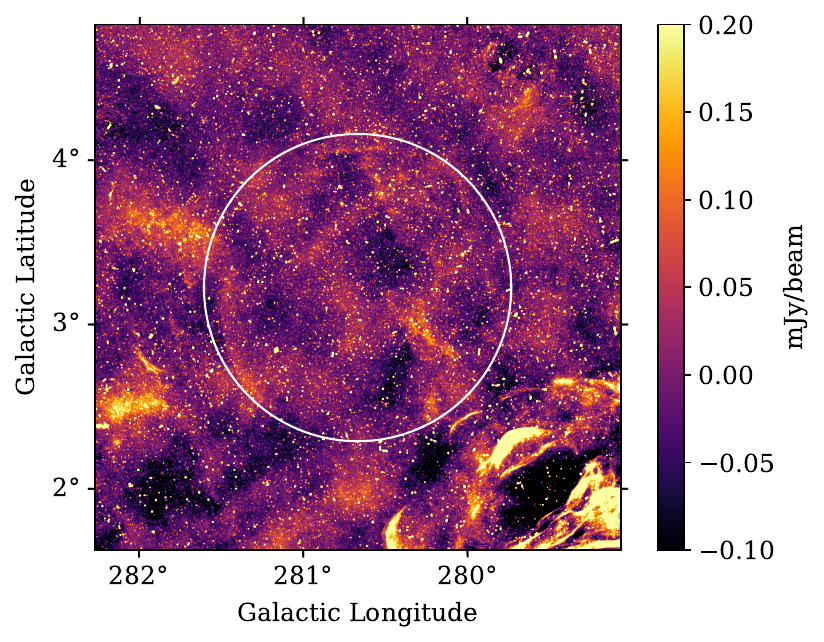}}
    \subfigure[G281.1$-$2.8]{\includegraphics[width=0.32\textwidth]{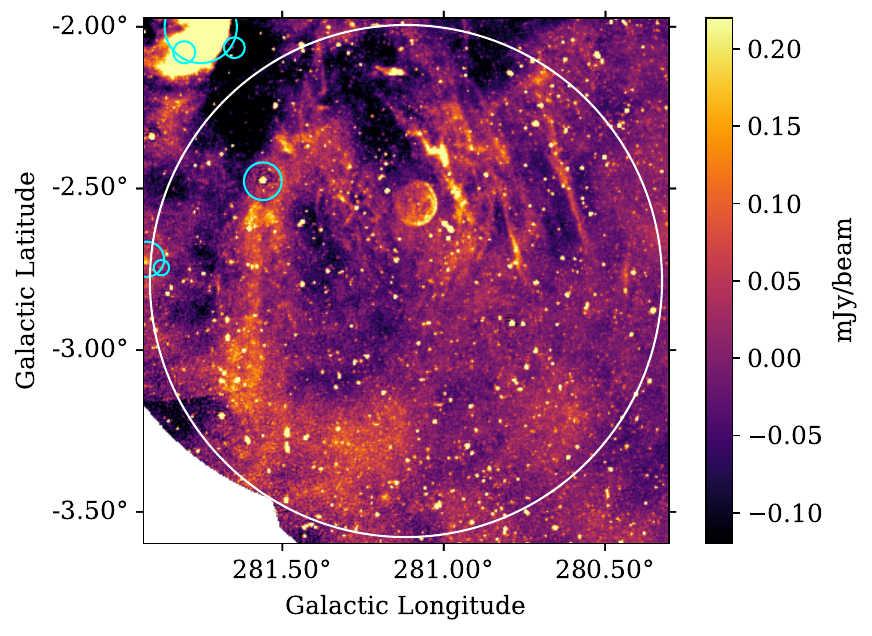}}
    \subfigure[G282.3$-$1.9]{\includegraphics[width=0.32\textwidth]{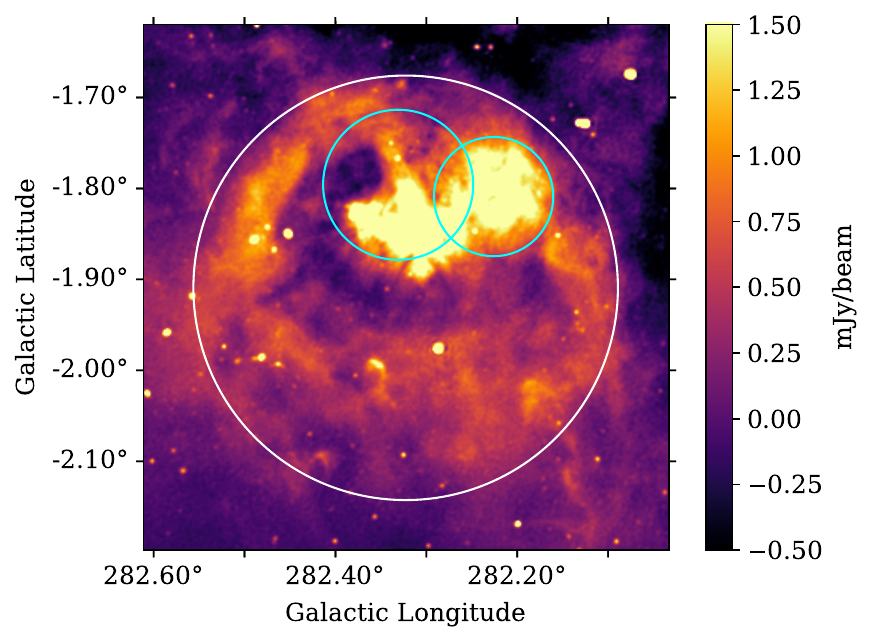}}
    \subfigure[G283.9+0.5]{\includegraphics[width=0.32\textwidth]{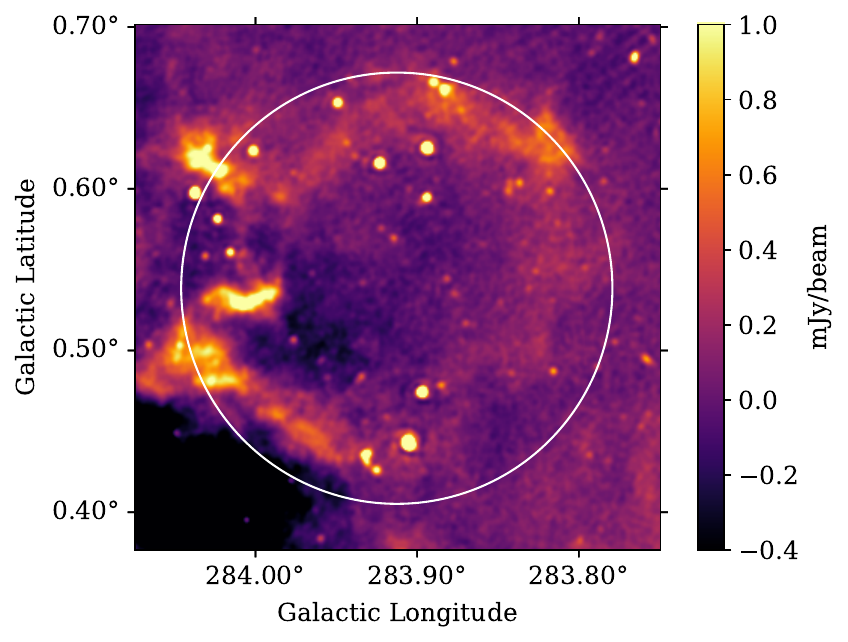}}
    \subfigure[G283.9$-$1.4]{\includegraphics[width=0.32\textwidth]{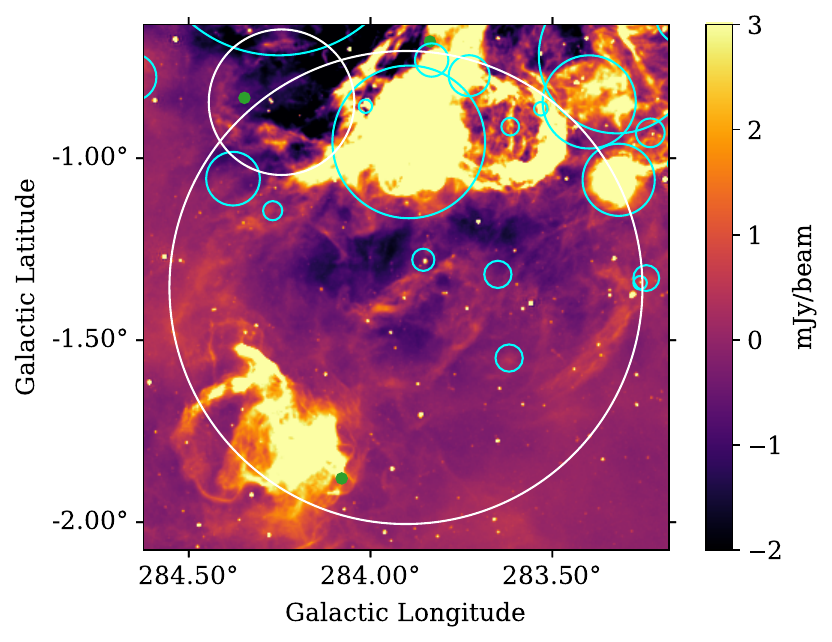}}
    \subfigure[G284.2$-$0.9]{\includegraphics[width=0.32\textwidth]{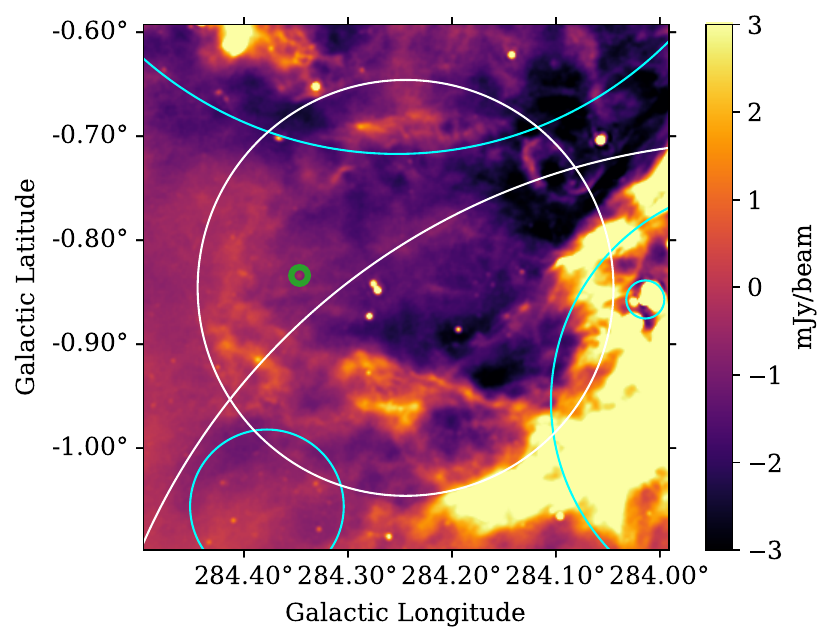}}
    \subfigure[G285.5$-$0.1]{\includegraphics[width=0.32\textwidth]{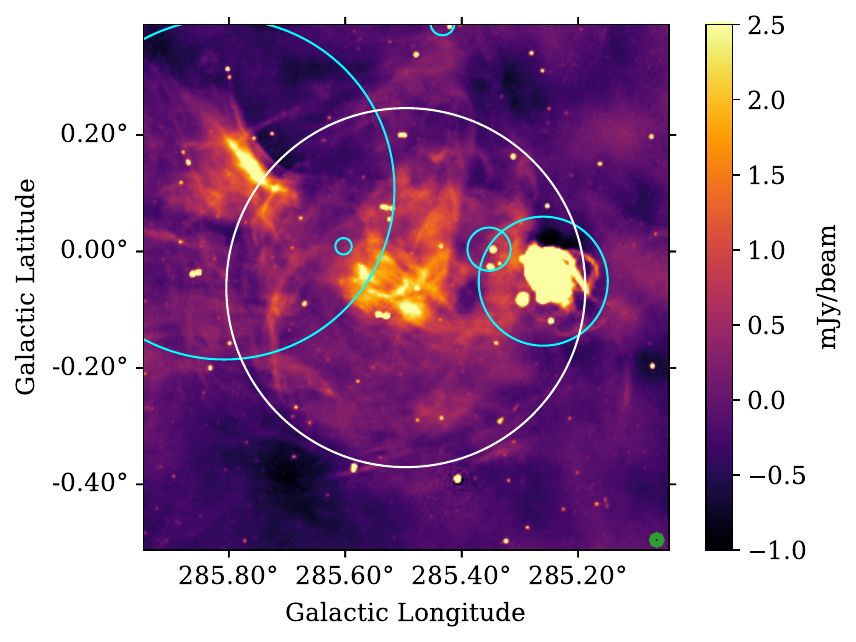}}
    \subfigure[G286.8+0.0]{\includegraphics[width=0.32\textwidth]{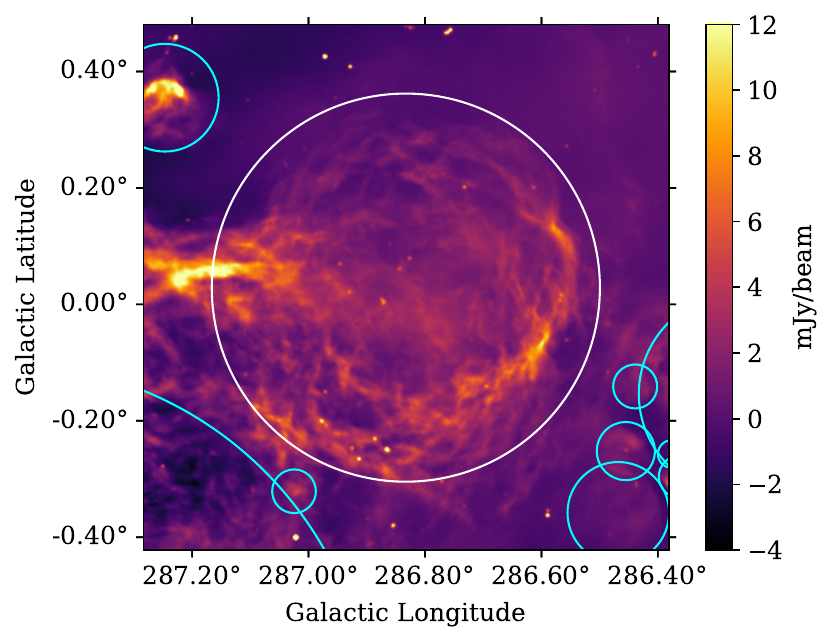}}
    \caption{Possible SNRs / Weak candidates.}
    \label{fig:weak_cands}
\end{figure}

\begin{figure}
    \centering
    \subfigure[G287.1$-$2.4]{\includegraphics[width=0.32\textwidth]{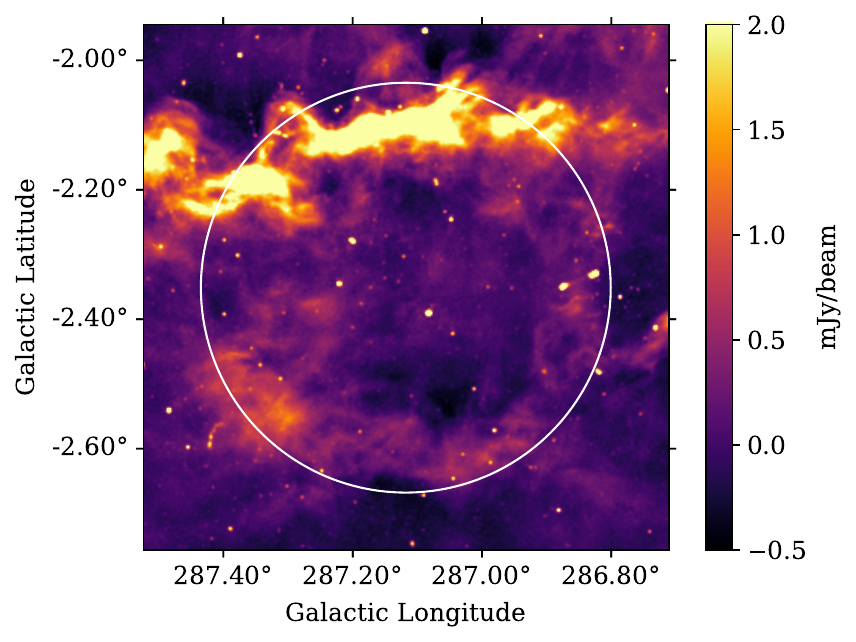}}
    \subfigure[G288.0+1.1]{\includegraphics[width=0.32\textwidth]{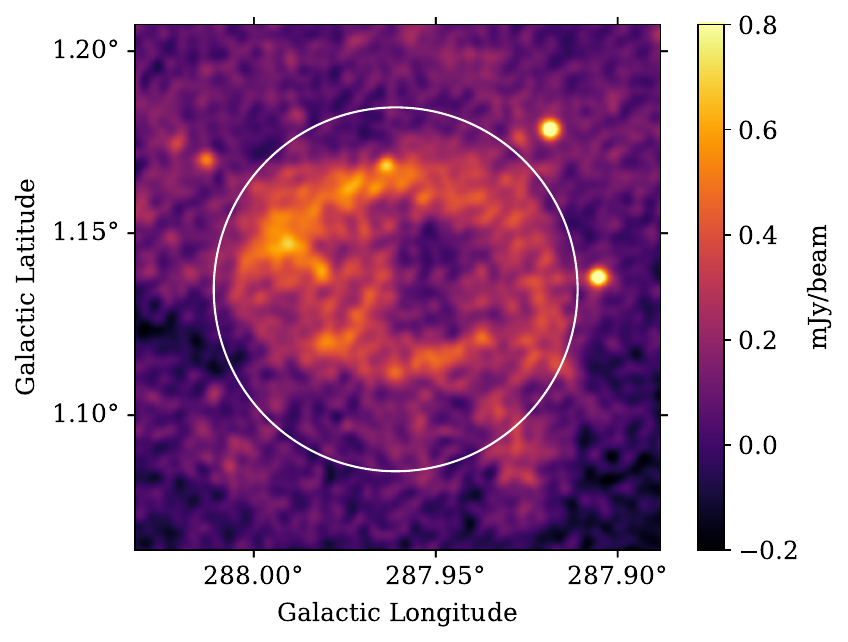}}
    \subfigure[G288.9$-$1.4]{\includegraphics[width=0.32\textwidth]{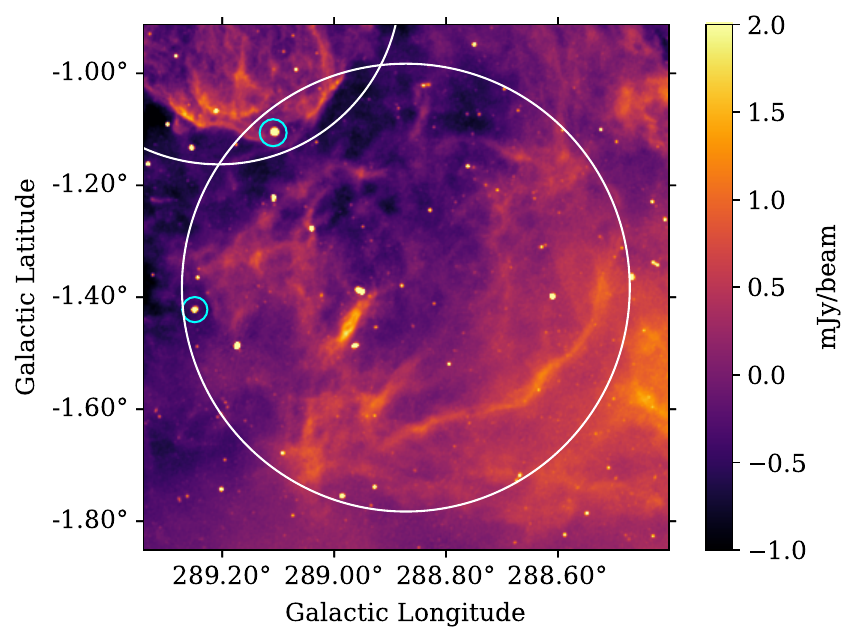}}
    \subfigure[G289.4$-$2.5]{\includegraphics[width=0.32\textwidth]{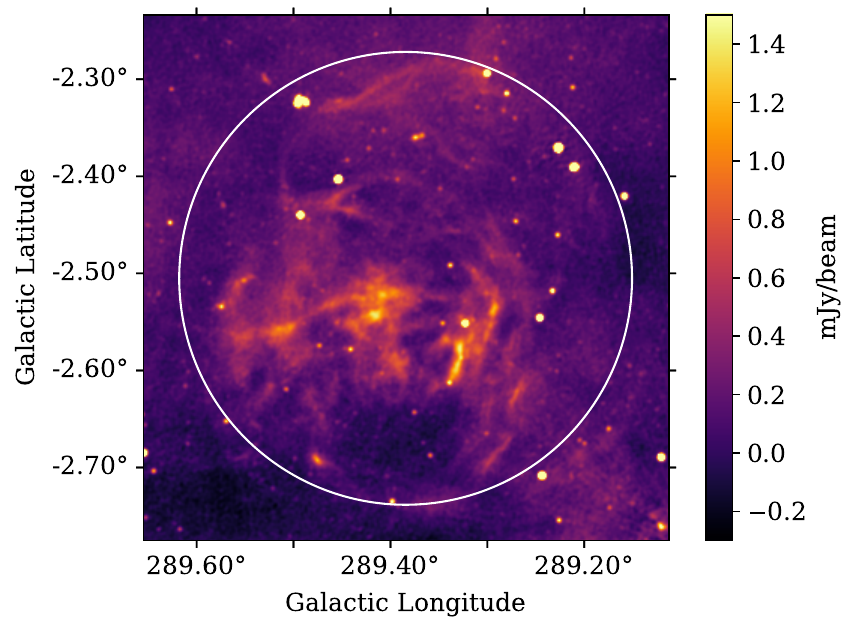}}
    \subfigure[G290.1+0.5]{\includegraphics[width=0.32\textwidth]{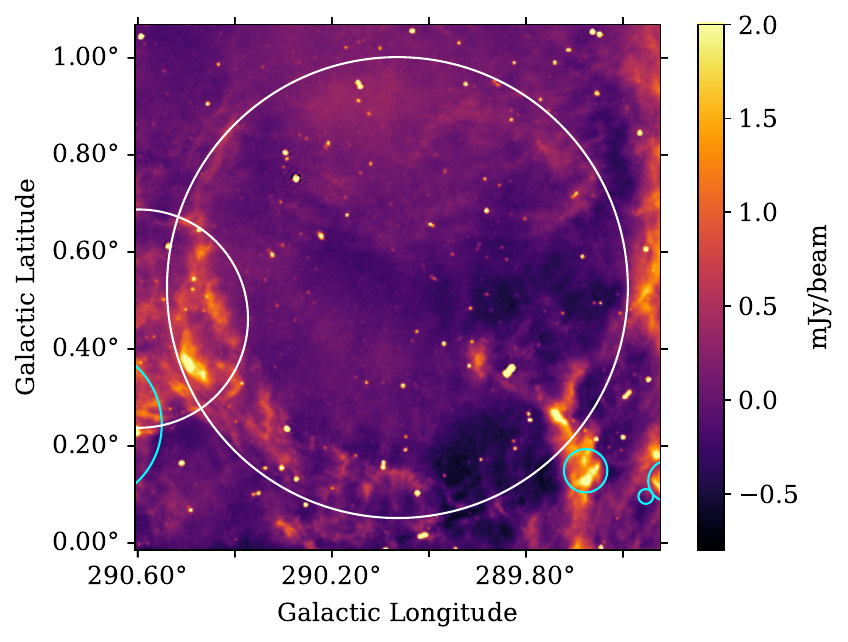}}
    \subfigure[G290.6+0.5]{\includegraphics[width=0.32\textwidth]{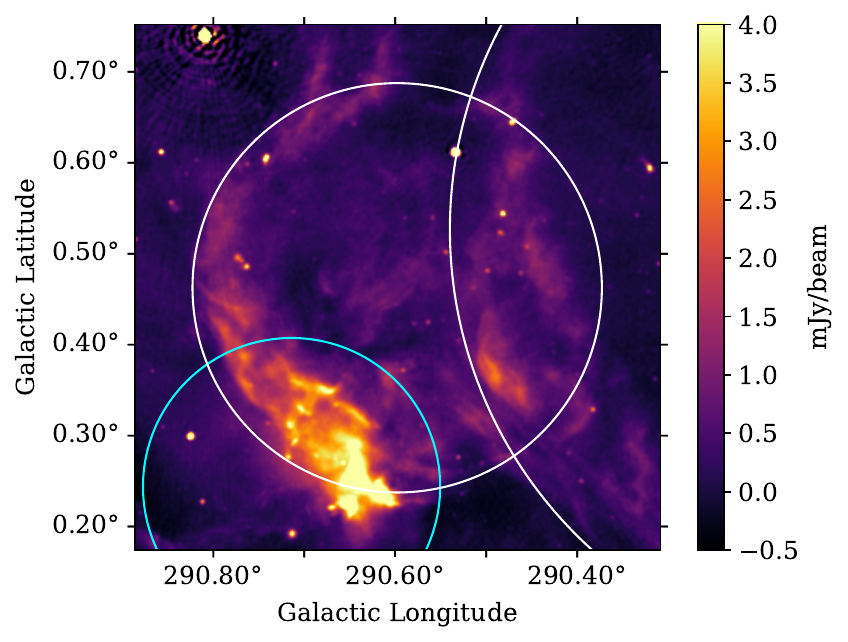}}
    \subfigure[G290.6$-$1.5]{\includegraphics[width=0.32\textwidth]{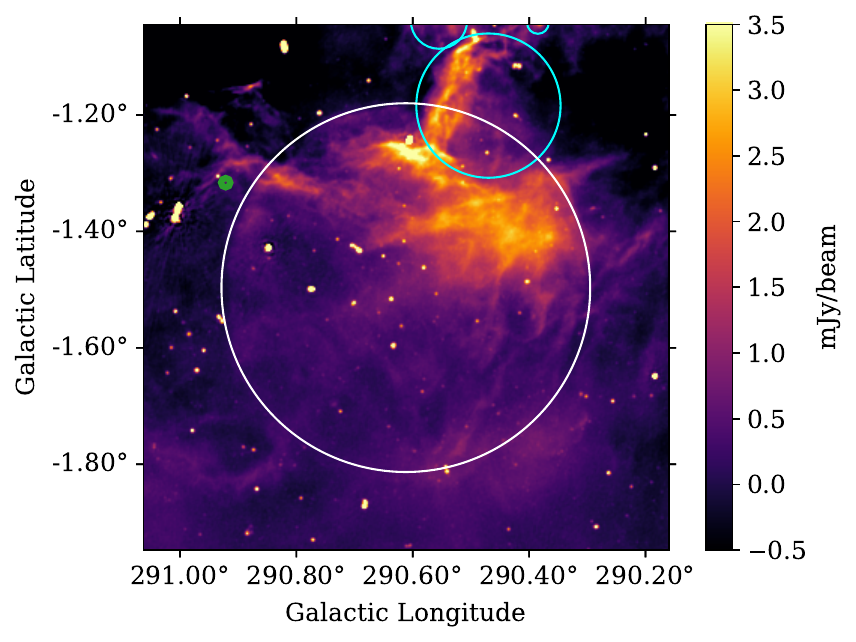}}
    \subfigure[G291.1$-$0.4]{\includegraphics[width=0.32\textwidth]{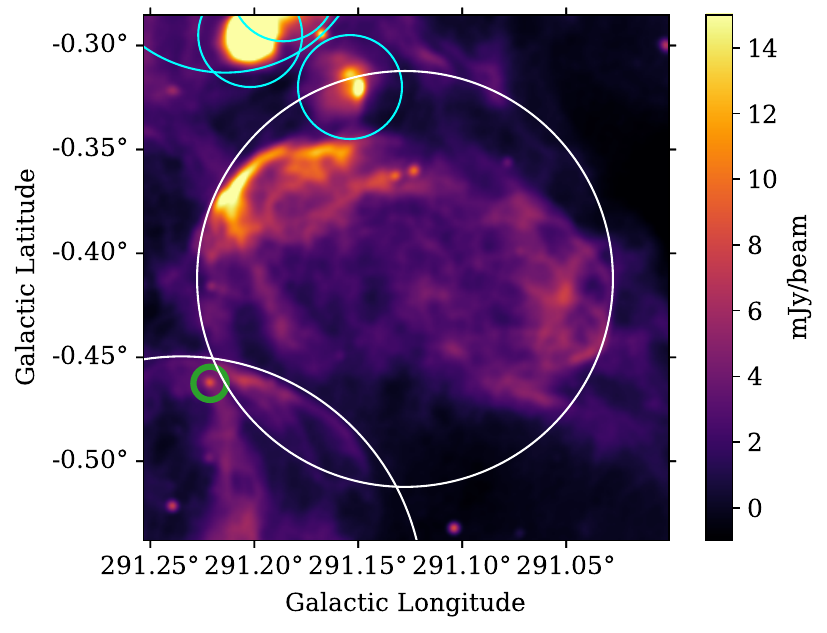}}
    \subfigure[G291.3+4.5]{\includegraphics[width=0.32\textwidth]{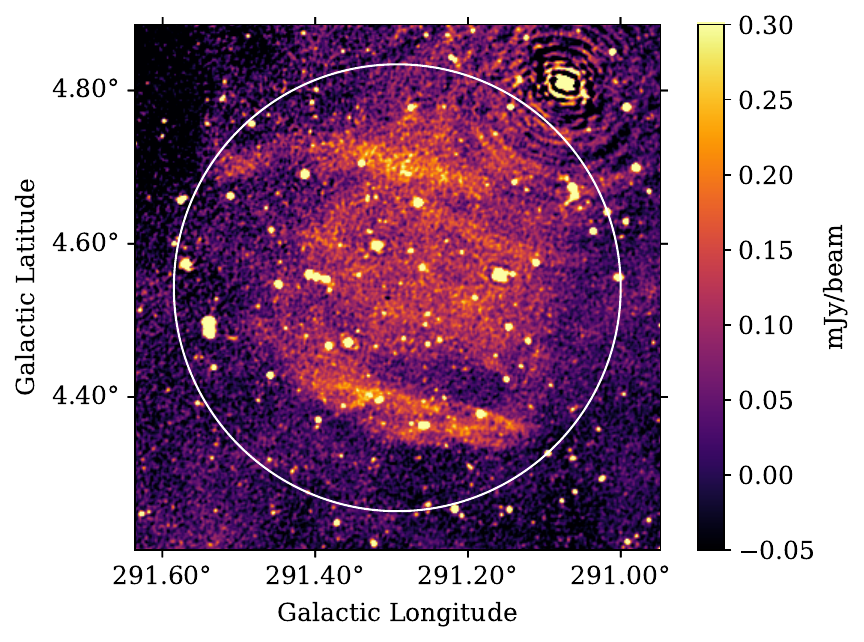}}
    \subfigure[G294.5$-$0.6]{\includegraphics[width=0.32\textwidth]{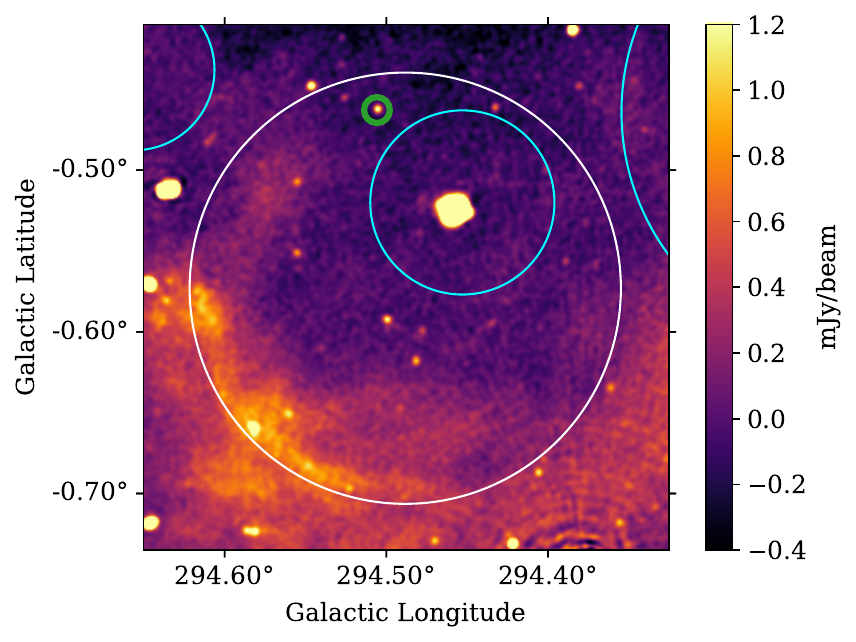}}
    \subfigure[G295.2$-$0.5]{\includegraphics[width=0.32\textwidth]{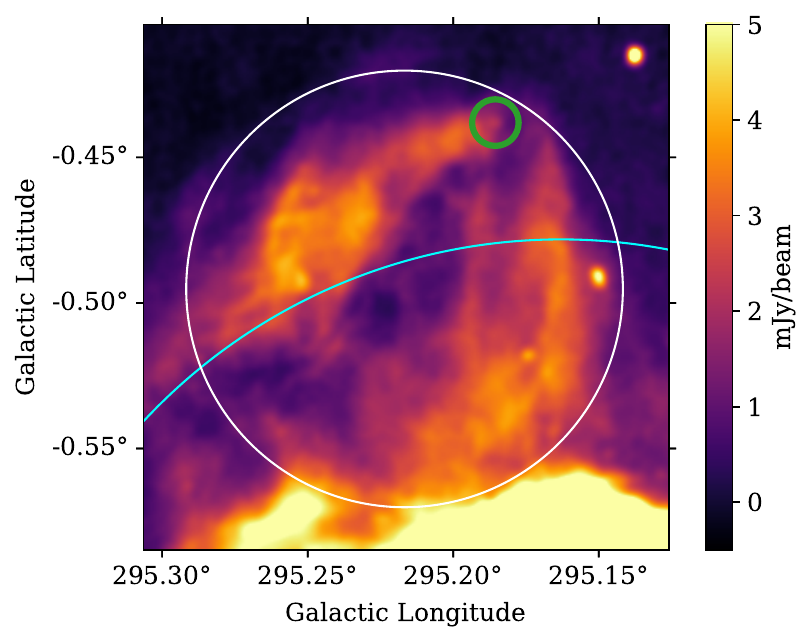}}
    \subfigure[G296.4$-$2.8]{\includegraphics[width=0.32\textwidth]{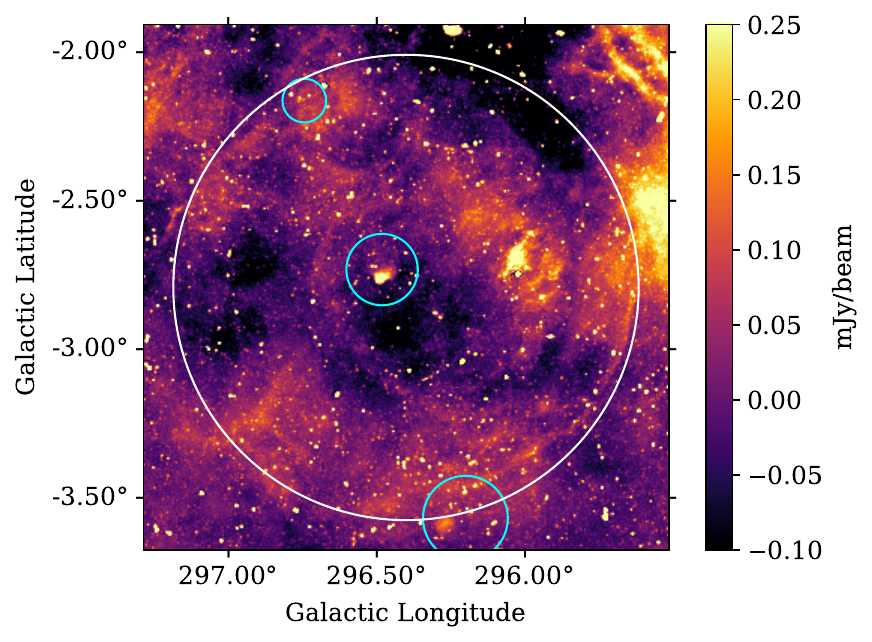}}
    \caption{Possible SNRs / Weak candidates.}
    \label{fig:weak_cands2}
\end{figure}

\begin{figure}
    \subfigure[G297.0$-$1.0]{\includegraphics[width=0.32\textwidth]{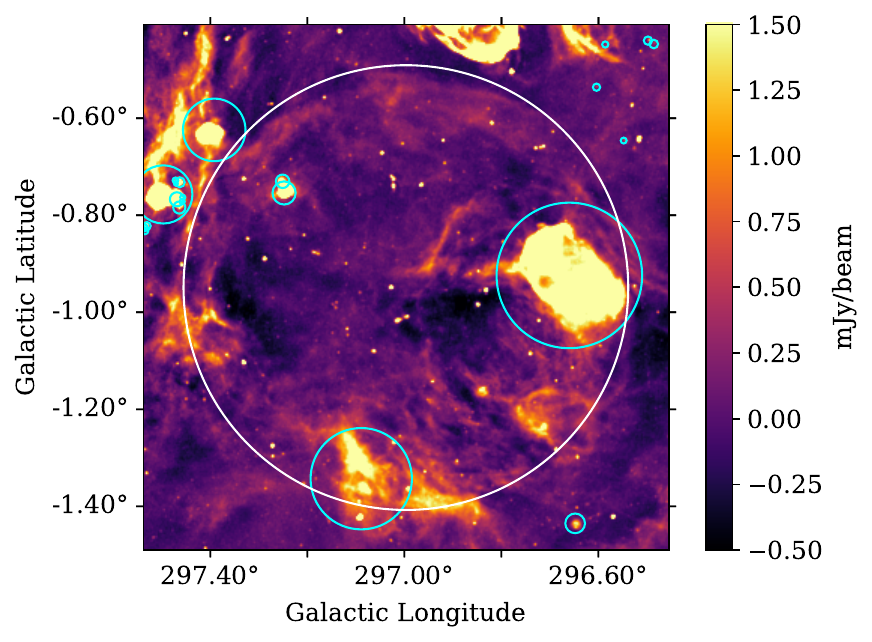}}
    \subfigure[G298.0+0.3]{\includegraphics[width=0.32\textwidth]{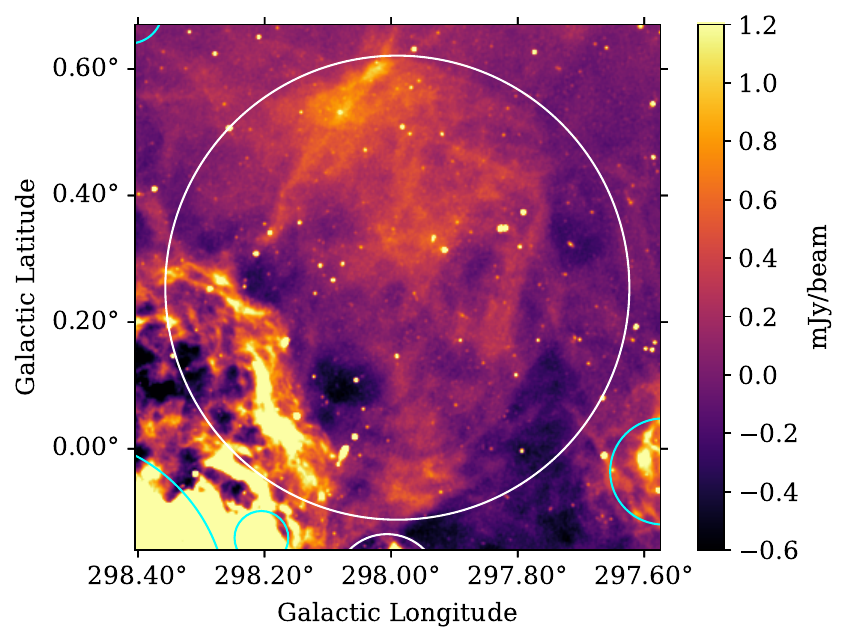}}
    \subfigure[G298.0+0.3 PI]{\includegraphics[width=0.32\textwidth]{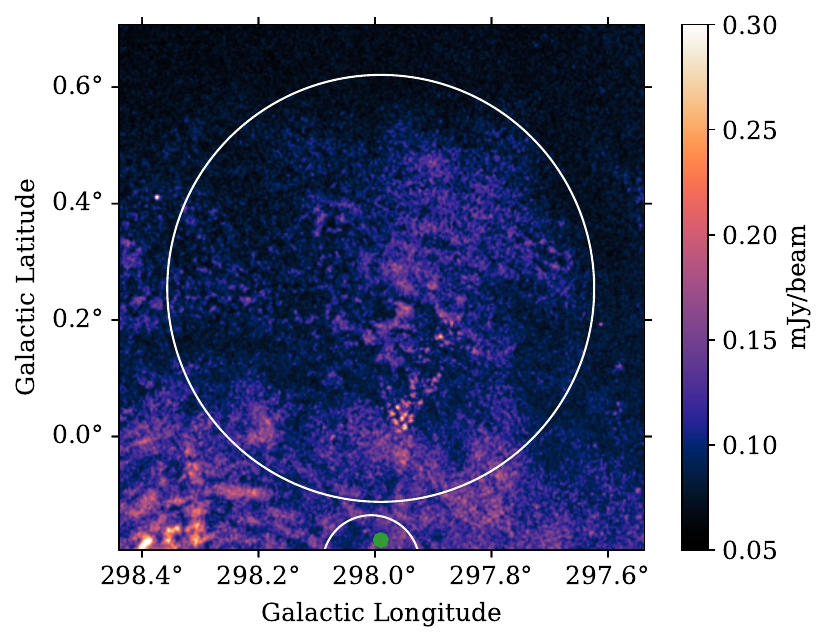}}
    \subfigure[G298.2$-$1.9]{\includegraphics[width=0.32\textwidth]{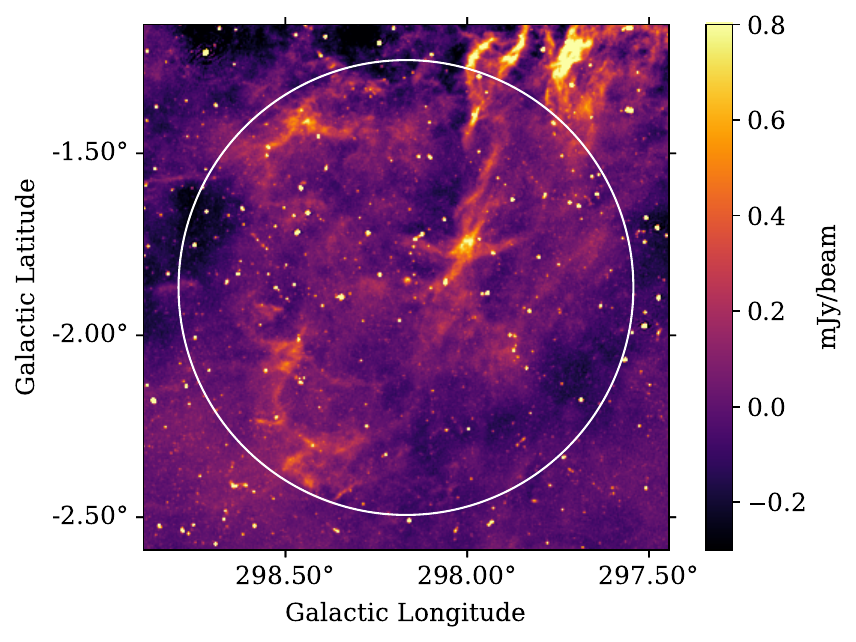}}
    \subfigure[G298.5+2.1]{\includegraphics[width=0.32\textwidth]{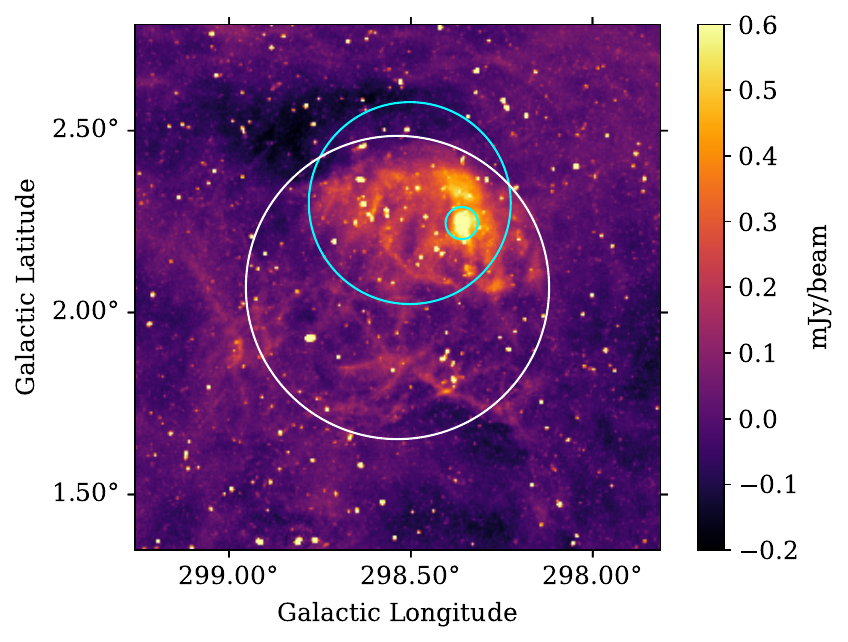}}
    \subfigure[G299.2+0.3]{\includegraphics[width=0.32\textwidth]{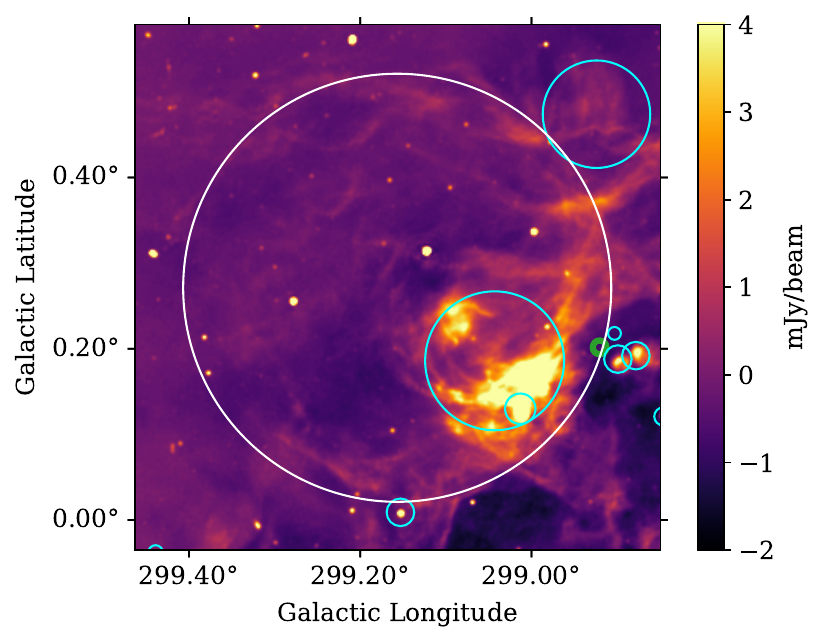}}
    \subfigure[G300.0$-$1.6]{\includegraphics[width=0.32\textwidth]{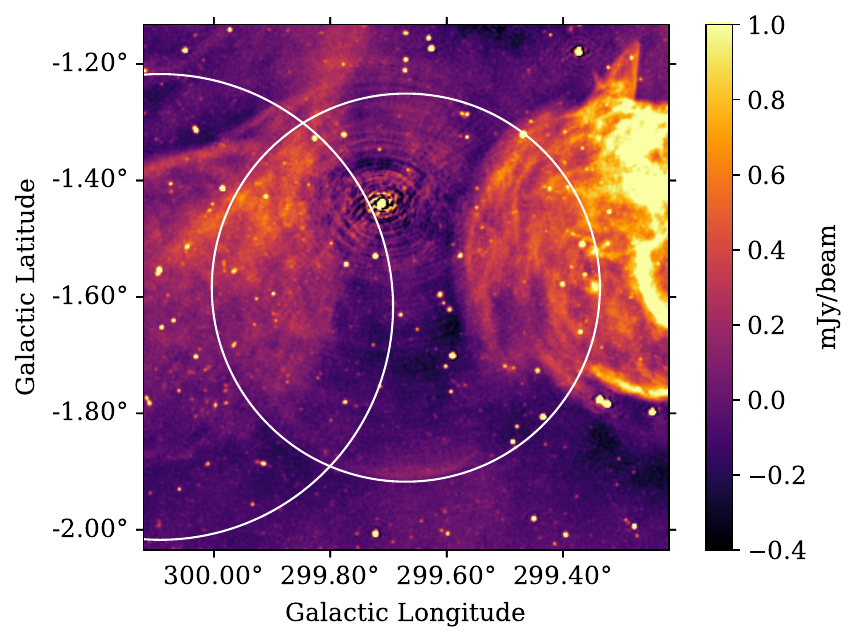}}
    \subfigure[G301.8$-$2.1]{\includegraphics[width=0.32\textwidth]{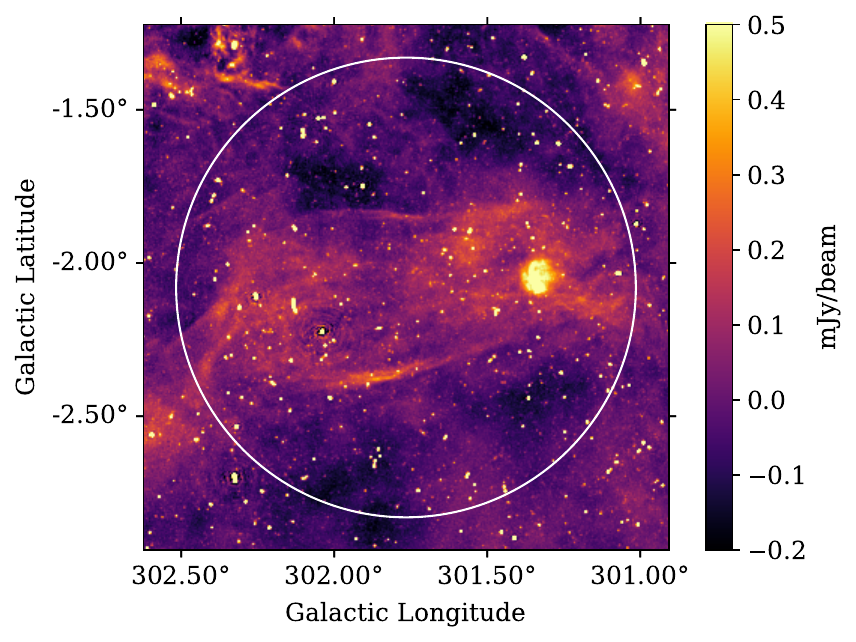}}
    \subfigure[G304.2$-$0.5]{\includegraphics[width=0.32\textwidth]{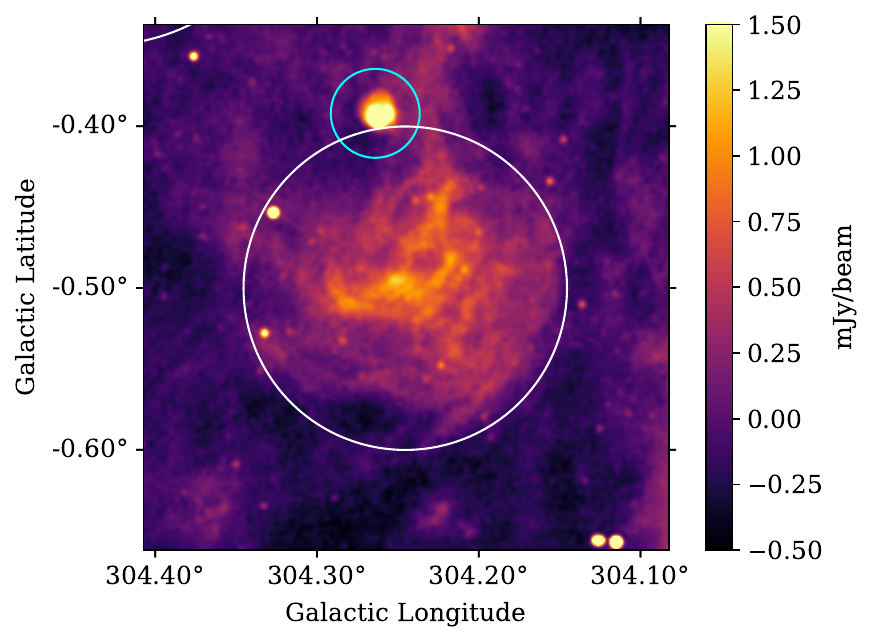}}
    \subfigure[G304.4$-$0.2]{\includegraphics[width=0.32\textwidth]{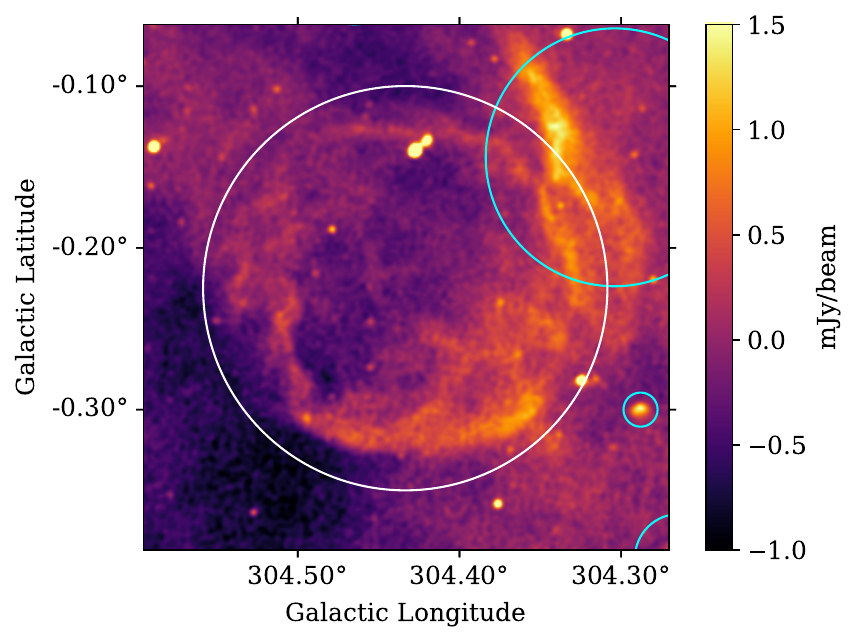}}
    \subfigure[G306.2$-$0.8]{\includegraphics[width=0.32\textwidth]{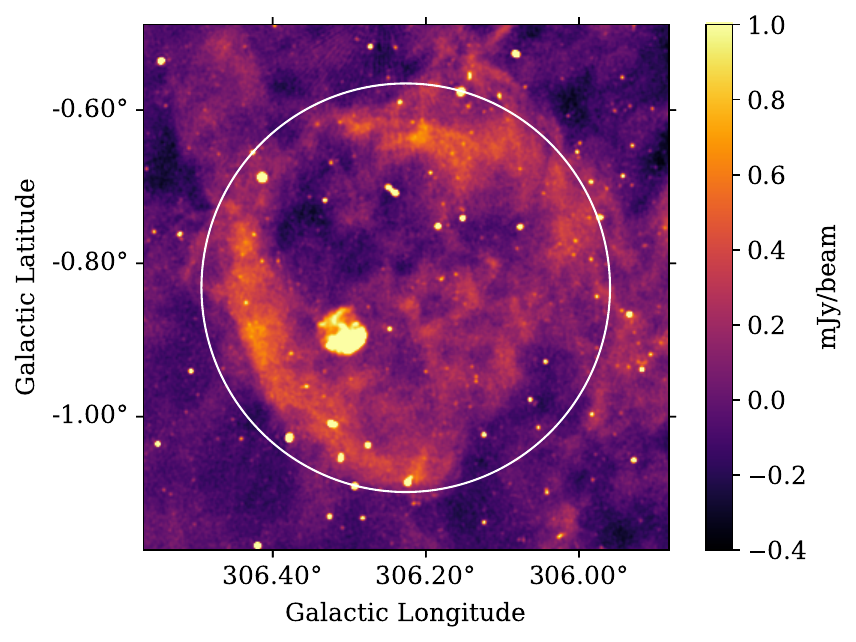}}
    \subfigure[G306.4+2.6]{\includegraphics[width=0.32\textwidth]{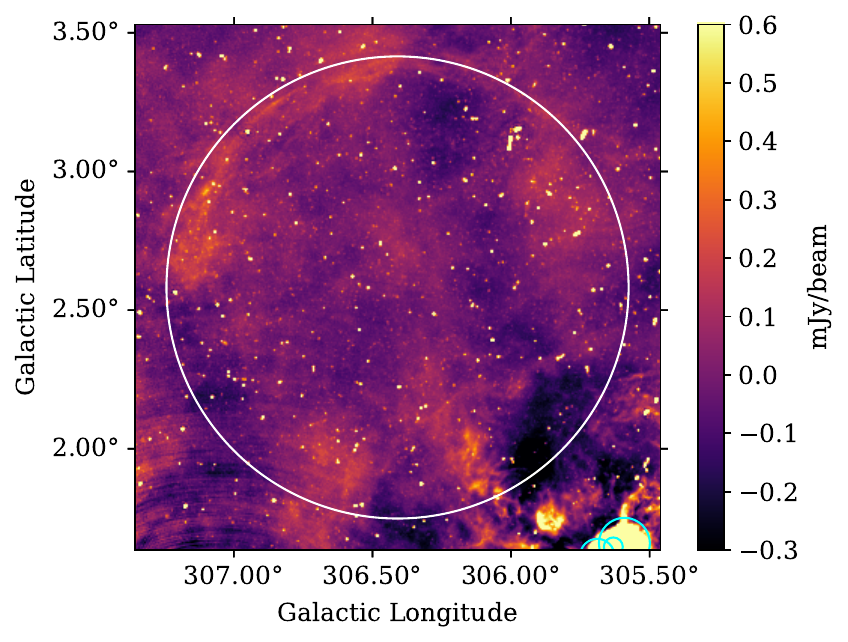}}
    \caption{Possible SNRs / Weak candidates.}
    \label{fig:weak_cands3}
\end{figure}

\begin{figure}
    \centering
    \subfigure[G307.1$-$0.7]{\includegraphics[width=0.32\textwidth]{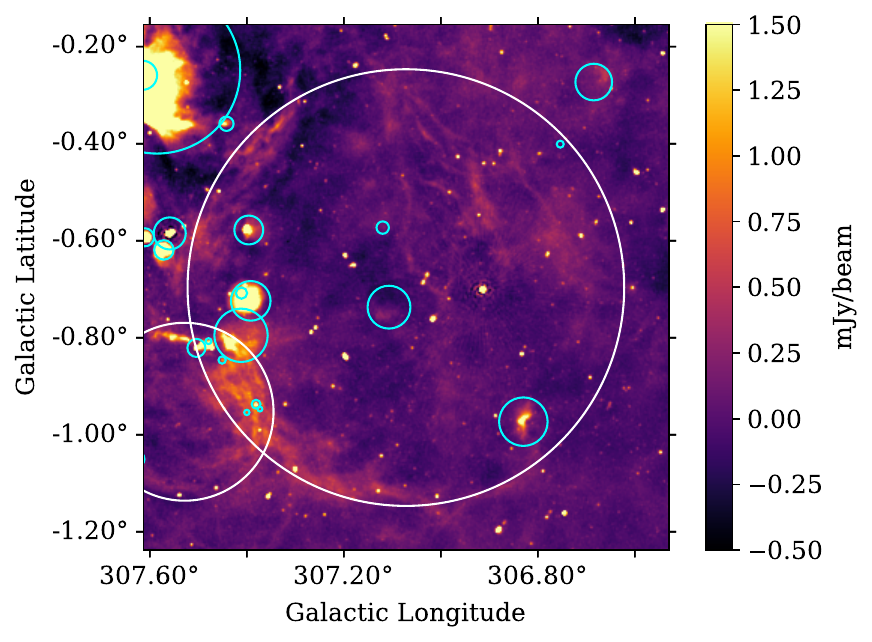}}
    \subfigure[G307.5$-$1.0]{\includegraphics[width=0.32\textwidth]{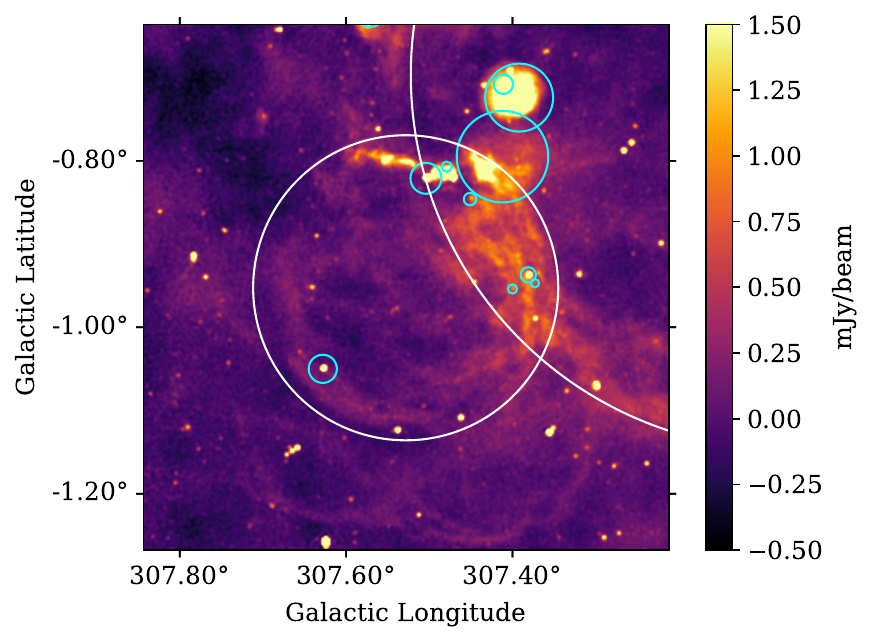}}
    \subfigure[G307.9+0.1]{\includegraphics[width=0.32\textwidth]{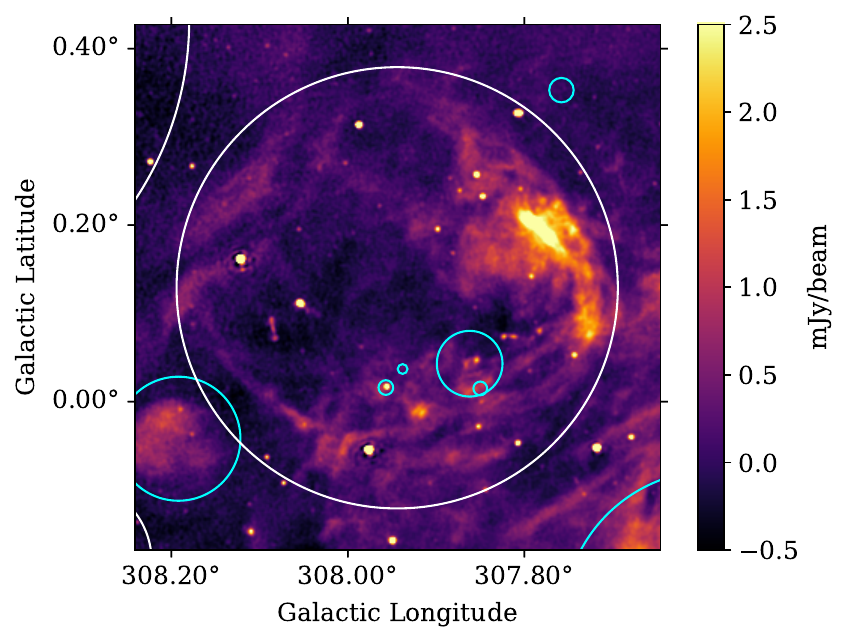}}
    \subfigure[G308.3$-$0.2]{\includegraphics[width=0.32\textwidth]{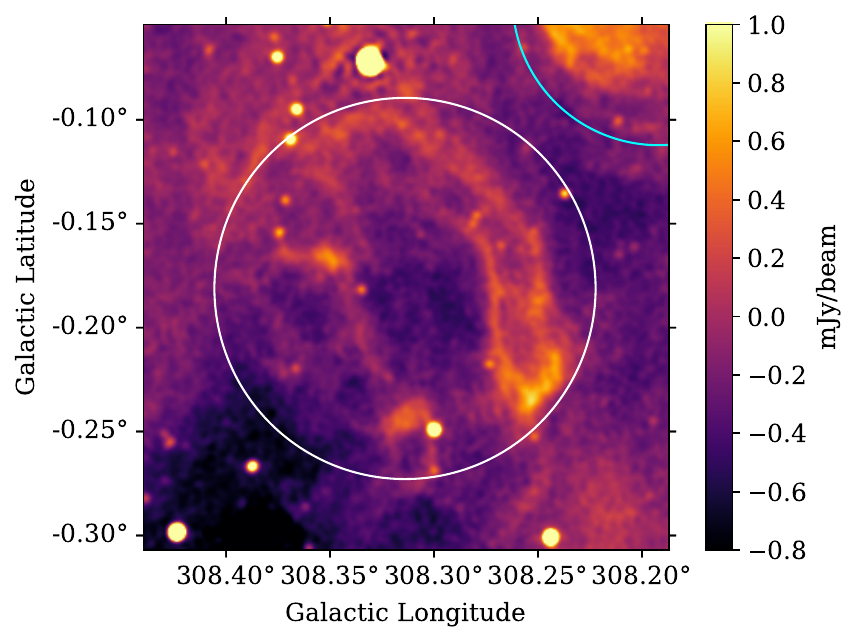}}
    \subfigure[G308.5+0.4]{\includegraphics[width=0.32\textwidth]{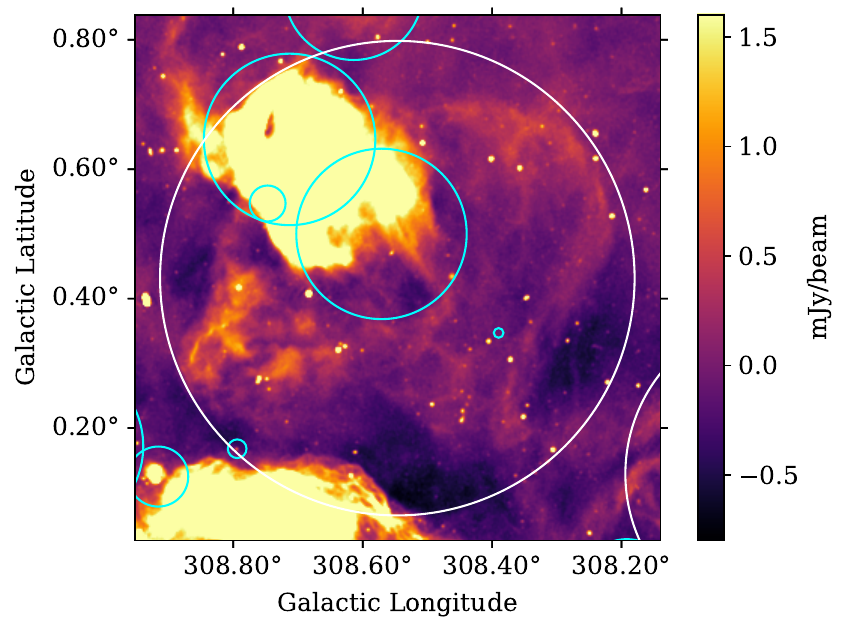}}
    \subfigure[G308.8$-$0.5]{\includegraphics[width=0.32\textwidth]{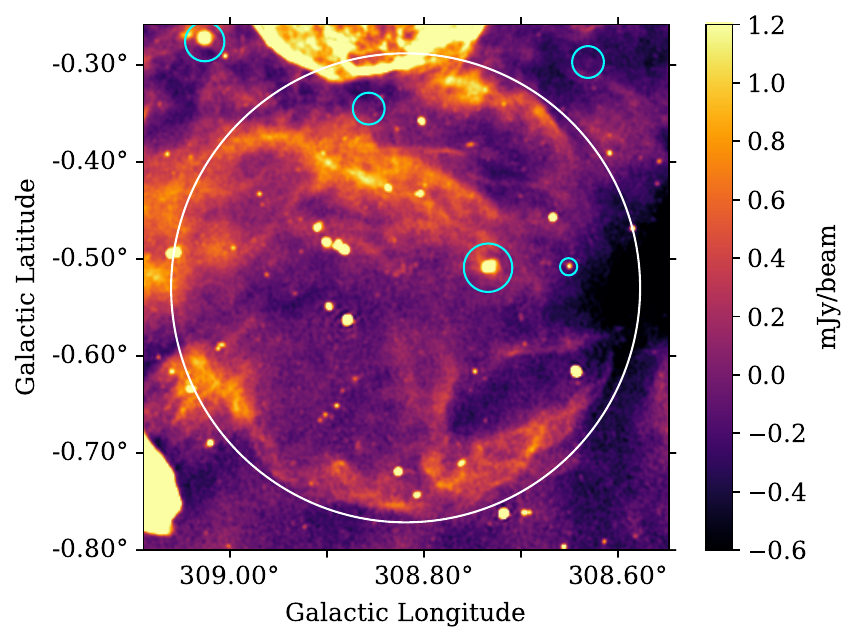}}
    \subfigure[G309.2$-$0.1]{\includegraphics[width=0.32\textwidth]{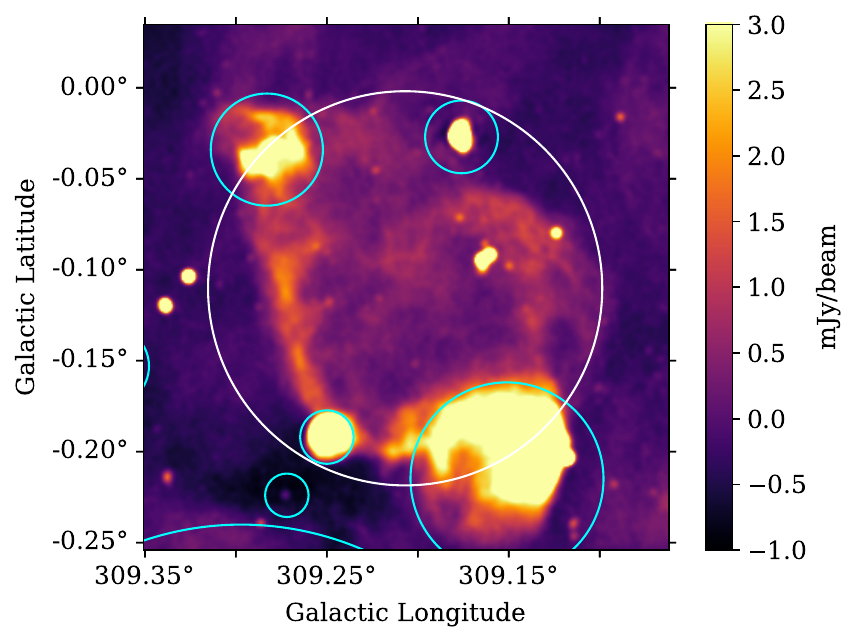}}
    \subfigure[G309.9$-$0.4]{\includegraphics[width=0.32\textwidth]{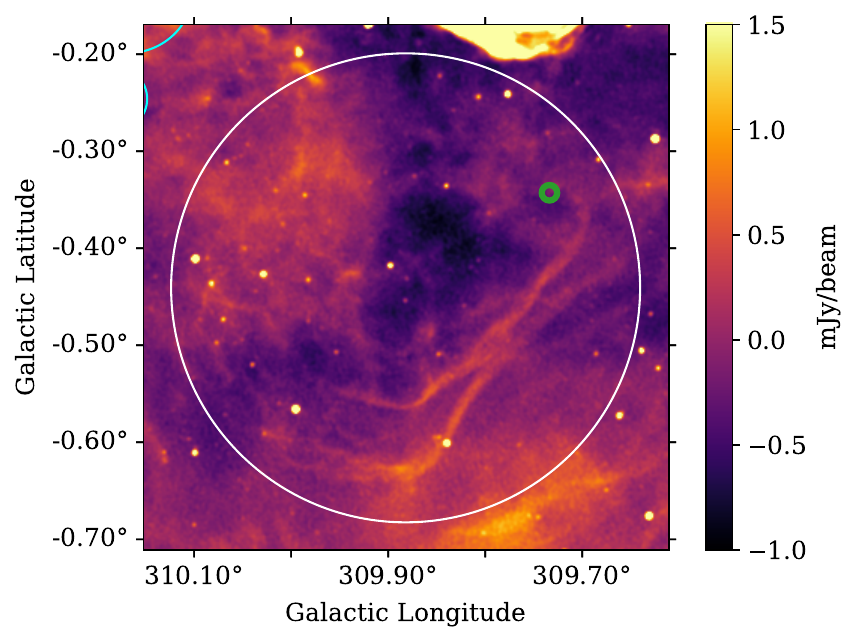}}
    \subfigure[G310.3+0.5]{\includegraphics[width=0.32\textwidth]{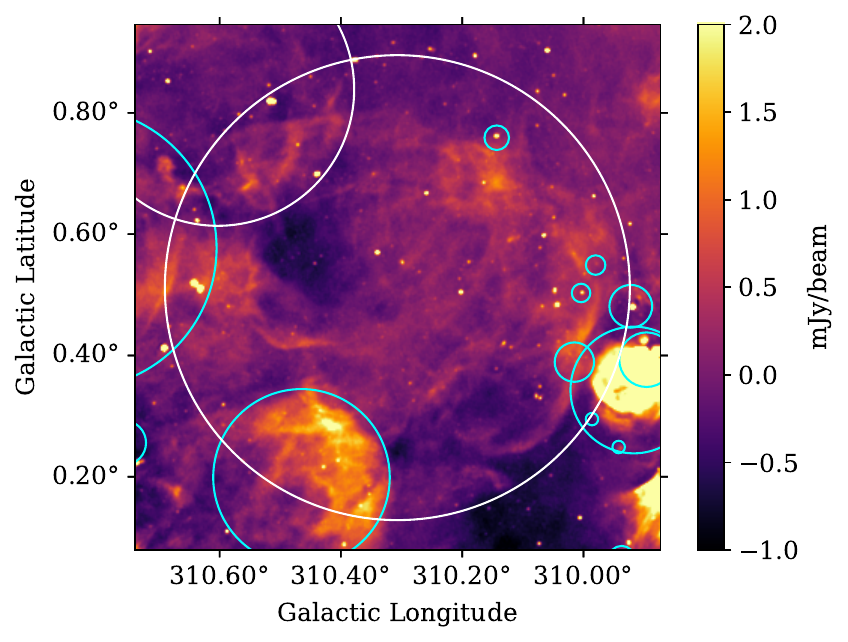}}
    \subfigure[G310.5$-$0.6]{\includegraphics[width=0.32\textwidth]{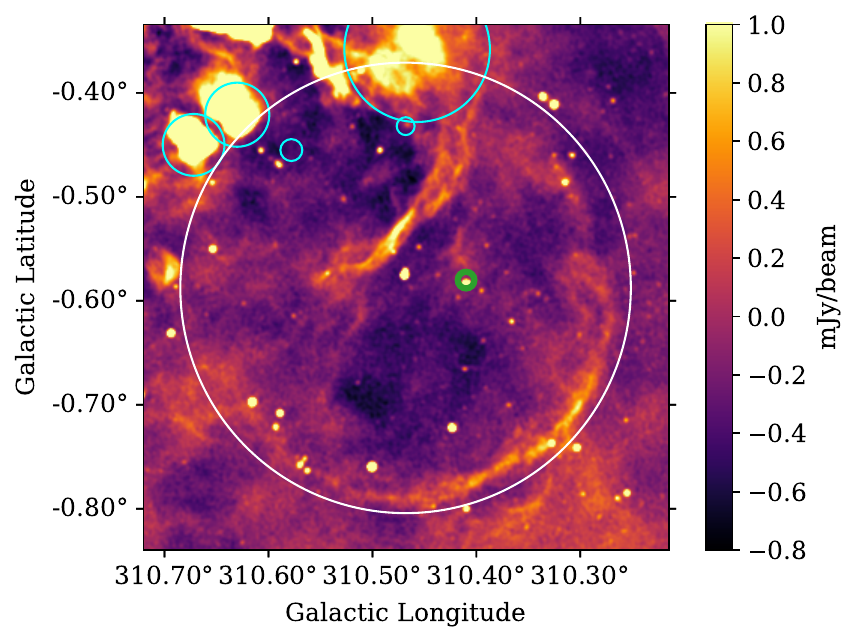}}
    \subfigure[G311.3+1.1]{\includegraphics[width=0.32\textwidth]{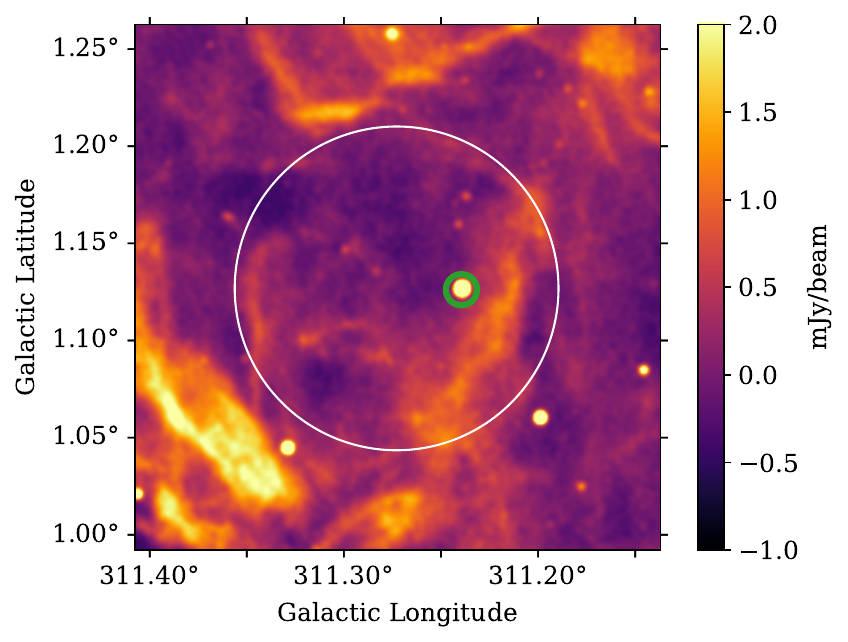}}
    \subfigure[G311.3+1.1 PI]{\includegraphics[width=0.32\textwidth]{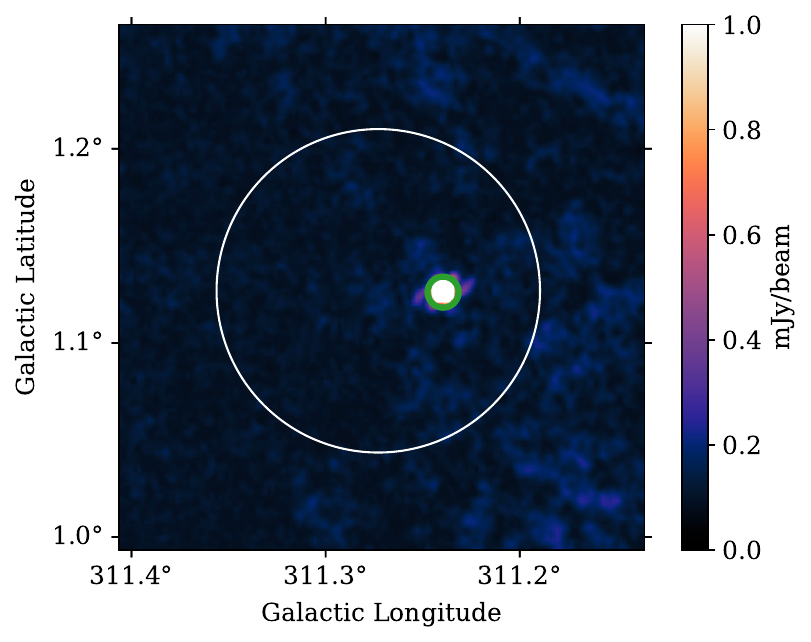}}
    \caption{Possible SNRs / Weak candidates.}
    \label{fig:weak_cands4}
\end{figure}

\begin{figure}
    \centering
    \subfigure[G311.5$-$0.1]{\includegraphics[width=0.32\textwidth]{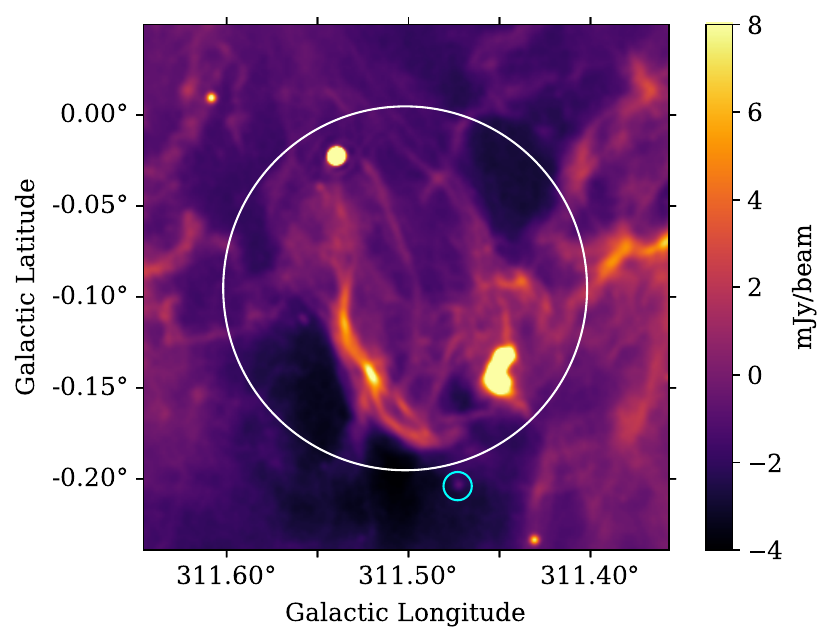}}
    \subfigure[G311.9+0.9]{\includegraphics[width=0.32\textwidth]{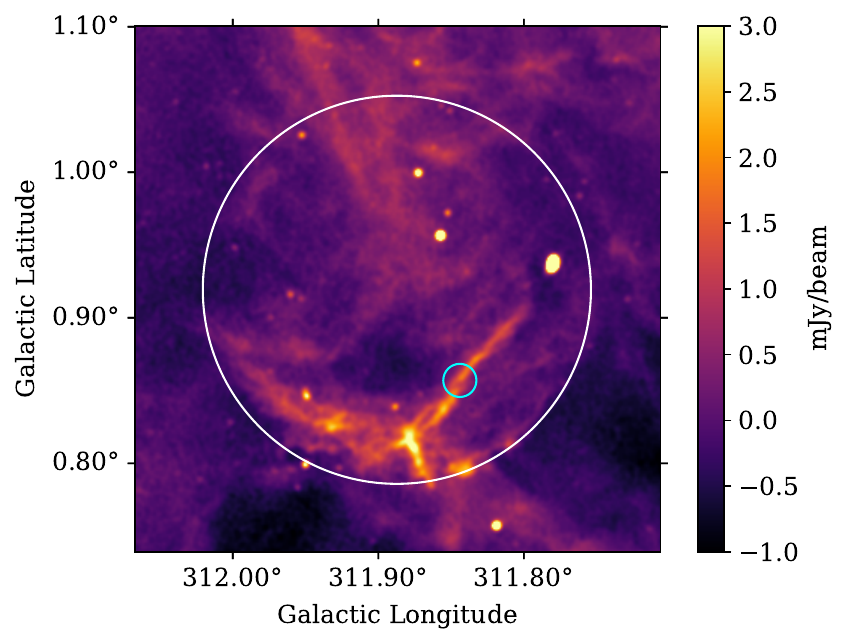}}
    \subfigure[G312.7+2.9]{\includegraphics[width=0.32\textwidth]{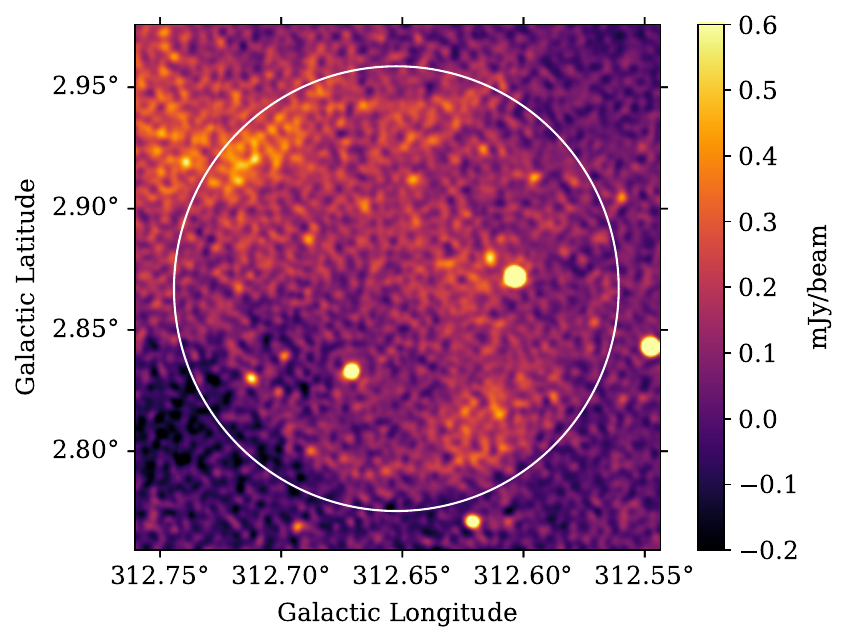}}
    \subfigure[G316.3$-$0.4]{\includegraphics[width=0.32\textwidth]{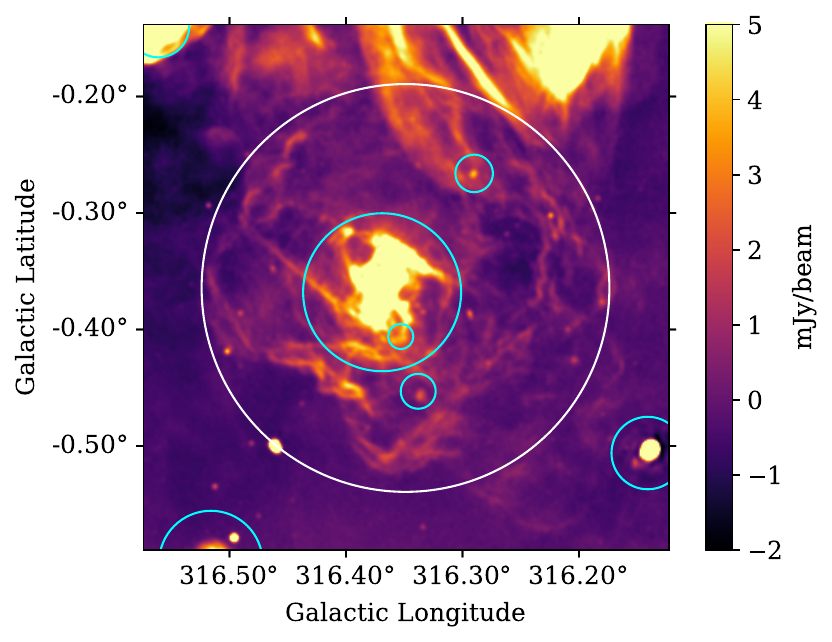}}
    \subfigure[G316.7+0.4]{\includegraphics[width=0.32\textwidth]{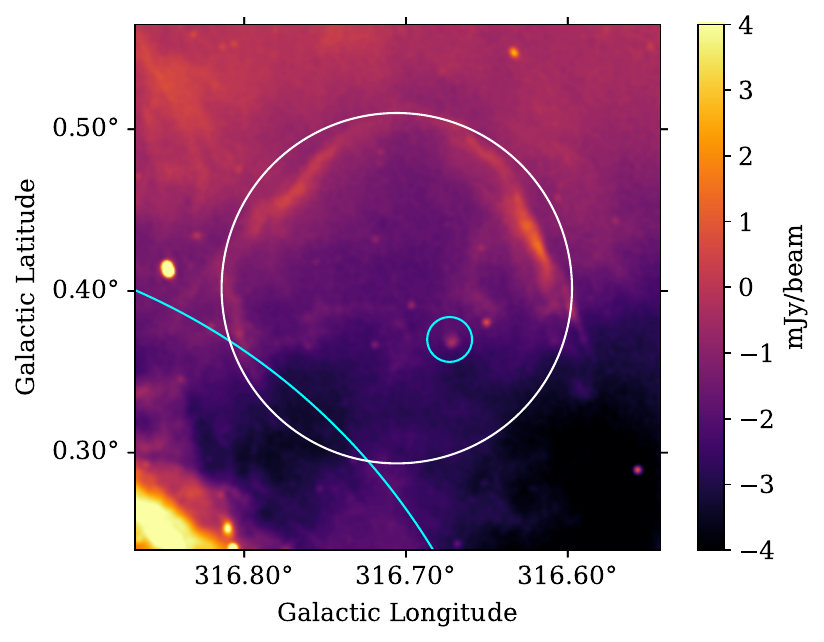}}
    \subfigure[G318.1$-$0.4]{\includegraphics[width=0.32\textwidth]{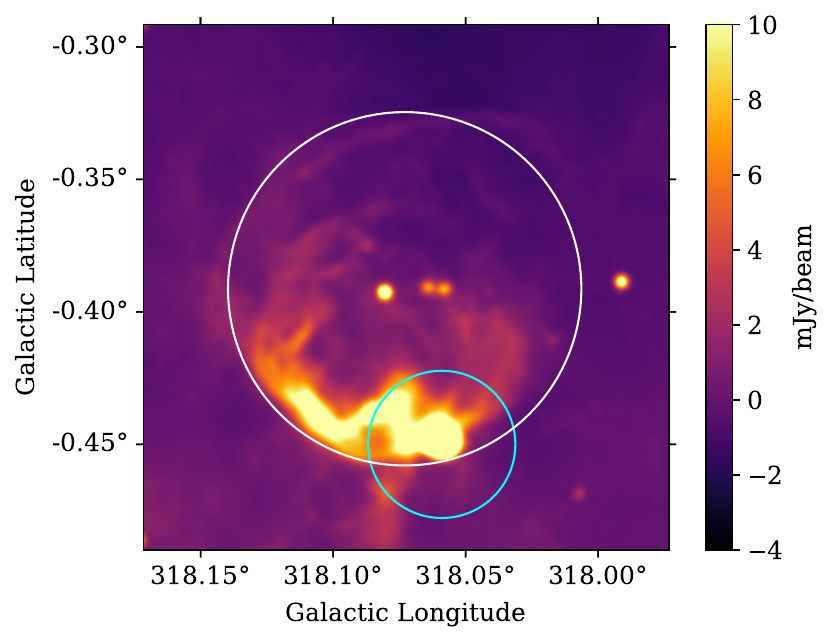}}
    \subfigure[G319.3$-$0.7]{\includegraphics[width=0.32\textwidth]{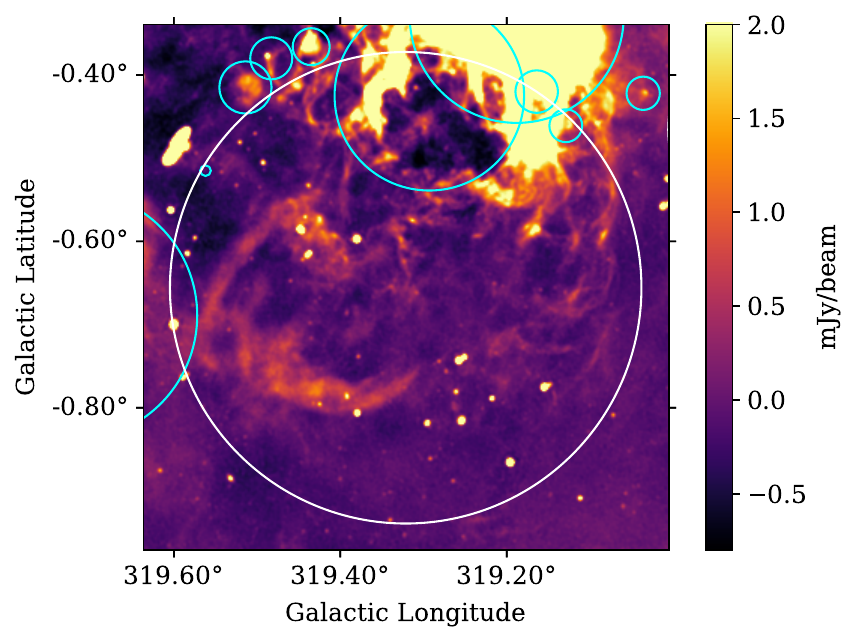}}
    \subfigure[G319.4+0.2]{\includegraphics[width=0.32\textwidth]{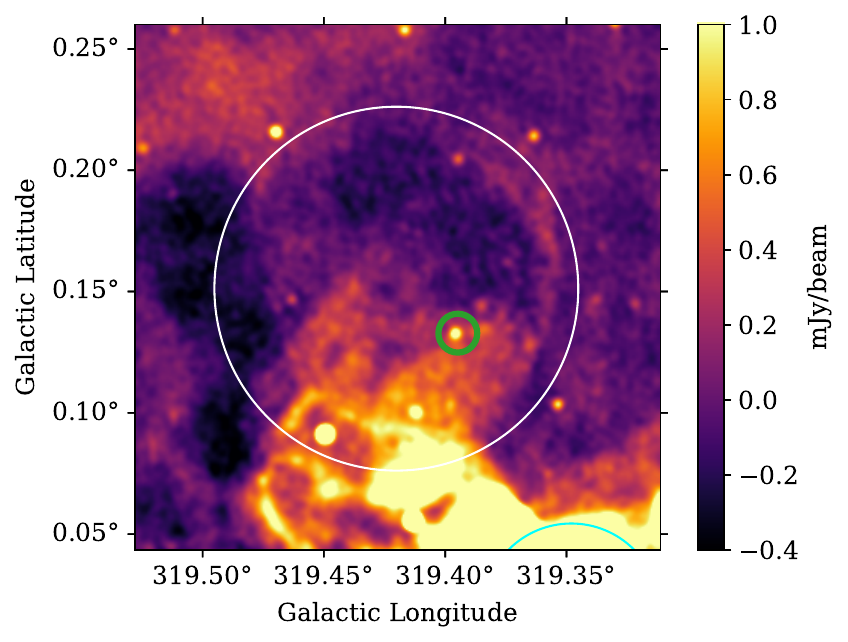}}
    \subfigure[G320.0$-$1.7]{\includegraphics[width=0.32\textwidth]{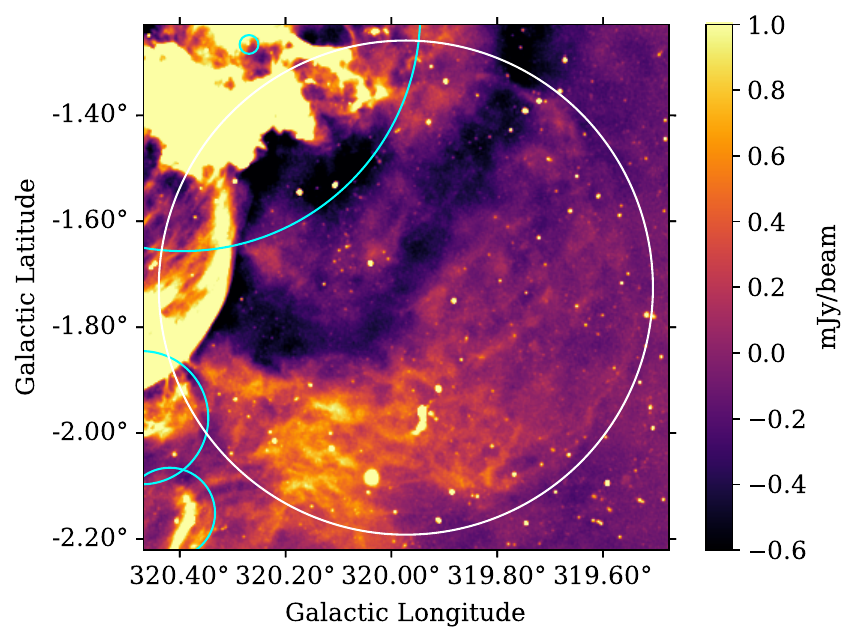}}
    \subfigure[G320.6$-$0.8]{\includegraphics[width=0.32\textwidth]{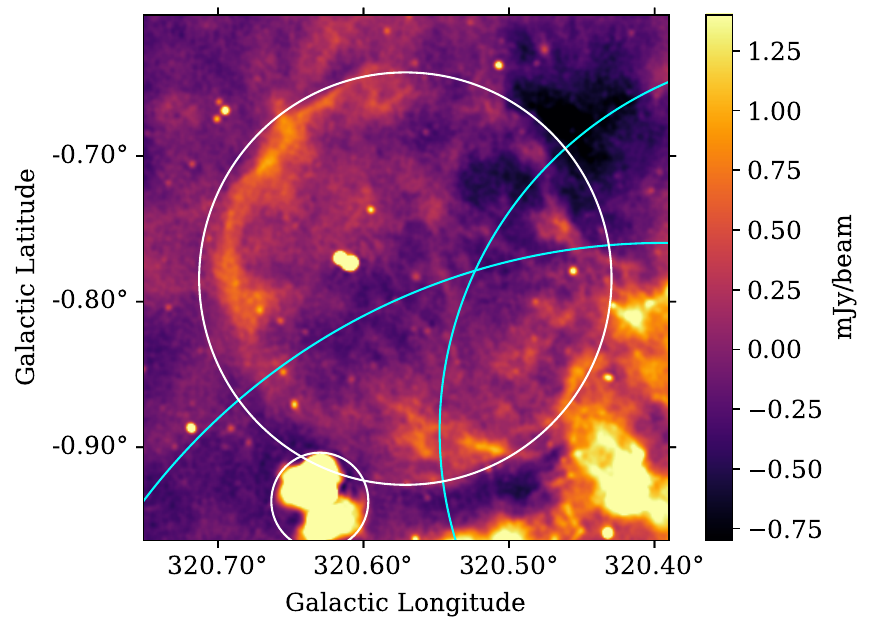}}
    \subfigure[G320.9$-$0.3]{\includegraphics[width=0.32\textwidth]{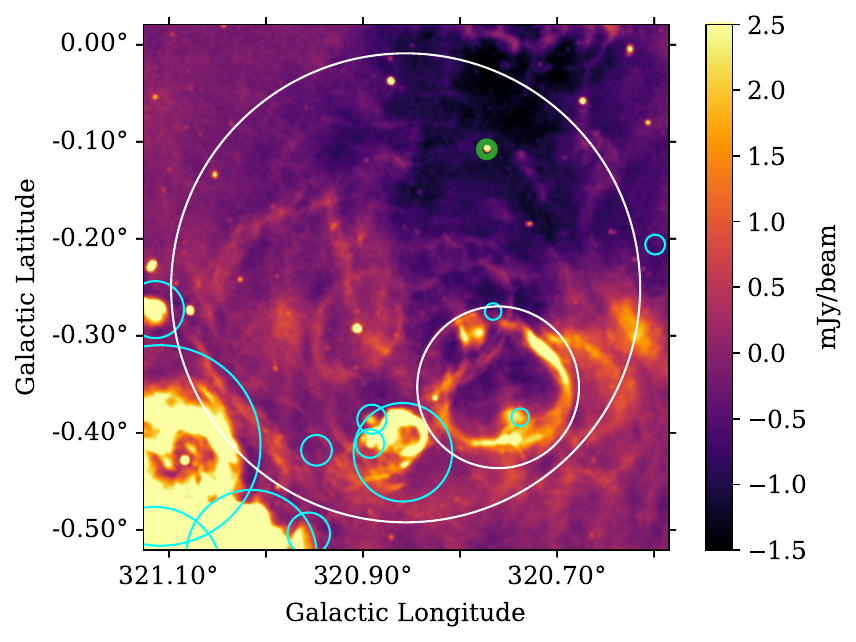}}
    \caption{Possible SNRs / Weak candidates.}
    \label{fig:weak_cands5}
\end{figure}

\bibliography{SNR_cat}{}
\bibliographystyle{aasjournal}

\end{document}